\documentclass[pdflatex,sn-mathphys-num]{sn-jnl}

\usepackage{epsfig}
\usepackage{float}
\usepackage[usenames,dvipsnames]{color}

\usepackage{amsmath,amssymb,amsfonts}
\usepackage{amsthm}
\usepackage{mathtools}      
\usepackage{mathrsfs}
\usepackage{eucal}
\usepackage{bbold}
\usepackage{bigints}
\usepackage{empheq}
\usepackage{diffcoeff}
\usepackage{upgreek}
\usepackage{siunitx}
\usepackage{wasysym}

\usepackage[utf8]{inputenc}
\usepackage[english]{babel}

\usepackage{graphicx}
\usepackage{float}
\usepackage{placeins}
\usepackage{rotating}

\usepackage{booktabs}
\usepackage{multirow}
\usepackage{tablefootnote}
\usepackage{manyfoot}

\usepackage{textcomp}
\usepackage{setspace}
\usepackage[normalem]{ulem}
\usepackage{xspace}

\usepackage{lineno}

\newcommand{\beq}{\begin{equation}}
\newcommand{\eeq}{\end{equation}}
\newcommand{\bea}{\begin{eqnarray}}
\newcommand{\eea}{\end{eqnarray}}

\newcommand{\sqrtsNN}{\sqrt{s_{\mathrm{NN}}}}

\graphicspath{{./Figures/}}

\newcommand{\figref}[1]{Fig.~\ref{#1}}
\newcommand{\Figref}[1]{Figure~\ref{#1}}

\definecolor{green}{rgb}{0,0.6,0}

\DeclareSIUnit[number-unit-product = {}]\AGeV{\ensuremath{\mathit{A}}\,GeV}
\DeclareSIUnit[number-unit-product = {}]\AGeVc{\text{\ensuremath{A}}\,GeV\!/\text{\ensuremath{c}}}
\DeclareSIUnit\GeVc{GeV\!/\text{\ensuremath{c}}}

\DeclareSIUnit\kB{\kilo\byte}
\DeclareSIUnit\MB{\mega\byte}
\DeclareSIUnit\GB{\giga\byte}
\DeclareSIUnit\TB{\tera\byte}
\DeclareSIUnit\kiB{\kibi\byte}
\DeclareSIUnit\MiB{\mebi\byte}
\DeclareSIUnit\GiB{\gibi\byte}
\DeclareSIUnit\TiB{\tebi\byte}

\DeclareSIUnit\kbit{\kilo\bit}
\DeclareSIUnit\Mbit{\mega\bit}
\DeclareSIUnit\Gbit{\giga\bit}
\DeclareSIUnit\Tbit{\tera\bit}
\DeclareSIUnit\kibit{\kibi\bit}
\DeclareSIUnit\Mibit{\mebi\bit}
\DeclareSIUnit\Gibit{\gibi\bit}
\DeclareSIUnit\Tibit{\tebi\bit}

\DeclareSIUnit[per-mode=symbol]\bps{\bit\per\second}
\DeclareSIUnit[per-mode=symbol]\kbps{\kilo\bit\per\second}
\DeclareSIUnit[per-mode=symbol]\Mbps{\mega\bit\per\second}
\DeclareSIUnit[per-mode=symbol]\Gbps{\giga\bit\per\second}
\DeclareSIUnit[per-mode=symbol]\Tbps{\tera\bit\per\second}

\DeclareSIUnit[per-mode=symbol]\Bps{\byte\per\second}
\DeclareSIUnit[per-mode=symbol]\kBps{\kilo\byte\per\second}
\DeclareSIUnit[per-mode=symbol]\MBps{\mega\byte\per\second}
\DeclareSIUnit[per-mode=symbol]\GBps{\giga\byte\per\second}
\DeclareSIUnit[per-mode=symbol]\TBps{\tera\byte\per\second}

\begin{document}

\title{CBM physics program and recent experimental developments}

\author{The CBM Collaboration\footnote{The list of CBM Collaboration members is available in Appendix A}}


\abstract{The Compressed Baryonic Matter (CBM) fixed-target experiment is under construction at the Facility for Antiproton and Ion Research. It aims to explore the phase structure of strong interaction (QCD) matter at high net-baryon densities and moderate temperatures using heavy-ion collisions in the center-of-mass energy range per nucleon pair $\sqrt{s_{\mathrm{NN}}} =$ 2.7--4.9\,GeV. Equipped with fast and radiation-hard detector systems and an advanced triggerless data acquisition scheme, CBM will collect data at interaction rates of up to 10 MHz by performing online reconstruction and event selection. This will facilitate measurements of rare probes not accessible so far in this energy range, including multi-strange hadron production and their flow coefficients, high-order net-baryon cumulants, dileptons, as well as production of double-strange hypernuclei. The CBM detector is outlined and selected physics performance results based on realistic simulations are discussed.
}


\maketitle

\section{Introduction}

The Compressed Baryonic Matter (CBM) experiment~\cite{CBM:2016kpk} is designed to explore the Quantum Chromodynamics (QCD) phase diagram~\cite{Andronic:2017pug,Du:2024wjm,Borsanyi:2025ttb} in the region of high baryochemical potential $\mu_B$ (net-baryon densities) with unprecedented precision and statistics. Measurements of heavy-ion collisions in the energy range $\sqrt{s_{\mathrm{NN}}} = 2.7 - 4.9$\,GeV will be the main focus of CBM, whereas proton beams will also be employed.
The richness of phenomena in this regime of hot and compressed QCD matter~\cite{Friman:2011zz,Bzdak:2019pkr} is unique and, despite earlier and ongoing efforts, remains not fully explored.
Precision measurements are envisaged with CBM for a full suite of observables, including:
i) the chemical freeze-out line ($T-\mu_B$)~\cite{Andronic:2017pug} via (multi-strange) hadron production; ii) the Equation of State (EoS) ~\cite{Sorensen:2023zkk} via particle production, collective flow, and vorticity; iii) (partial) chiral symmetry restoration and the  temperature of the fireball via dilepton production~\cite{Rapp:1999ej,Salabura:2020tou,Geurts:2022xmk};
iv) the critical point~\cite{Fischer:2026uni} via high-order net-baryon cumulants~\cite{Braun-Munzinger:2026krf}. 
The discovery of doubly-strange hypernuclei, whose production is expected to be enhanced in the CBM energy range~\cite{Andronic:2010qu,Balassa2023}, will be within reach for CBM.

The installation of the CBM experimental infrastructure commenced in 2023 and will proceed through 2027, culminating in the installation of the detectors and the start of global commissioning. The CBM experiment will operate at the SIS100 synchrotron, the central accelerator of the FAIR (Facility for Antiproton and Ion Research) complex in Darmstadt. First beams from SIS100 and the start of CBM operations are expected in late 2028.

\section{The CBM Experiment at FAIR}

The FAIR facility will provide unique opportunities for research in nuclear, hadron, atomic and plasma physics. Within this broad scientific program, the study of compressed baryonic matter is one of its major research pillars. The research program will begin with high-intensity beams from the SIS100 synchrotron, providing protons up to $\sqrtsNN = 7.6$ GeV, gold nuclei up to \SI{4.9}{\GeV}, and ions with $Z/A = 0.5$ up to \SI{5.5}{\GeV}, as part of the FAIR Modularized Start Version. Future upgrades may further extend these capabilities to higher beam energies and intensities. The extracted beams delivered to the CBM experimental cave will reach intensities up to \num{e11} protons and \num{e9} Au ions per second, with excellent beam quality characterised by a halo level below $10^{-5}$ beyond 5 mm from the beam axis and spill intensity fluctuations below 50\% down to time scales of several tens of nanoseconds~\cite{FAIR:operation_modes}. Segmented targets with a nuclear-interaction probability of 1\% are envisaged.

\begin{figure}[htb]
    \centering
    \includegraphics[width=\linewidth]{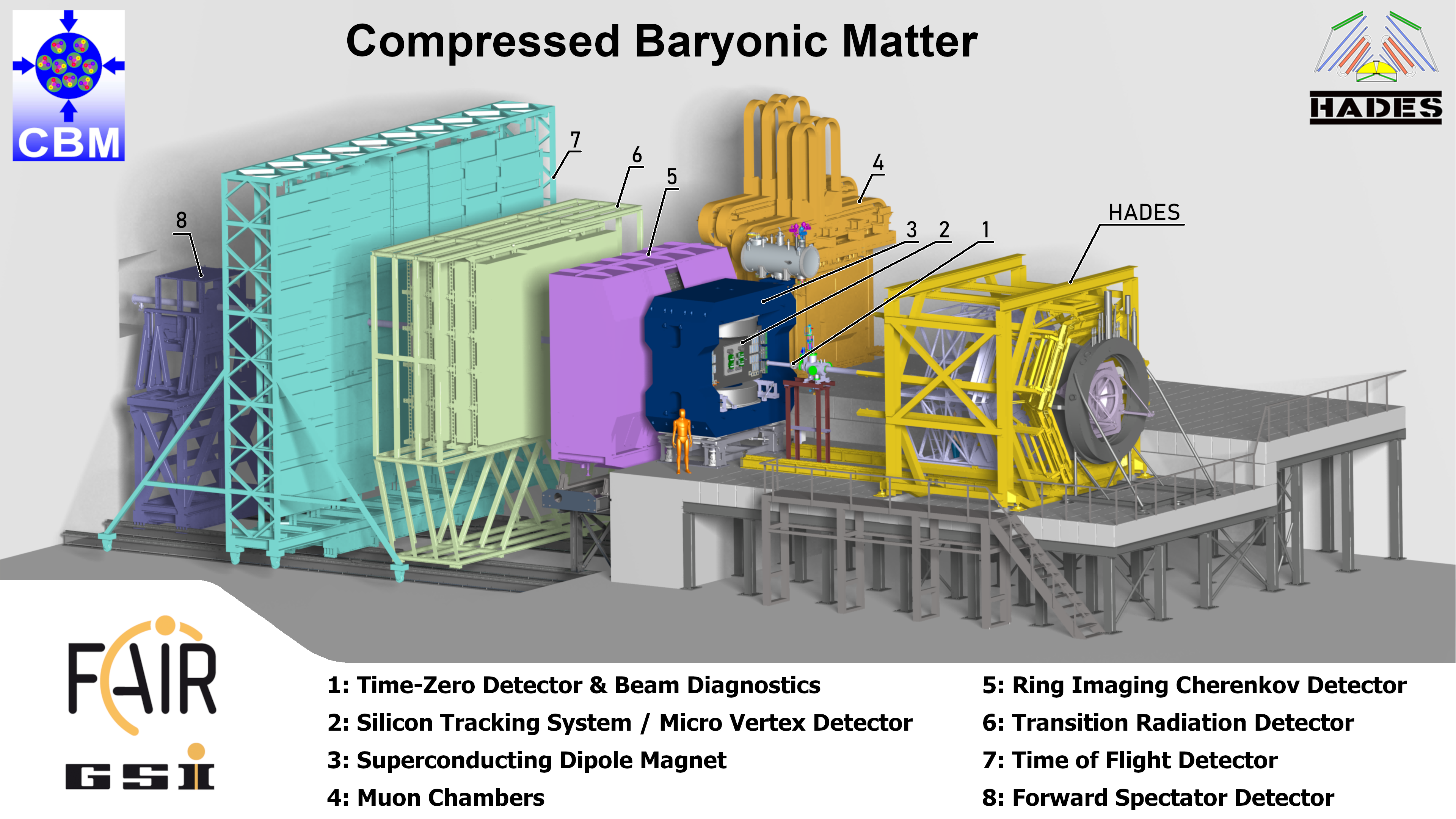}
    \caption{Schematic view of the FAIR experimental cave hosting the CBM experiment, with its detector components marked and described in the legend. The independent HADES experiment, planed for measurements at the lower end of the FAIR energy range, is shown too.
    }
    \label{fig:cbm}
\end{figure}

CBM's experimental strategy relies on high-rate measurements of rare probes in nucleus--nucleus, proton--nucleus, and proton--proton collisions over a broad range of beam energies. 
To identify and reconstruct, alongside the more abundant hadrons, the rare probes of hot and dense QCD matter, the CBM experiment, pictured in\,\figref{fig:cbm}, will employ a continuous, self-triggered readout of all the detectors combined with the fast online event reconstruction and selection on a high-performance computing farm. This represents a shift away from traditional hierarchical trigger systems toward real-time, software-based data selection~\cite{cbm:time-based,cbm:online-ev-reco, Friese_2017}, an approach increasingly adopted in modern high-rate experiments.

\subsection{Superconducting Dipole Magnet}

The experimental setup (see \figref{fig:cbm}) is built around a $\sim$2\,m long large-aperture dipole magnet, which provides a field integral of 1\,Tm along straight trajectories within $\pm0.5$\,m around the magnet centre. This configuration ensures excellent momentum resolution for charged-particle tracking at FAIR energies. The magnet gap, with a height of 147\,cm and a width of 330\,cm, provides the aperture required to accommodate the Silicon Tracking System, with its acceptance of $\pm25^{\circ}$ vertically and $\pm30^{\circ}$ horizontally with respect to the nominal target position. The total weight of the magnet is about 160\,t. The yoke is equipped with field clamps to reduce the stray field in the downstream region of the Ring-Imaging Cherenkov photodetector to an acceptable level below 10\,mT.

Cooling of the coils is performed by a liquid-helium thermosiphon system based on the natural circulation of helium. The flow is driven by the density difference between the heated fluid, containing a gaseous component in the return line, and the colder fluid in the supply line. The circulation can be additionally supported by a heater in the return line. A passive quench protection system in the form of cold rectifier diodes—cryogenic bypass diodes that allow the current to circumvent a quenched coil—is foreseen for the CBM magnet coils. Each coil is subdivided into four sections, with each section protected by four pairs of back-to-back diodes. This redundant diode configuration accounts for two possible diode failure scenarios in both current directions. The mechanical support structure of the coils must securely fix the coils and compensate for the Lorentz force of about 3\,MN (equivalent to a weight of about 300\,t), while simultaneously minimizing heat transfer between the cold mass (at liquid-helium temperature) and the coil cryostat at room temperature. This support structure is realized using six tension rods and twelve bushings. The tension rods fix the coils in position, while the bushings compensate the compressive component of the Lorentz force.

The design and construction of the magnet is performed by Bilfinger Nuclear and Energy Transition (BNET, W\"{u}rzburg) in collaboration with GSI. At the time of the publication, the magnet has successfully passed its final design reviews, marking the transition from design to production. Manufacturing of the yoke, coil, and magnet support structures is ongoing, with the completion expected in 2026.

\subsection{Beam Monitoring: T0 and HALO detectors}
The Beam Monitoring System (BMON), installed upstream of the target, comprises a fast T0 detector and a beam halo (HALO) detector. The primary goal of the former is to provide a precise start-time reference for time-of-flight measurements, while the latter continuously monitors the beam quality and stability.
The T0 detector is based on radiation-hard CVD diamond sensors with active area of $10\times10$\,mm$^2$ and thickness of 60\,\textmu m. The sensors have been extensively investigated with respect to their radiation hardness and timing performance \cite{bmon1,bmon2}.
The system is designed to achieve a time resolution better than 50\,ps. To mitigate radiation-induced performance degradation, we employ a dedicated amplification system originally developed for LGAD sensors, which significantly extends the operational lifetime of diamond detectors \cite{Pietraszko:2025kxv}.
The T0 system must also operate efficiently at beam intensity of up to 10$^7$ particles/s, and maintain stable performance over extended running periods. The HALO detector, arranged around the beam rather than in-beam, monitors the transverse beam profile and halo intensity with sufficient time and position resolution to ensure that the halo fraction remains below 0.001\% at 0.5\,cm from the beam core. Both subsystems are realized using polycrystalline CVD diamond sensors, which provide excellent timing performance, high rate capability, and the radiation hardness necessary for long-term operation under intense heavy-ion and proton beams.

\subsection{Tracking detectors}

Within the magnetic-field volume, a sequence of precision tracking detectors provides the space points required for the reconstruction of charged-particle trajectories over a distance of approximately one metre downstream of the target.

\subsubsection{Micro-Vertex Detector}
 
Closest to the interaction point and still inside the target vacuum, the Micro-Vertex Detector (MVD) employs 288 thin, large-area monolithic active pixel sensors arranged in four equidistant planar detector stations, with the first station placed only 8\,\unit{\centi\metre} downstream the target~\cite{tdr:mvd}. The detector is optimized for achieving excellent vertexing precision of better than 100\,\unit{\micro\metre} along the beam axis by minimizing the material budget per station to $0.3$--$0.5\%\,X_0$~\cite{Matejcek:2025fdh}.\par
\begin{figure}[htb]
    \centering   \includegraphics[width=0.57\linewidth]{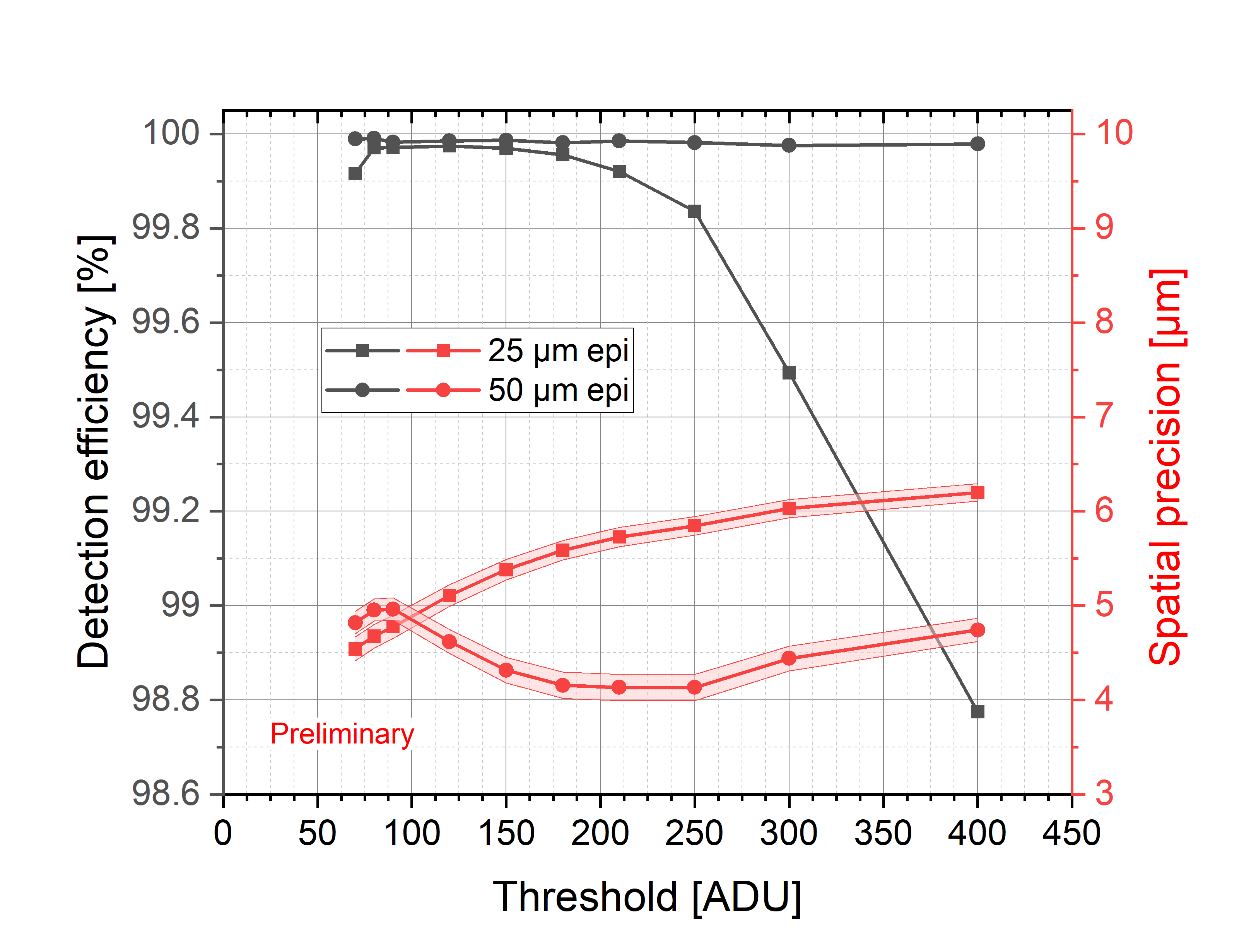}
    \caption{Detection efficiency and spatial precision of minimum ionizing particles measured with MIMOSIS-2.1, employing pixel candidates referred to as AC p-stop with 25 and 50\,\unit{\micro\metre} thick active medium (epi)~\cite{Deveaux:2025sxl}.}
    \label{fig:MVD-1}
\end{figure}

 
The final sensor called MIMOSIS~\cite{Deveaux:2025sxl, Darwish:2025ega} will combine a spatial precision of $\sim$5\,\unit{\micro\metre} and a time resolution of $\sim$5\,\unit{\micro\second}, featuring 1024$\times$504 top-biased, fully-depleted pixels of $27\times30$\,\unit{\micro\square\metre} size and relying on the TowerJazz 180\,\unit{\nano\metre} CMOS imaging process. 
Figure~\ref{fig:MVD-1} presents detection efficiency and spatial precision of minimum ionizing particles measured with the sensor prototype MIMOSIS-2.1. MIMOSIS has to withstand a peak rate of  80\,\unit{\mega\hertz\per\centi\square\meter} and a combined radiation dose of $5\,\mathrm{MRad}$ and $10^{14}\,\mathrm{n_{eq}}\,\mathrm{cm^{-2}}$, where $\mathrm{n_{eq}}$ refers to the 1\,MeV neutron-equivalent fluence in silicon.

The 50\,\unit{\micro\metre} thin sensors are mounted on highly heat conductive Thermal Pyrolytic Graphite carriers to  evacuate the $\sim$70\,\unit{\milli\watt\per\centi\square\meter} dissipated heat to actively cooled heat sinks located outside the detector acceptance.\par
The MVD performance is essential for the reconstruction of displaced decay vertices of open-charm hadrons and weakly decaying hyperons. The proximity of the MVD to the target region enables the reconstruction of low-momentum track fragments and short track segments. This capability, among others, provides a substantial reduction of the combinatorial background in the dielectron analysis arising from incompletely reconstructed photon conversions and Dalitz decays.\\

\subsubsection{Silicon Tracking System}
The Silicon Tracking System (STS), located downstream of the MVD detector, is the core tracking device of the CBM experiment for the reconstruction of charged particle trajectories~\cite{Heuser:54798,CBM:2025mnp,CBM:2025voh}. Eight tracking stations provide space points along the particle trajectories, enabling precise track reconstruction and accurate momentum determination.

The STS employs highly segmented, double-sided silicon microstrip sensors read out by fast, self-triggering front-end electronics. It achieves a spatial hit resolution of about 15~\textmu m and a time resolution of approximately 5~ns. 
The fundamental building block of the detector is the STS module, which decouples the silicon sensor from the readout electronics. The heat-dissipating front-end electronics, together with their powering and cooling infrastructure, are placed at the detector periphery using ultra-low-mass cables and interconnection technologies to minimize the material budget in the active volume. Up to ten modules are mounted on a common lightweight carrier structure, referred to as a ladder, realized with carbon-fiber components.

\begin{figure}[htb]
\centering
\includegraphics[width=0.48\linewidth]{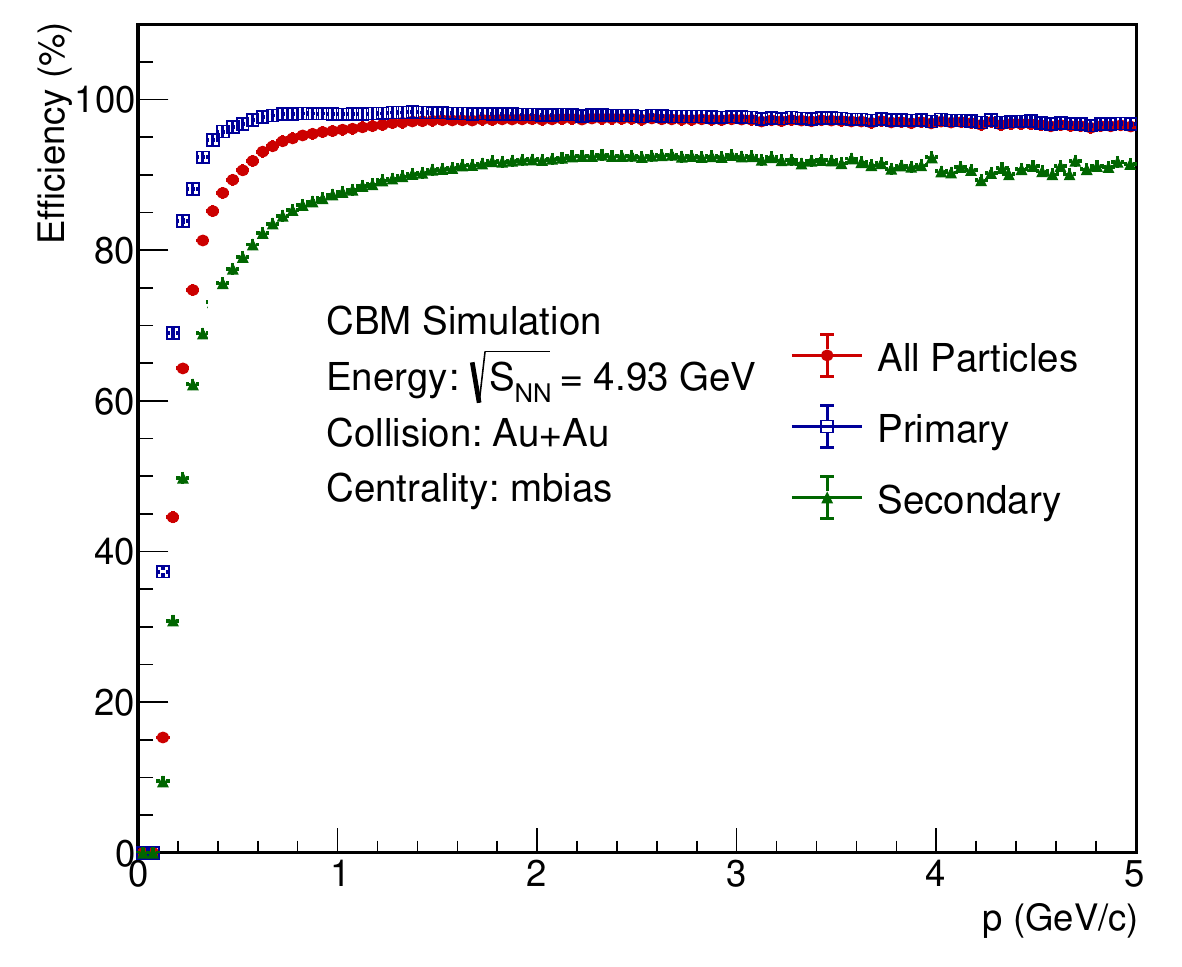}
\includegraphics[width=0.48\linewidth]{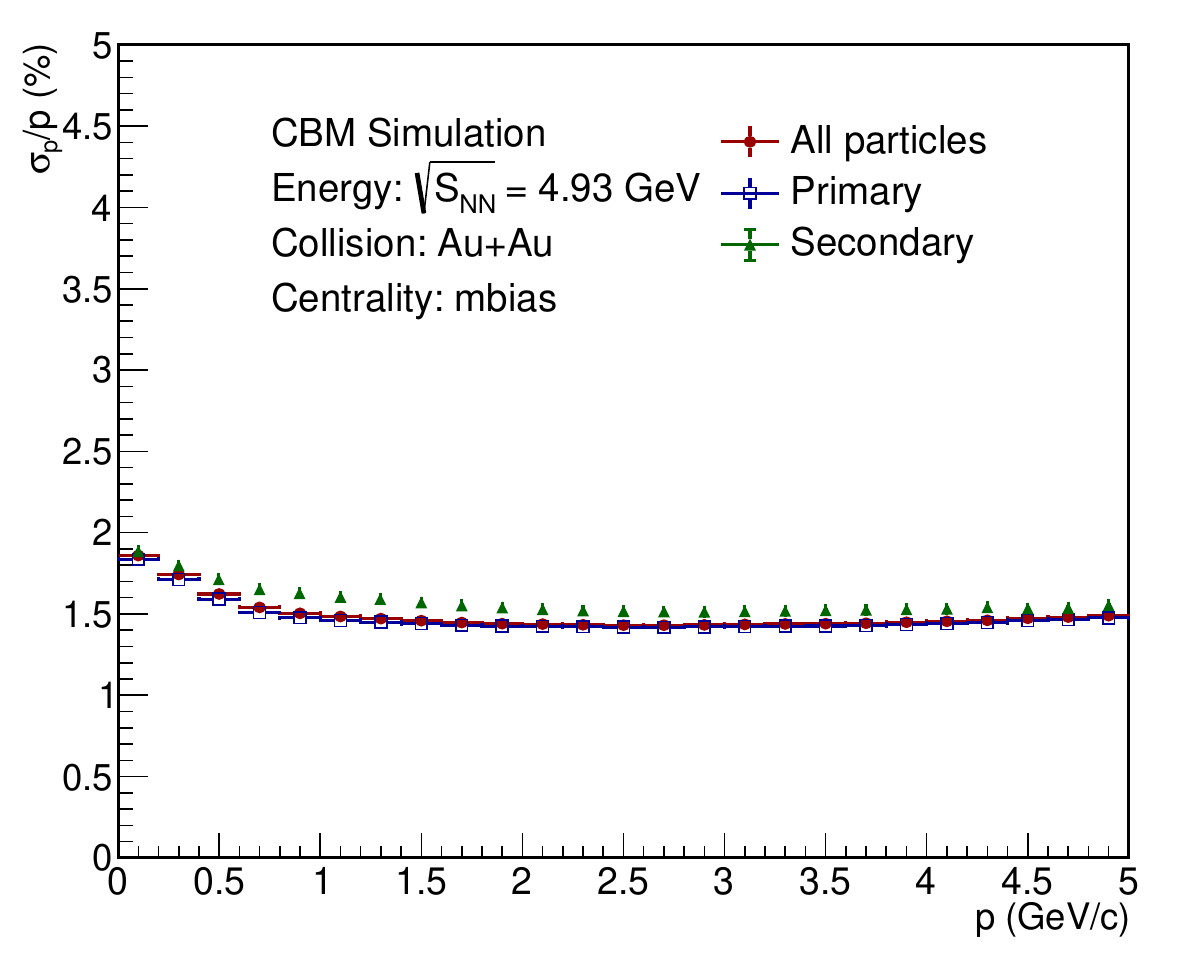}
\caption{The momentum dependence of the STS tracking efficiency (left) and momentum resolution (right) for all charged particles, and separately for primary and secondary particles in Au+Au collisions at $\sqrt{s_{NN}} = 4.9$~GeV.} \label{fig:sts-mom}
\end{figure}

The STS consists of eight tracking stations comprising 20 mechanical half-units. Each half-unit integrates ladders, readout and power electronics, and cooling on C-shaped aluminum support frames. In the experimental setup, the half-units are arranged symmetrically on both sides of the beam axis and distributed between 30~cm and 105.5~cm downstream of the target. The low-mass layout results in a momentum resolution better than 2\%, as shown in \figref{fig:sts-mom}.

\begin{figure}[htb]
\centering
\includegraphics[width=0.62\linewidth]{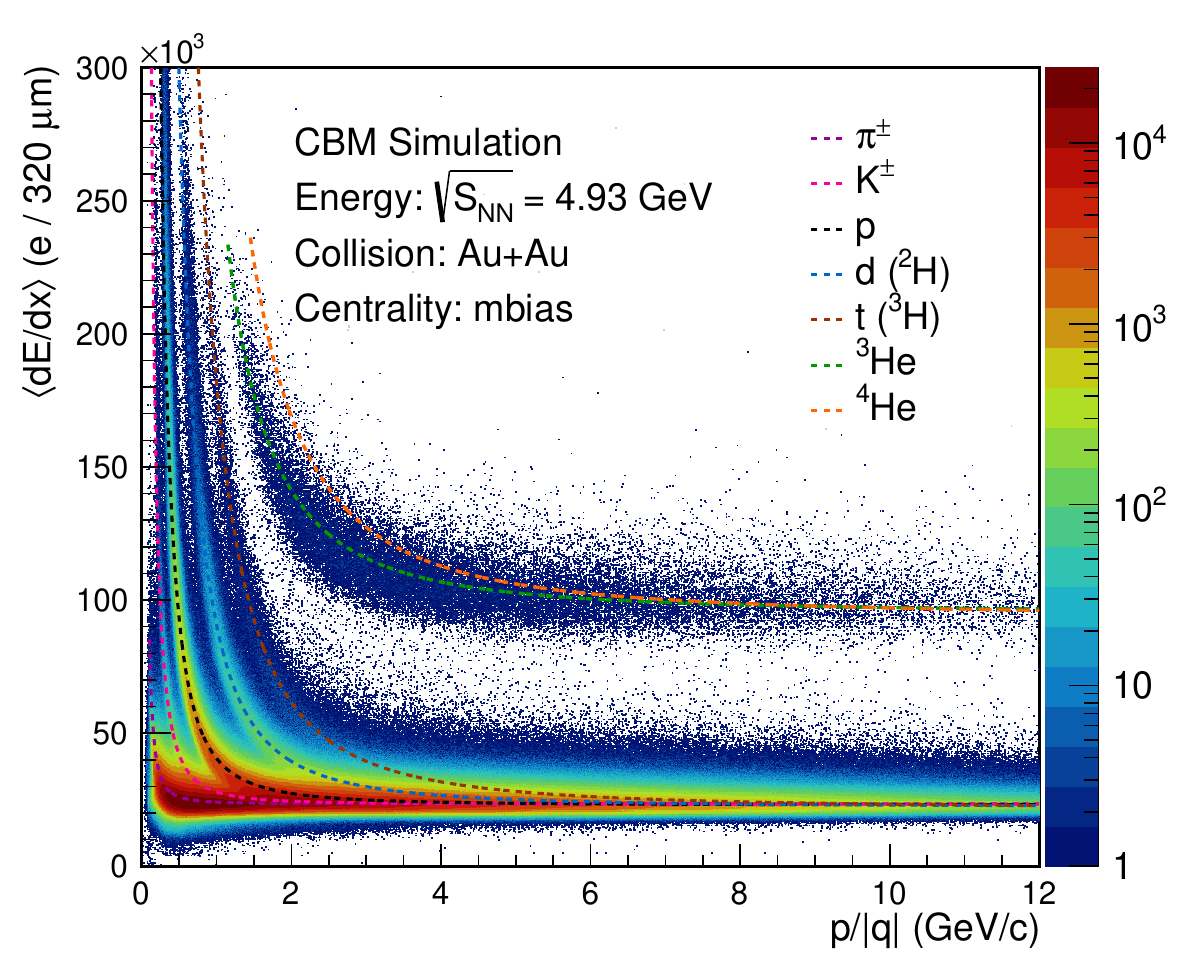}
\caption{Particle identification via specific energy loss measured in the STS as a function of momentum. The energy loss is obtained by normalizing the layer-averaged cluster charge (in electrons) to the effective path length in the silicon sensor, determined from the particle’s incident angle. The dashed curves indicate the expected mean values.
}
\label{fig:pid-sts}
\end{figure}

The tracking performance has been evaluated for Au+Au collisions at $\sqrt{s_{NN}} = 4.9$~GeV using fully realistic simulations. These include detector noise and are performed in time-based mode by combining UrQMD reaction products with beam–target interaction backgrounds, transported through the setup with GEANT4 software package. The detector parameters, such as dead time and thresholds, correspond to a validated configuration established in dedicated beam tests \cite{cbmSTS2025}. For primary high-momentum tracks, the efficiency exceeds 97\%, demonstrating the excellent performance of the STS under realistic operating conditions.
In addition, the STS measures the energy deposition of charged particles in each layer, enabling particle identification at low momenta (see \figref{fig:pid-sts}).

\subsection{Particle Identification Detectors}
Several complementary detector systems are combined to achieve particle identification over a broad momentum range. In the following subsections we briefly introduce these detector systems and outline their respective roles in the CBM experiment.
\subsubsection{Ring Imaging Cherenkov detector}
Electrons and positrons are identified using a Ring Imaging Cherenkov detector (RICH) located downstream of the magnet~\cite{TRB:2023asr}. The detector comprises an approximately $80\,\mathrm{m}^3$ gas radiator volume filled with CO$_2$, focusing mirrors, and photon detectors based on multi-anode photomultiplier tubes, providing high granularity and UV sensitivity. The signals are read out by newly developed electronics based on the \emph{DiRICH} family~\cite{TRB:2017xkj}, achieving an average timing precision of about $225\,\mathrm{ps}$.

\begin{figure}[htb]
    \centering \includegraphics[width=0.58\linewidth]{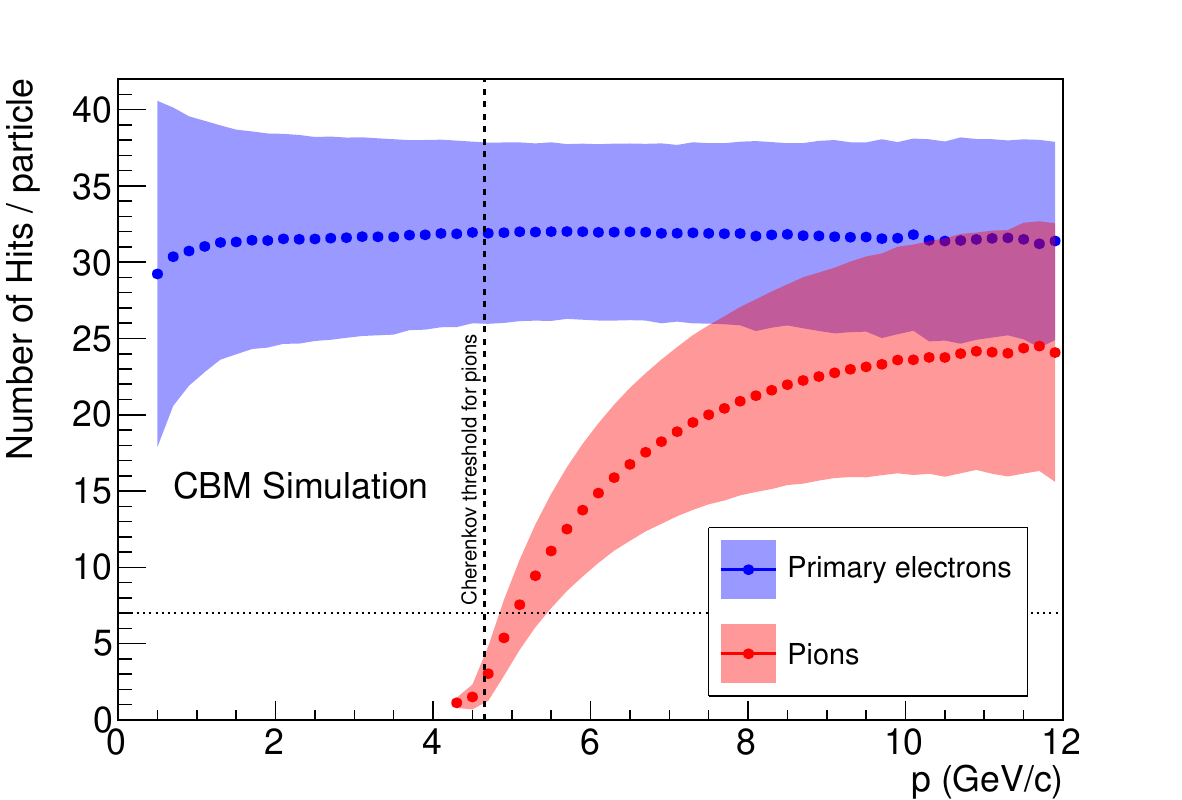}
    \caption{Number of measured hits per particle in the RICH detector versus momentum. The reconstruction threshold of 7 hits/particle is indicated by the dashed line.}
    \label{fig:rich-perf}
\end{figure}

The CBM RICH enables electron identification up to momenta of approximately \SI{8}{\GeVc} with a pion suppression factor exceeding 100. \Figref{fig:rich-perf} shows the number of hits recorded per particle that emits Cherenkov light in the RICH detector. The broad distribution of hit multiplicities arises from geometrical variations in the track path length through the radiator. On average, however, particles produce significantly more than seven hits, enabling efficient and robust ring reconstruction. Even low-momentum electrons generate a sufficient number of hits, while, due to dispersion effects, a small fraction of pions also produce hits below the nominal Cherenkov threshold.

\subsubsection{Transition Radiation Detector}
\begin{figure}[htb]
\includegraphics[width=0.5\linewidth]{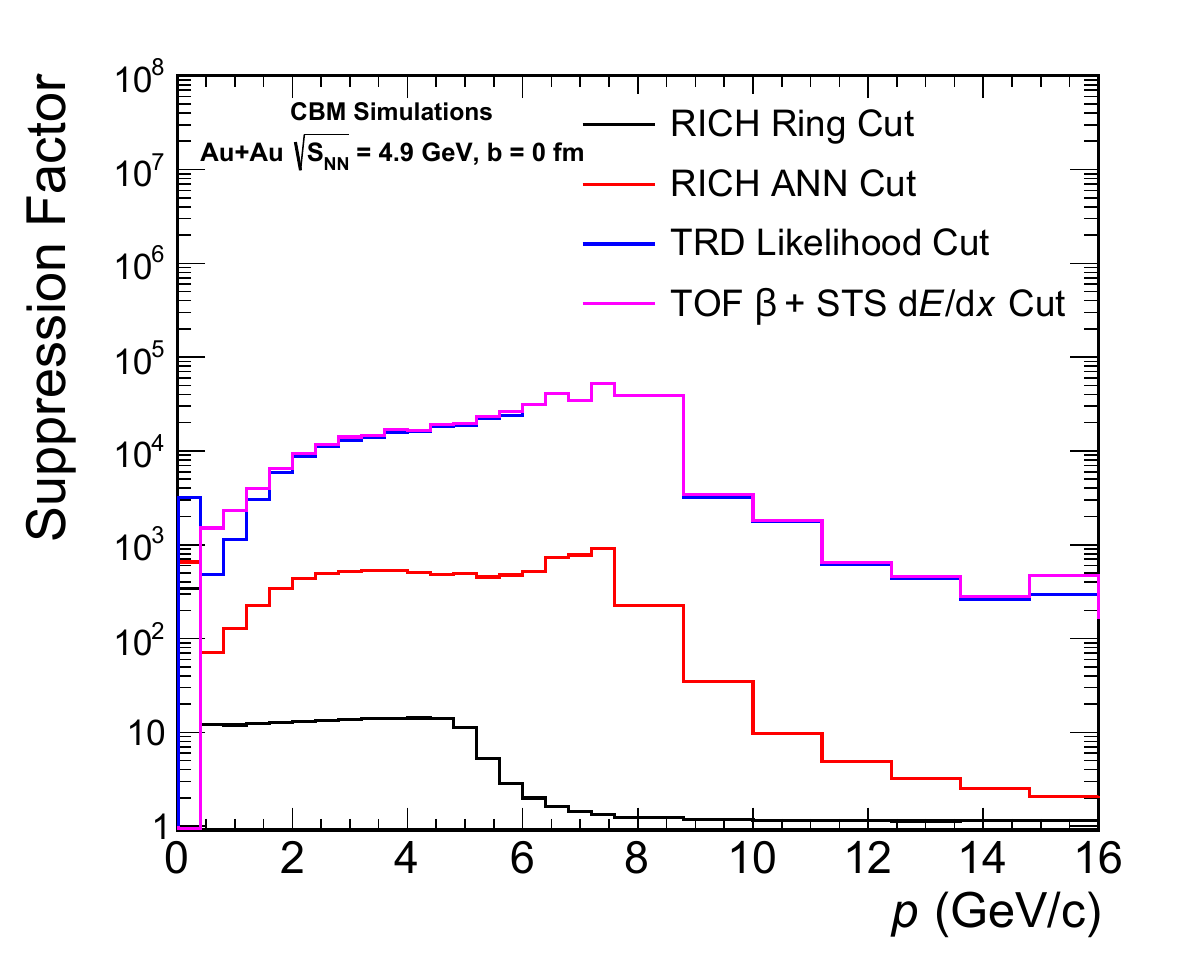}
\includegraphics[width=0.5\linewidth]{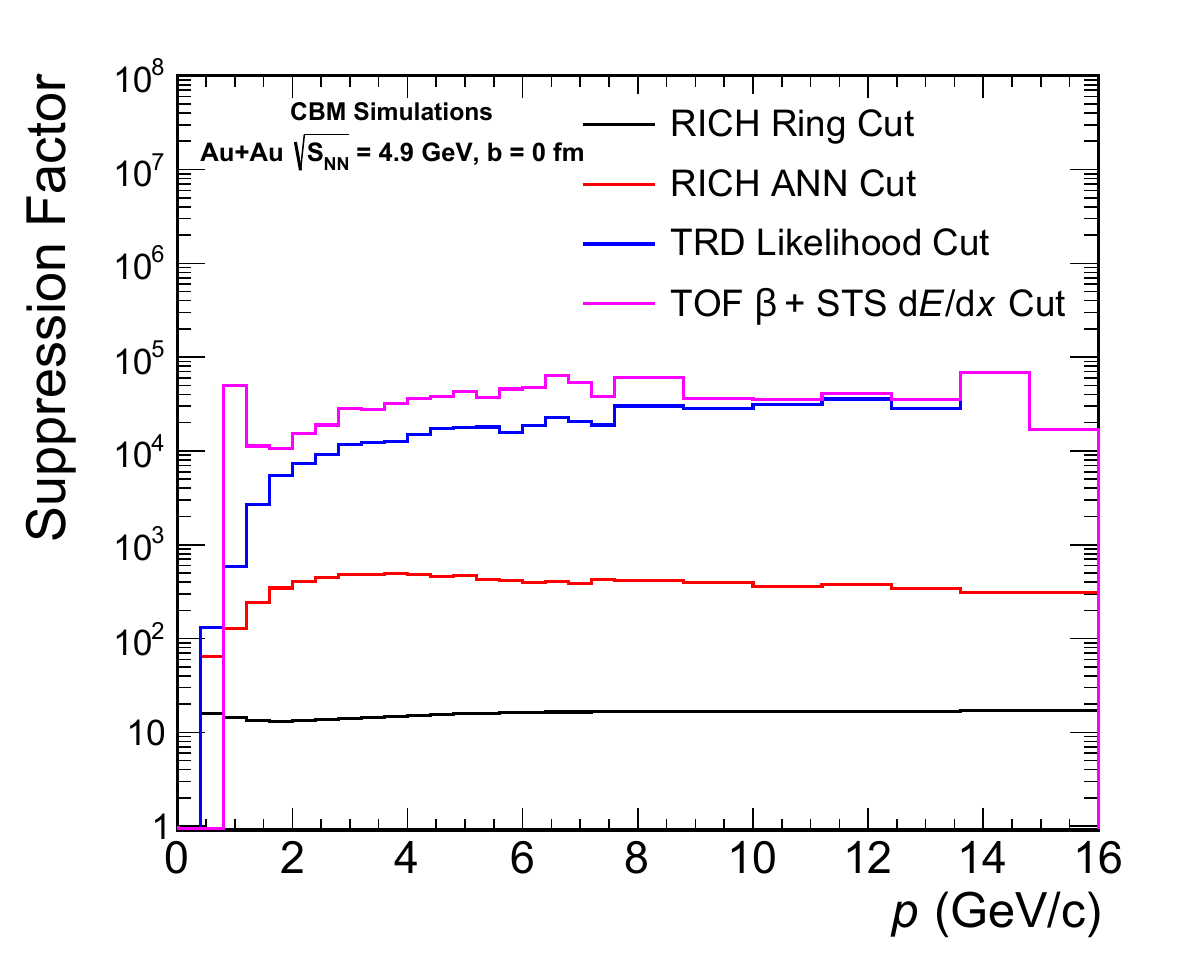}
\caption{Pion (left) and proton (right)  suppression factors for Au+Au collisions at $\sqrt{s_{NN}} = 4.9$~GeV with various combinations of CBM subdetectors\,\cite{MeyerAhrens:2025}. For the RICH detector the simple selection based on the ring radius and an Adaptive Neural Network (ANN) approach utilizing all ring parameters\,\cite{Subramani:2025} are shown.} \label{fig:had-suppr}
\end{figure}

The Transition Radiation Detector (TRD) provides additional electron-hadron discrimination and tracking capabilities. It consists of four layers of large-area multi-wire proportional chambers (MWPCs) operated with a Xe--CO$_2$ gas mixture. The primary task of the TRD is the separation of electrons with momenta above 1\,GeV/$c$, thereby enhancing the electron capabilities of the CBM RICH detector and extending the electron identification within CBM towards high transverse momenta (see \figref{fig:had-suppr}). This identification must be achieved with a pion suppression factor of about 20 at an electron efficiency of 90\%, enabling the measurement of dielectrons over a broad invariant mass range, from below the $\rho$ and $\omega$ resonances up to beyond the $J/\psi$ mass, while maintaining a favorable signal-to-background ratio.

In addition to electron identification, the TRD provides valuable tracking and particle identification information through measurements of the specific energy loss. This capability is particularly important for the identification of nuclear fragments, such as separating deuterons and $^4$He nuclei, which cannot be achieved by time-of-flight measurements alone.

These requirements are fulfilled by MWPCs filled with a Xe/CO$_2$ gas mixture in combination with an appropriate radiator. The default MWPC design for the CBM TRD features a symmetric amplification region with a total thickness of $3.5 + 3.5$\,mm, followed by a 5\,mm drift region to enhance the absorption probability of transition radiation photons in the active gas volume. This geometry ensures efficient and fast signal formation and readout, with characteristic time scales below 200\,ns per charged-particle track. The detector performance is further optimized by minimizing the material budget between the radiator and the gas volume.

The innermost region of the TRD stations employs a modified detector design, referred to as TRD2D. In addition to the baseline TRD functionality, TRD2D enables two-dimensional spatial position reconstruction within a single detector layer, thereby facilitating efficient track seeding via the TRD. This feature, combined with good acceptance at small polar angles, allows for measurements of transverse momentum spectra down to values of approximately 100\,MeV/$c$. 
The main difference with respect to the standard TRD design is the replacement of rectangular readout pads with a combination of triangular-shaped pads and quadrilateral overlapping readout elements. The pairing of pads into rectangular or parallelogram-shaped readout channels is performed by dedicated front-end electronics ASICs. This detection concept achieves a position resolution across pads at the level of 100\,\textmu m and provides improved energy resolution.

\subsubsection{Time of Flight}
The key component of hadron identification in CBM is a high-resolution Time-of-Flight (TOF) system based on multi-gap resistive plate chambers (MRPCs), covering an active area of about \SI{120}{\m\squared}. The primary task of the TOF system is the precise measurement of the arrival time of charged particles, allowing, in combination with track information from the STS, reliable particle identification. The TOF system is designed to operate over the full range of reaction rates and beam energies foreseen in the CBM physics program, both for high-multiplicity heavy-ion collisions and high-rate proton–proton interactions. To achieve sufficient separation power, particularly for charged kaon identification, flight paths of up to 10\,m and an overall system time resolution of about 80\,ps are required. These requirements result in a TOF wall with an overall size of approximately $12 \times 10$\,m$^{2}$. To reach the target system time resolution, the active area must be instrumented with timing detectors that provide an intrinsic time resolution better than 60\,ps per detector and a detection efficiency exceeding 95\%.

\begin{figure}[htb]
\centering
\includegraphics[width=0.6\linewidth]{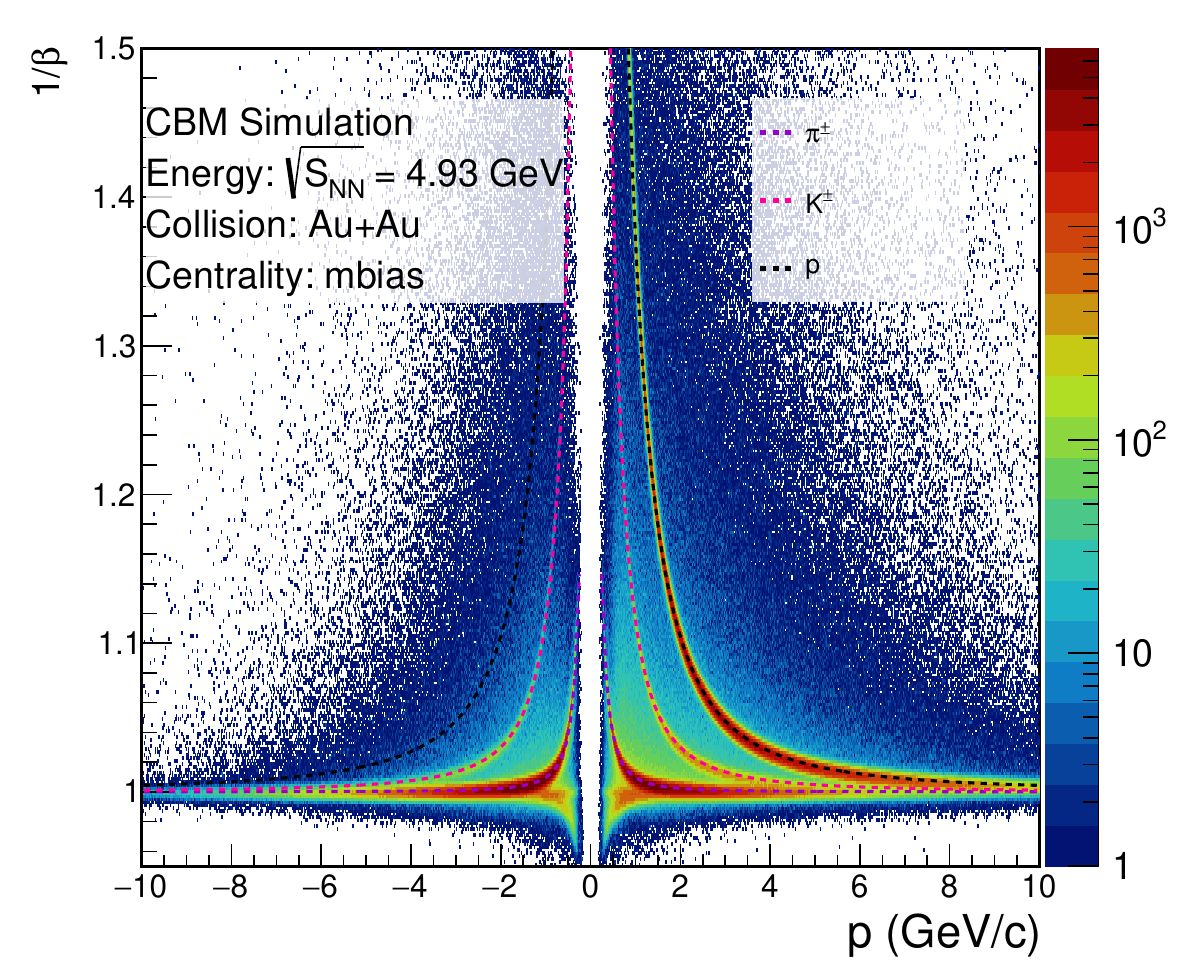}
\caption{Particle identification via inverse velocity 1/$\beta$ measured in the TOF as a function of momentum $p$. The inverse velocity is determined from the time-of-flight measurement and the flight path obtained from tracking. The dashed curves indicate the expected mean values.}
\label{fig:pid-tof}
\end{figure}

The CBM TOF wall is composed of MRPCs with strip readout. The regions exposed to intermediate and high particle fluxes, ranging from 1\,kHz/cm$^{2}$ to 30\,kHz/cm$^{2}$, are equipped with MRPCs based on low-resistivity glass. In contrast, the low-rate regions of the TOF wall, with particle rates below 1\,kHz/cm$^{2}$, employ MRPCs using float glass. To maintain an occupancy below 5\%, the detector granularity varies between 4\,cm$^{2}$ and 50\,cm$^{2}$ across the TOF wall to adapt to the local rate conditions. The expected performance of the TOF system is illustrated in \figref{fig:pid-tof}, where the inverse velocity $(1/\beta)$ is shown as a function of the reconstructed particle momentum.

Particle identification is performed in a combination of detectors, both for hadrons and in particular for electrons. This is illustrated in \figref{fig:had-suppr}, where the pion and proton suppression factors are shown as a function of the total momentum for various combinations of CBM subdetectors.
Rejection factors in excess of $10^4$ are demonstrated. 

\subsubsection{Muon Chambers}

Muons are identified in the Muon Chamber System (MuCh), where charged-particle tracks are measured downstream of a sequence of absorbers and tracking stations. This configuration enables muon identification with high purity, even at high particle multiplicities. The MuCh detector of the CBM experiment is interchangeable with the RICH detector, allowing CBM to operate in either muon or electron mode.

The MuCh detector comprises a series of segmented absorbers with detector stations installed between them. Each station consists of a triplet of detector layers. In the first two stations, large-area triple-GEM detectors with pad readout will be employed~\cite{Dubey:2013hva, Ghosh:2025hmc}. The pad dimensions gradually increase from $3\times3$\,mm$^2$ up to $10.6\times10.6$\,mm$^2$. The maximum expected particle rate in the first MuCh station is about 400\,kHz/cm$^{2}$ for minimum-bias Au+Au collisions at an interaction rate of 10\,MHz and a beam energy of \SI{10}{\AGeV}~\cite{Agarwal:2025kle}.

The detector technology for MuCh Stations~3 and~4 is currently under evaluation, with chamber concepts based either on straw tubes or on Micro-Pattern Gaseous Detectors (MPGDs) being considered.

\subsection{The Forward Spectator Detector}

The Forward Spectator Detector (FSD) will be installed 12~m downstream of the target in the projectile-spectator region. It covers an active area of $132\times110$~cm$^2$ and consists of 356 plastic-scintillator pads, each 2~cm thick and read out by a photomultiplier tube. The detector is assembled from 30 removable $22\times22$~cm$^2$ modules. Three pad sizes are used, $4.4\times4.4$, $5.5\times5.5$, and $7.3\times7.3$~cm$^2$, with the finest segmentation closest to the beam pipe, where the local particle density is highest. The modular construction also permits the most exposed scintillator pads to be replaced between running campaigns.

Based on FLUKA simulations of 11\,$A$GeV Au+Au collisions at a beam intensity of $10^9$ ions/s, corresponding to a 10~MHz interaction rate, and two months of running at a 100\% duty factor, the absorbed dose reaches about 5~kGy near the beam pipe and the integrated one-MeV-neutron-equivalent fluence at the FSD plane reaches approximately $5\times10^{12}$~cm$^{-2}$. This fluence precludes practical operation of uncooled SiPMs because irradiation would increase their dark current by orders of magnitude~\cite{Tolba:2022oxn}; vacuum photomultiplier tubes are therefore used. The radiation-sensitive readout electronics are placed at the base of the support, about 5~m below the detector, where the integrated neutron-equivalent fluence is below $10^{10}$~cm$^{-2}$ over the reference two-month period. The FSD segmentation is largely driven by the expected occupancy. The maximum simulated local density is 0.012 hits per event per cm$^2$, corresponding to 120~kHz/cm$^2$ at the design interaction rate of 10~MHz, while the highest expected signal rate is about 2~MHz per channel.
The PMT signals are processed by the DiRICH readout electronics, which record the hit time and provide a charge-sensitive observable through time over threshold (ToT). In dedicated high-rate tests, the DiRICH electronics processed channel rates up to 2.2~MHz.

The FSD is not a conventional zero-degree calorimeter and does not measure the complete spectator energy. Its signal is formed mainly by projectile-spectator protons and charged nuclear fragments with $Z>1$, whose trajectories are separated by the CBM dipole field according to their charge-to-mass ratio and transverse momentum. From the multiplicity, positions, azimuthal distribution, and energy deposition of the detected charged spectators, the FSD provides a first-order spectator-plane estimate and a centrality estimator largely independent of particle production at midrapidity.

The performance studies use DCM--QGSM events with statistical multifragmentation~\cite{Bondorf:1995ua,Baznat:2019iom}, followed by Geant4 transport through the full CBM geometry, including the FSD and the downstream conical beam pipe with an angle of 5 mrad. At 11\,$A$GeV, the first-order event-plane correction factor reaches approximately 0.3, and pion and proton $v_1$ are reconstructed with a relative systematic uncertainty of about 10--20\%, depending on centrality and kinematics.
The FSD centrality estimator is based on the multiplicity and energy deposition of detected spectator protons and charged fragments. It provides centrality determination for approximately the 0--40\% most central collisions. At 11\,$A$GeV, the relative impact-parameter resolution is 20--30\%. The performance improves at 2\,$A$GeV because of the larger contribution from charged fragments.

The FSD does not detect spectator neutrons, and charged spectators with charge-to-mass ratios close to that of the primary beam may remain inside the beam-pipe aperture. The quantitative effects of the unmeasured signal component on impact-parameter resolution, centrality-bin migration, and volume fluctuations are currently being investigated. Comparisons with alternative fragmentation models and cross-calibration with independent observables will be required to assess the associated model dependence.

\subsection{CBM Infrastructure}
The CBM experiment will be installed in a dedicated underground cave designed to accommodate high-intensity, high-energy heavy-ion beams. The cave is equipped with a massive beam dump. Civil construction of the cave and building shell has been completed, providing stable conditions for the subsequent installation phases. Current activities focus on the installation of major infrastructure elements, particularly the upstream platform and the rail system required for the positioning of downstream detectors (TRD, TOF, and FSD). In parallel, preparatory work for the implementation of the Technical Building Infrastructure (TBI) is ongoing, including detailed engineering and planning of services such as power distribution, cooling, and gas systems. At the same time, integration activities are underway to define detector placement and infrastructure routing within the cave. Detector installation is foreseen to follow in the subsequent phase, with the overall schedule targeting CBM readiness for beam commissioning in 2028 and beam delivery from SIS100 shortly thereafter.

\subsection{Data Acquisition and Online Event Selection}

A defining feature of the CBM experiment is its capability to operate at interaction rates of up to \num{e7} events per second, a regime that has not been reached in previous nuclear collision experiments. This requirement places stringent demands on detector design, front-end electronics, and in particular on the data acquisition concept. For minimum-bias Au+Au collisions, the typical raw event size is about \SI{50}{\kB} and the peak rate is \SI{10}{\MHz}. This results in an instantaneous data stream of about \SI{500}{\GBps}. To ensure feasible operation, this data volume must be reduced in real time by at least two orders of magnitude before storage. Such reduction is only possible through online identification of physics signatures associated with rare probes. Typical examples include displaced decay vertices from hyperons, which require at least partial online reconstruction up to the track level. These complex selection tasks cannot be handled by conventional hardware triggers or FPGA-based logic alone and must instead be performed in software on CPUs and GPUs. Furthermore, the high particle multiplicity in heavy-ion collisions limits the effectiveness of simple low-level trigger criteria, making it necessary to process the full data stream in software.

These constraints led to the development of a fully free-streaming data acquisition architecture that does not require a hardware trigger~\cite{ONLINE_TDR}. The front-end electronics of all detector systems operate autonomously in a self-triggered mode, generating time-stamped hit messages whenever signals exceed predefined thresholds. This continuous data stream is aggregated and transported to the online processing system, where event building, full reconstruction and data selection are performed in software. The software trigger determines which data are retained for permanent storage without sending any feedback signal to the detector readout; as a result, system performance is limited by data throughput rather than latency.

In this self-triggered scheme, the association of detector signals with individual physical collisions relies entirely on precise timestamps generated by the front-end electronics. Sub-nanosecond synchronization of all detector components is ensured by a central Timing and Fast Control (TFC) system, which distributes a common clock and time reference to all readout nodes via a hierarchical optical link network.

The interface between the custom detector optical links and standard server equipment is provided by the Common Readout Interface (CRI), a custom FPGA-based PCIe card~\cite{ONLINE_TDR}. Each CRI terminates GBT links~\cite{Moreira:1235836} from the detector readout boards, reformats the incoming hit data into fixed time-interval units called microslice containers, and transfers them via DMA to the main memory of the server. The full CBM readout system will comprise approximately 220 CRI boards.

At the highest interaction rates, the required online processing power corresponds to approximately \num{1e6}~HepSpec06, a standard high-energy-physics CPU benchmark unit used to express the aggregated performance of large computing clusters. Since this capacity is only needed during relatively short data-taking periods each year, and due to the use of specialized detector interfaces, the compute infrastructure is organized into two parts. The First-level Event Selector (FLES) entry node cluster, which is located in the server room above the experimental area, hosts the CRI interface cards and, using a high-bandwidth InfiniBand RDMA network for all-to-all communication between nodes, assembles timeslice containers comprising data from the full experiment over defined time intervals. The FLES entry node cluster is designed to handle total input data rates of up to \SI{1}{\TBps}; its nodes perform no reconstruction, apart from organizing and time-ordering the data, and their number is primarily determined by detector connectivity and data throughput requirements. The subsequent processing stage is carried out on a large compute farm located in the Green IT Cube, situated approximately \SI{1}{\km} from the experimental hall. The timeslice data produced by the entry nodes are transferred to these processing nodes, where online reconstruction and data selection are performed.

The readout and data acquisition concept has been validated in the FAIR Phase-0 experiment mCBM~\cite{mCBM_TDR}, which operated the complete data path from self-triggered front-end readout through timeslice building to online processing using prototype detector systems and a subset of the CRI and FLES entry stage hardware. The reconstruction of the $\Lambda$ baryon in the mCBM setup was recently demonstrated~\cite{CBM:2026kqm}.

\section{Physics Performance}\label{sec:phys}

We present a selection of physics performance results obtained with Monte Carlo simulations with transport models\,\cite{Bass:1998ca} 
PHQMD v5.2\,\cite{Aichelin:2019tnk} and UrQMD v4.0\,\cite{Bleicher:2022kcu} and the final CBM detector configuration 
implemented for particle transport in GEANT3. All the results are for Au+Au collsions at $\sqrtsNN = 4.9$\,GeV.
Most of the results are event-based simulations, while the time-based mode corresponding to the continuous data readout in CBM is currently being refined.
Also in preparation is the inclusion of residual detector misalignment and miscalibration as well as a more accurate simulation of the beam-induced background.

\subsection{Hadron observables}

\begin{figure}[htb]
\includegraphics[width=0.99\linewidth]{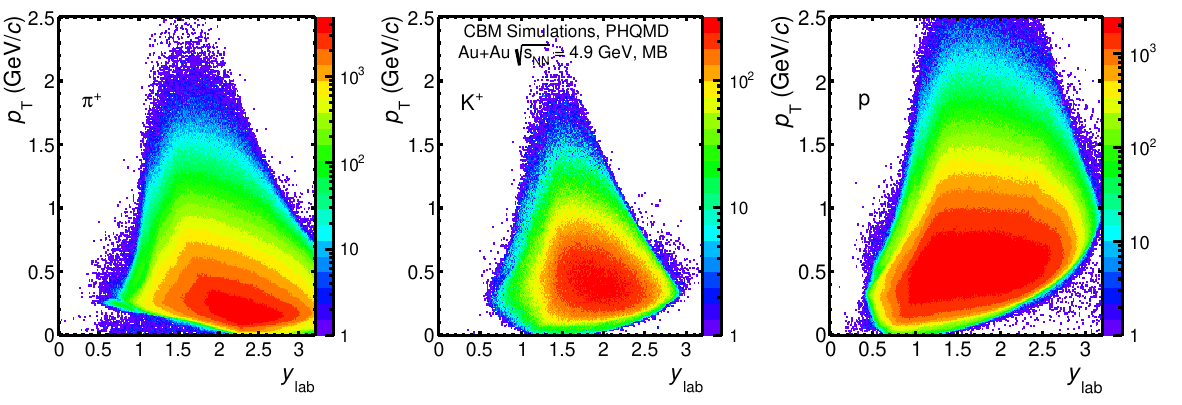}
\caption{The acceptance in transverse momentum and laboratory rapidity for pions, kaons, protons for Au+Au collisions at $\sqrt{s_{NN}} = 4.9$~GeV (the center-of-mass has  $y_{\mathrm{lab}}=1.62$).}
\label{fig:acceptance}
\end{figure}

The broad acceptance of CBM for pions, kaons and protons, see Fig.~\ref{fig:acceptance}, 
is the basis for the study of various observables, from collective flow to event-by-event fluctuations. It serves as well the efficient reconstruction of phase space distributions of a multitude of hadrons decaying into these particles. 

The production of strange hadrons is a very important probe of the medium created in heavy-ion collisions and has therefore been measured since the beginnings of the heavy-ion programs \cite{Rafelski:1982pu,Koch:1982ij,Koch:1986ud}.  However, in the near-threshold energy region those measurements were restricted to particles with single strangeness and rare probes, such as multi-strange hyperons or hypernuclei, were not accessible up-to-now \cite{Blume:2005ru,Blume:2011sb,Piasecki:2023hoo}.  The high rate capabilities of CBM will allow to fill these gaps with very high statistical accuracy.  Most measurements will therefore be dominated by systematic uncertainties and a careful analysis and minimization of systematic effects is essential.  

\begin{figure}[htb]
\includegraphics[width=0.47\linewidth]{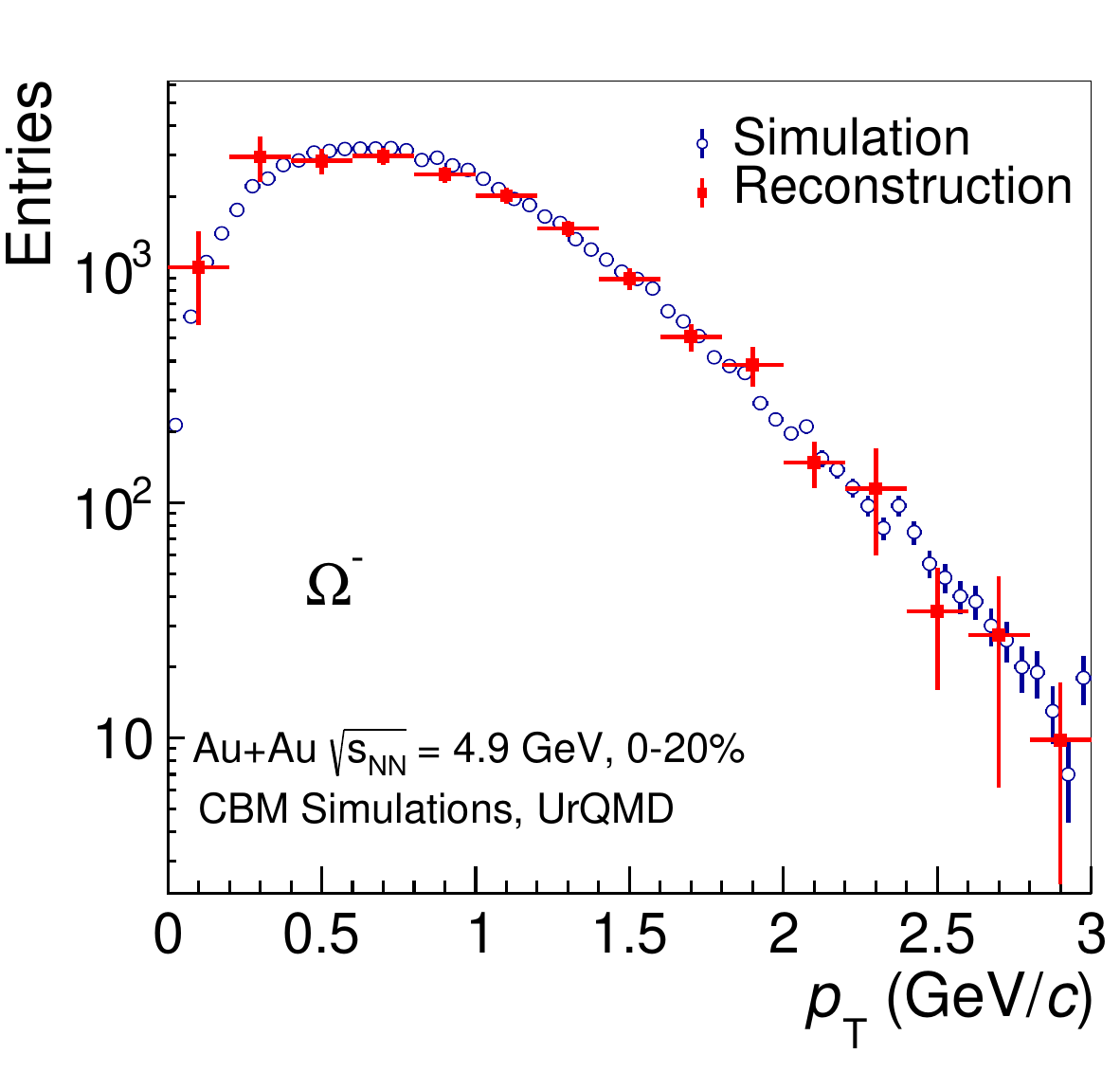}
\includegraphics[width=0.47\linewidth]{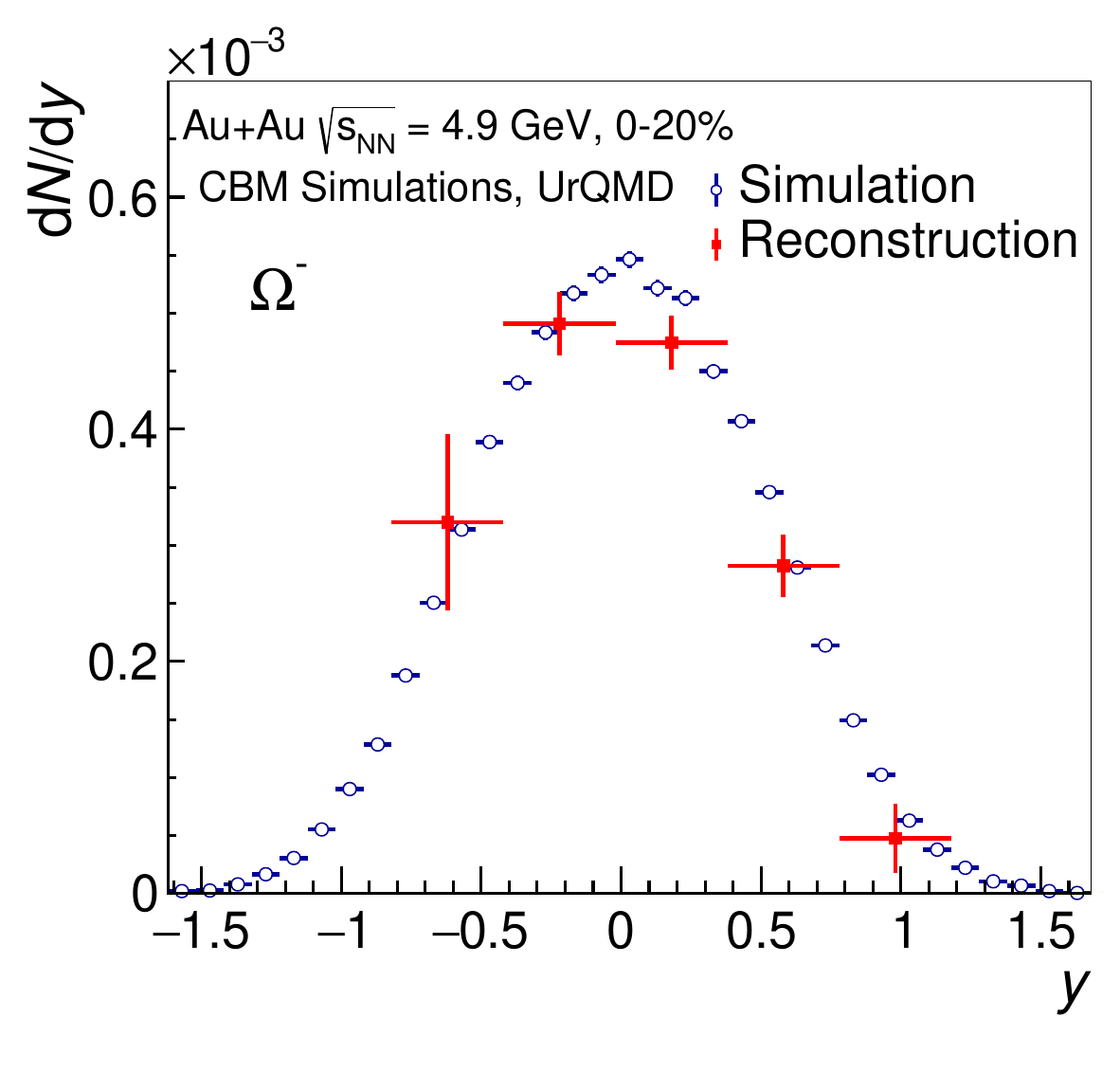}
\caption{The simulated and reconstructed (in the $\Lambda \text{K}^-$ channel) transverse momentum spectrum integrated over all rapidities (left panel) and rapidity distribution (right panel) of $\Omega^-$ baryons in Au+Au collisions at $\sqrtsNN = 4.9$~GeV. The statistical errors correspond to 10$^7$ events in the 0-20\% centrality, generated by the UrQMD v4.0 model.}  
\label{fig:Omega}
\end{figure}

As an example for the possibilities for a rare hyperon probe, \figref{fig:Omega} presents the $\Omega^{-}$ reconstruction performance in terms of $p_T$ and rapidity distributions, obtained via the weak decay topology $\Omega^{-} \rightarrow \Lambda + \textrm{K}^{-}$. Note that the shown signal is based only on $10^7$ simulated central Au+Au events. The foreseen event statistics during a normal data taking campaign will thus yield several orders of magnitude more signal counts, thus facilitating a significantly more precise multi-differential measurement and flow studies.

\begin{figure}[H]
\centering\includegraphics[width=0.45\linewidth]{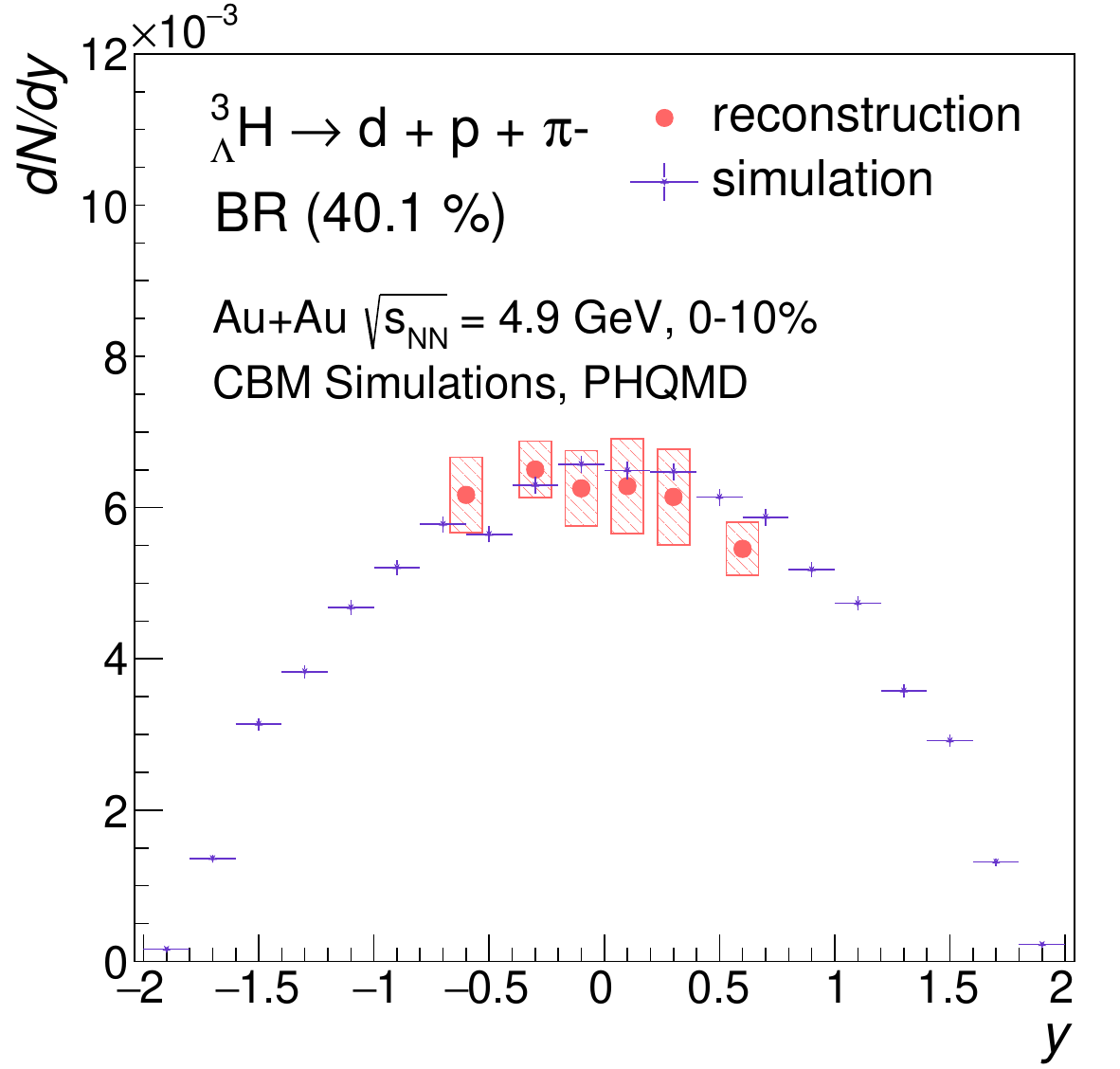}
\caption{Rapidity distribution of hypertritons in 10\% most central Au+Au collisions at $\sqrtsNN = 4.9$~GeV as reconstructed via their 3-body-decay $^3_{\Lambda}\textrm{H} \rightarrow \textrm{d} + \textrm{p} + \pi^{+}$ (red symbols), together with an estimation of the achievable systematic uncertainties (red boxes). The blue symbols represent the MC data, generated with the PHQMD v5.2 model \cite{Aichelin:2019tnk,Glassel:2021rod}.  For this study $10^{6}$\,events have been simulated, while the uncertainties shown for the data points correspond to the expectation for $10^{9}$\,events.}
\label{fig:H3L-rap}
\end{figure}

An important topic in the CBM physics program will be the measurement of hypernuclei as they provide relevant information on the hyperon-nucleon and hyperon-hyperon interaction and can help constrain the nuclear equation-of-state (a review of the current status can be found in \cite{qcdatfair2025}).  Due to the high baryon density, high abundances of hypernuclei are expected in the energy range accessible with the SIS100 accelerator \cite{Andronic:2010qu,Reichert:2022mek}.  Of particular interest will be double-$\Lambda$ hyperons, e.g. $_{\Lambda\Lambda}^{6}\textrm{He}$.  As these hypernuclei will require the reconstruction of their three-body weak decay (e.g. ${}^6_{\Lambda\Lambda}\textrm{He} \rightarrow {}^5_{\Lambda}\textrm{He} + \textrm{p} + \pi^{-}$ and subsequently ${}^5_{\Lambda}\textrm{He} \rightarrow {}^4\textrm{He} + \textrm{p} + \pi^{-}$), instead of the less demanding two-body decay, the hypertriton decay $_{\Lambda}^{3}\textrm{H} \rightarrow \textrm{d} + \textrm{p} + \pi^{-}$ has been investigated as a case study. \Figref{fig:H3L-rap} presents the reconstructed $_{\Lambda}^{3}\textrm{H}$ rapidity distribution in comparison with the simulation input. 
The $p_T$ spectra are analysed in intervals of rapidity, corrected for acceptance and efficiency and extrapolated into unmeasured $p_T$ regions using fits with Boltzmann and Blast-Wave functions\,\cite{Schnedermann:1993}.
A very good agreement can be achieved, well within the estimated systematic uncertainties.  These are to a large extent determined by the extrapolation into the low-$p_T$ region not covered by the detector acceptance.

\subsection{Dileptons}

Dileptons provide unique observables for the characterization of the hot and dense medium produced in relativistic nuclear collisions. They are produced in a variety of processes and emitted in the various stages of the fireball evolution. Since dileptons interact only electromagnetically, they carry undistorted information about the dense interior of the collision zone and serve as a penetrating probe of the dense phases of QCD matter produced in nuclear collisions. Depending on the appropriately selected pair mass window, dileptons are particularly sensitive to the source temperature, medium modification of the vector mesons, fireball lifetime, baryon density due to coupling to baryonic resonances, and possible restoration of chiral symmetry~\cite{Salabura:2020tou,Rapp:2016xzw,Jung:2016yxl}. In particular, the invariant mass distribution of the thermal dileptons above $m_{\text{inv}} > 1.5$ GeV/$c^2$, once the contribution from the other dilepton sources are suitably subtracted, provides a direct estimate of the mean fireball temperature~\cite{Rapp:2014hha}. The selected large mass window makes the extracted temperature less sensitive to the late phases of fireball evolution. Moreover, being directly extracted from the Lorentz invariant mass spectra, the estimated source temperature is not blue-shifted by the collective velocity of the fireball. This method was first employed by NA60 Collaboration at SPS, and yielded a source temperature of $T_{\text{slope}} = 205 \pm 12$ MeV in $\sqrt{s_{NN}} = 17.3$ GeV In+In collisions~\cite{NA60:2008dcb}. Subsequently, the HADES Collaboration extracted a fireball temperature $T_{\text{slope}} = 71.8 \pm 2.1$ MeV from the acceptance-corrected dielectron excess yield in $\sqrt{s_{NN}} =2.42$ GeV Au+Au collisions~\cite{HADES:2019auv}. On the other hand, results from the HADES Collaboration for low-mass vector mesons measured in $\sqrtsNN = 2.61$ Ar+KCl collisions showed that the strength of low mass dilepton yield is due to coupling to the baryons~\cite{HADES:2011nqx}, a feature later utilized to explain dilepton yields at higher beam energies. To date, no dilepton measurements have been performed in heavy-ion collisions at $\sqrtsNN$ between 3 and 10 GeV. Hence, dilepton measurements from low to intermediate mass range at highest net baryon densities as accessible by CBM, in the collision energy range between HADES and SPS, is of great interest. CBM plans to perform multi-differential systematic measurements of dilepton production in both dielectron ($e^{+}e^{-}$) and dimuon ($\mu^{+}\mu^{-}$) decay channels.\\
Extensive performance simulations are carried out to test the sensitivity of the CBM electron (muon) setup to the measurement of dielectron (dimuon) spectra. In the results presented here, UrQMD v4.0 is used to model 0-20\% most central Au+Au collisions at $\sqrt{s_{NN}} = 4.9$ GeV. To adequately describe the dilepton distributions, the relevant decay channels are simulated with the PLUTO~\cite{Frohlich:2007bi} generator, embedded into the UrQMD events at an enhanced rate and subsequently scaled down to the expected yields calculated with the thermal model. In addition to the dileptons produced by thermal radiation, decay channels of the $\omega$ and $\phi$ mesons are generated with PLUTO, while $\eta$ and $\pi^0$ are already sufficiently produced by UrQMD.

As all the dilepton signals are very rare, effective methods for reduction of the background are crucial. In the dielectron case, the main contributions to this background are misidentified hadrons on the one hand, and electrons and positrons originating from photon conversions inside the target or detector material on the other. Sufficient hadron rejection can be achieved by using the two subdetectors dedicated to electron identification, RICH and TRD, as well as information from TOF and STS (see \figref{fig:had-suppr}). While the hadron rejection in this analysis is still based on conventional cut-based selections, it has been demonstrated in Refs.~\cite{Subramani:2025,MeyerAhrens:2025} that the signal-to-background ratio can be further improved using machine learning methods.
As stated above, electrons and positrons from photon conversions are a major contribution to the background. While the ones originating in the detector material can be rejected on the basis of the track extrapolation to the primary vertex, in particular making use of the exceptional vertex resolution of the MVD, this is not possible for photon conversions occurring still inside the target. To alleviate this issue, CBM plans to use a segmented target approach, consisting of multiple slices with a thickness of a few tens of $\mu$m each, thus making it possible to disentangle the collision vertex and the origin of the conversion electron. However, the data used in this analysis is simulated based on a monolithic $250$~$\mu$m gold target and therefore needs to be adjusted in order to account for the increased possible rejection with a segmented target. As the amount of photon conversions scales approximately linearly with the target thickness (see Ref.~\cite{Galatyuk:2009}), a reduction of the resulting electron tracks by $80$~$\%$ is deemed a reasonable estimate and implemented in the analysis using Monte Carlo information. 

For the dimuon measurements in the SIS100 energy domain, the major challenge is to identify, particularly the soft muons, in an environment of high particle multiplicities. Each central Au+Au collision produces on average more than 700 hadrons which need to be suppressed by the hadron absorbers. The particle identification in the muon setup starts with the reconstruction of charged particle tracks in the STS, followed by extrapolation of the reconstructed tracks through the downstream detectors, namely MuCh, TRD and TOF. The selection of muon track candidates include a cut on the position of the primary collision vertex, cuts on the number of the associated hits and the quality of the tracks in STS, MuCh and TRD, as well as a cut on the particle mass based on momentum and time-of-flight of the reconstructed tracks. The TOF mass cut has been found to be particularly useful for rejection of punch-through hadrons. 

\begin{figure}[htb]
\includegraphics[width=0.49\linewidth]{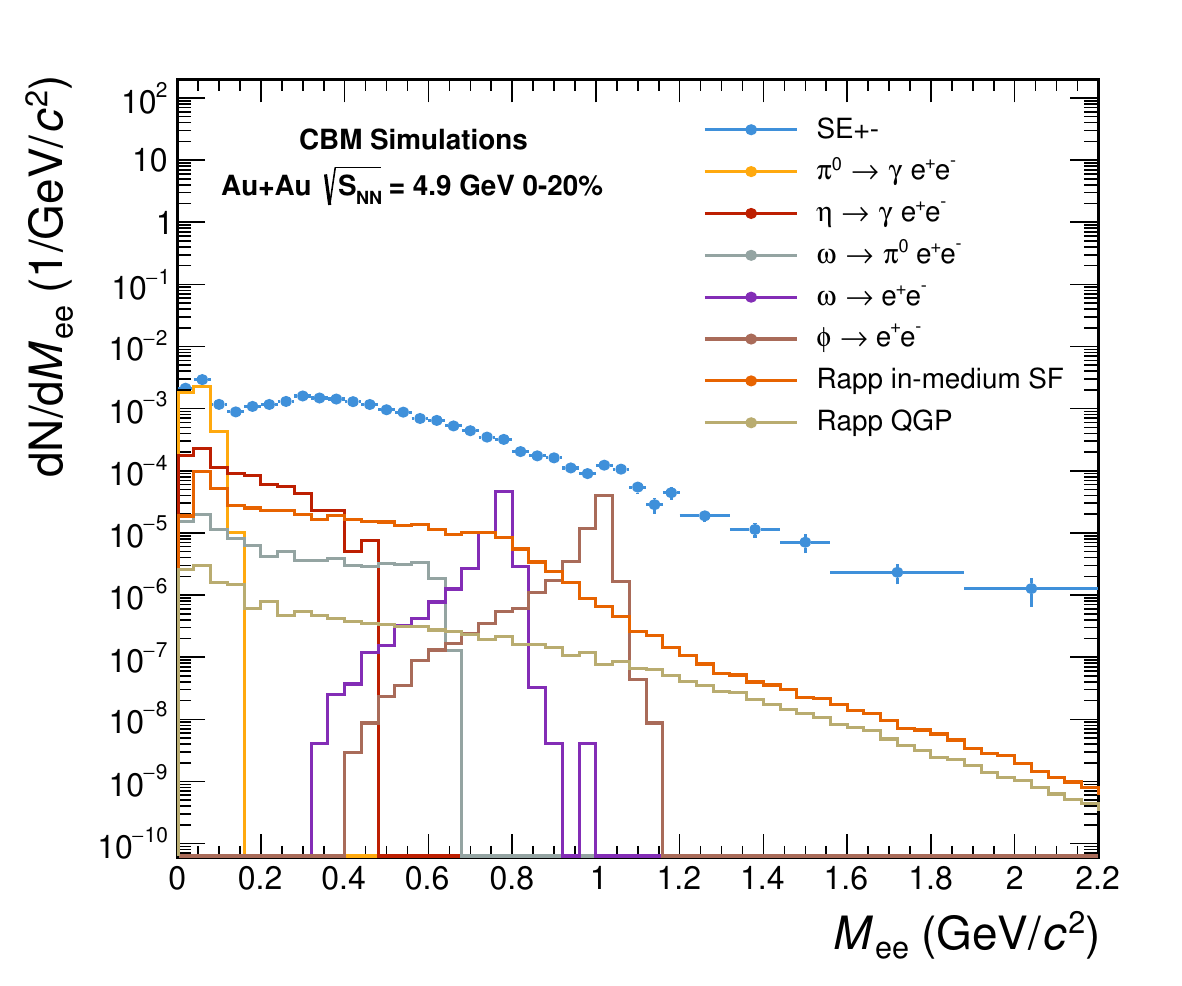}
\includegraphics[width=0.49\linewidth]{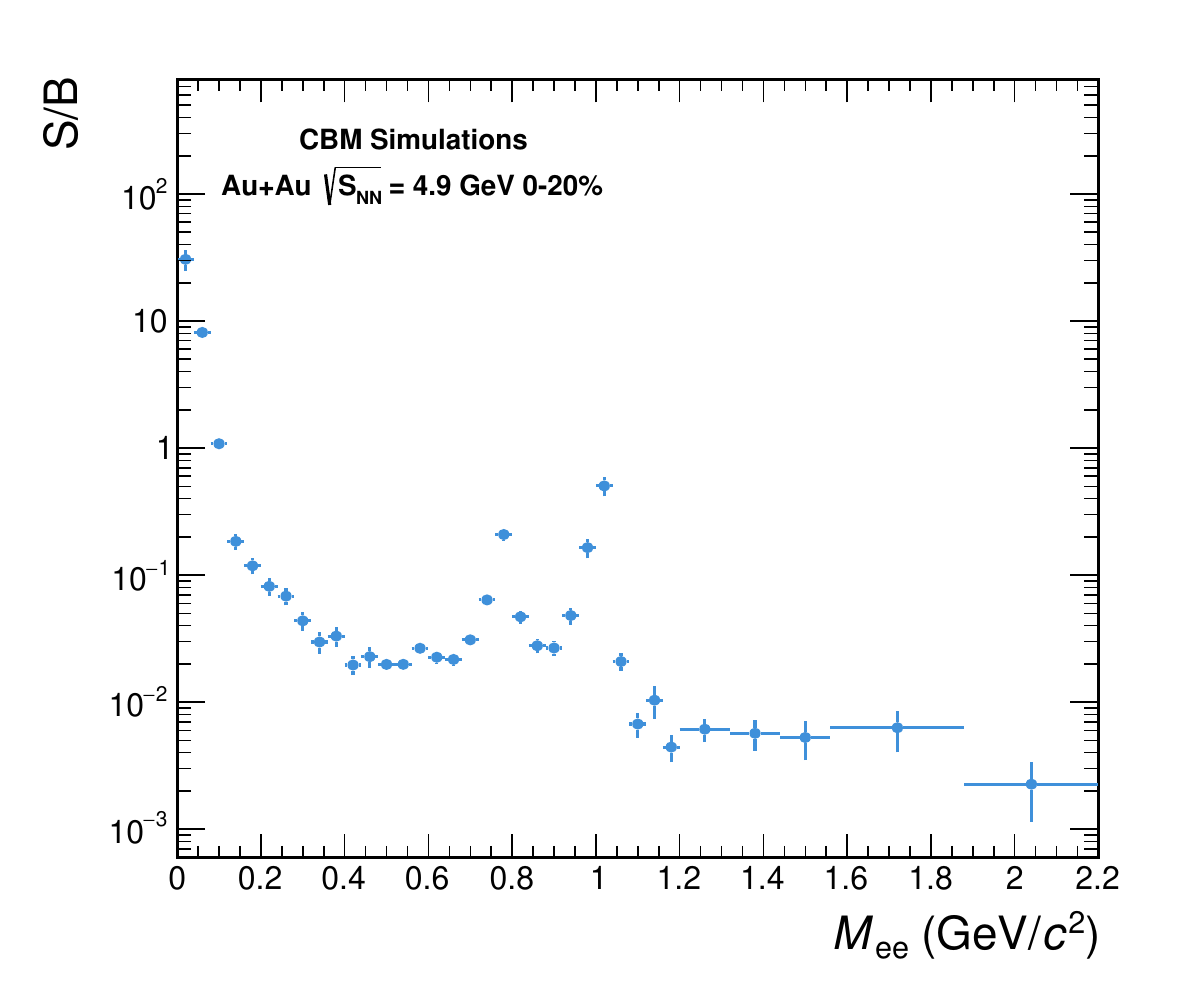}
\caption{Left: invariant mass spectra of the reconstructed dielectrons ($e^{+}e^{-}$) in 0-20\% central Au+Au collisions at $\sqrt{s_{NN}} = 4.9$ GeV. $\text{SE}{+-}$ denotes the full combinatorics, namely all the same-event opposite-sign pairs. The signals are shown based on MC truth. Right: Signal-to-Background (S/B) ratio. }
\label{fig:dielectron}
\end{figure}

\begin{figure}[htb]
\includegraphics[width=0.49\linewidth]{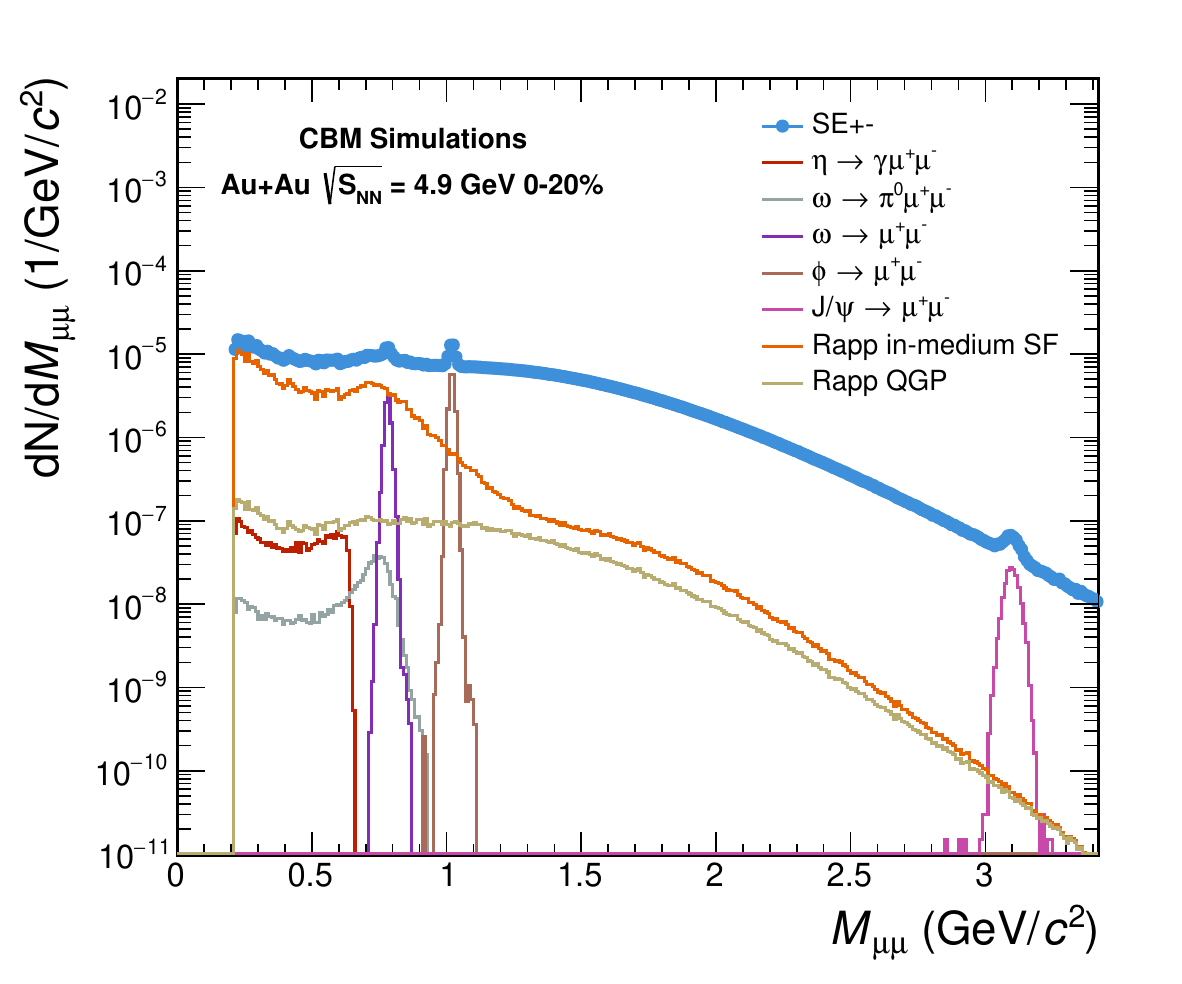}
\includegraphics[width=0.49\linewidth]{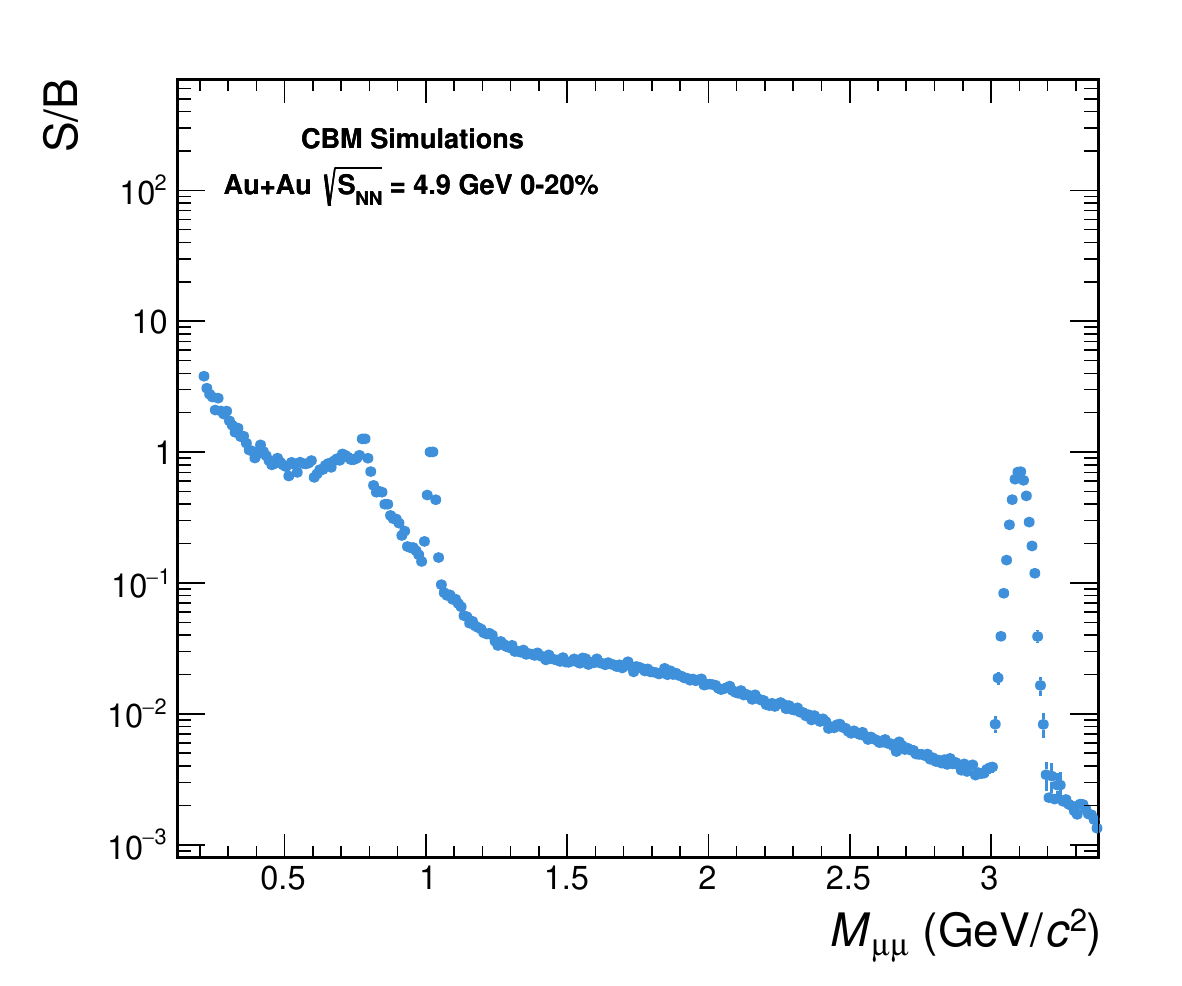}
\caption{Left: invariant mass spectra of the reconstructed dimuons ($\mu^{+}\mu^{-}$) in 0-20\% central Au+Au collisions at $\sqrt{s_{NN}} = 4.9$ GeV. $\text{SE}{+-}$ denotes the full combinatorics, namely all the same-event opposite-sign pairs. The signals are shown based on MC truth. Right: Signal-to-Background (S/B) ratio. }
\label{fig:dimuon}
\end{figure}

\Figref{fig:dielectron} and \Figref{fig:dimuon} show the reconstructed dielectron and dimuon spectra, respectively, with $\text{SE}{+-}$ denoting all the same-event opposite signed pairs of dilepton track candidates after all the selection cuts, therefore including signal pairs as well as combinatorial background, normalized by the number of events. The individual contributions of signal pairs from the different dilepton sources are shown, as identified with Monte Carlo information. For the muon channel, to overcome large statistical fluctuations, the combinatorial background is calculated via the event mixing technique, by combining the oppositely charged uncorrelated muon track candidates, irrespective of their event tags. The individual contributions from different dimuon signals, after suitably weighting by their multiplicity and branching ratios are added to the combinatorial background to obtain the equivalent same event (SE+-) oppositely charged reconstructed dimuon spectrum, shown up to the $J/\psi$ mass region. To additionally increase the statistics of the reconstructed dimuon signals without further increasing the computing time and the simulated data volume, the following scaling techniques have been adopted. Firstly, pure dimuon signal events without embedding into background events are used. The increase in efficiency for this highly idealistic case without the presence of additional background tracks is determined using the dimuon continuum originating from the thermal radiation. The ratio between the thermal dimuons reconstructed in the absence (pure signal) and presence (embedded signal) of the background events simulated with UrQMD shows an increase in efficiency of about a factor of two over the entire pair mass range for the pure signal events. 
Thus, to  model the reduced reconstruction efficiency in the realistic case (i.e. in the presence of background) the pure-signal events are weighted down by the corresponding efficiency.
This method is used for all dimuon signals, vastly decreasing the file size and computing times.  
The results of our studies clearly confirm CBM's capability of measuring thermal dilepton radiation, for both the dielectron and dimuon channels, which thus will offer a unique confidence on the systematic uncertainties.
In the SIS100 energy domain, a rather small contribution from the Drell-Yan process and a negligible part from open-charm decays will be advantageous for the extraction of the thermal dilepton signal in the intermediate mass regime. The signal-to-background ratio of about $10^{-2}$ in both the $e^{+}e^{-}$ and  $\mu^{+}\mu^{-}$ channels shows, however, the challenge of this measurement, where the precision of the background subtraction determines the systematic uncertainties of the extracted physics, namely the in-medium $\rho$ spectral function and the average temperature of the fireball. The close to unity signal-to-background ratio for the J/$\psi \rightarrow \mu^{+}\mu^{-}$ reconstruction indicates the promising possibility to detect charmonium production at sub-threshold energies.

\subsection{Collective flow} \label{sec:flow}

Collective flow \cite{Danielewicz:1985hn, Ollitrault:1992bk, E895:1999ldn} is one of the key manifestations of the collective behavior of strongly interacting matter formed in heavy-ion collisions. In particular, the Fourier coefficient $v_2$ of the azimuthal distribution of particles, often referred to as elliptic flow, quantifies the second harmonic modulation of particle emission with respect to the reaction plane in heavy-ion collisions. It arises due to the initial spatial anisotropy of the overlap region of the colliding nuclei, which is converted into a momentum anisotropy through pressure gradients and collective behavior during the system’s collective expansion. As a result, $v_2$ provides important information about the early-stage dynamics and the transport properties (such as viscosity) of the hot and dense medium created in relativistic heavy-ion collisions. At FAIR energies, where the system evolves at high net-baryon density and moderate temperatures, elliptic flow is expected to be especially sensitive to the transport properties and the equation of state of dense QCD matter. Precise measurements of the elliptic flow for identified particles therefore constitute a cornerstone of the CBM physics program. The study of elliptic flow at CBM will also provide crucial information on the degree of collectivity and thermalization achieved at available energies around 4.9 GeV. Model calculations \cite{Aichelin:2019tnk,Steinheimer:2025trr} predict sizeable flow signals already at these energies, with a strong dependence on the early-stage dynamics and the treatment of hadronic and partonic interactions. Reliable reconstruction of flow observables requires excellent tracking performance, momentum resolution, and control of non-flow effects. In this respect, systematic comparisons of flow-related observables obtained from pure Monte Carlo simulations and after full detector reconstruction are essential to validate the experimental sensitivity. Elliptic flow, accompanied by other flow harmonics like directed flow $v_1$, triangular flow $v_3$, and higher components, and by femtoscopic measurements (see \ref{sec:femto}), will enable a coherent and complementary characterization of the space–time evolution and collective dynamics of baryon-rich matter.

\begin{figure}[htb]
\begin{center}
\includegraphics[width=0.58\linewidth]{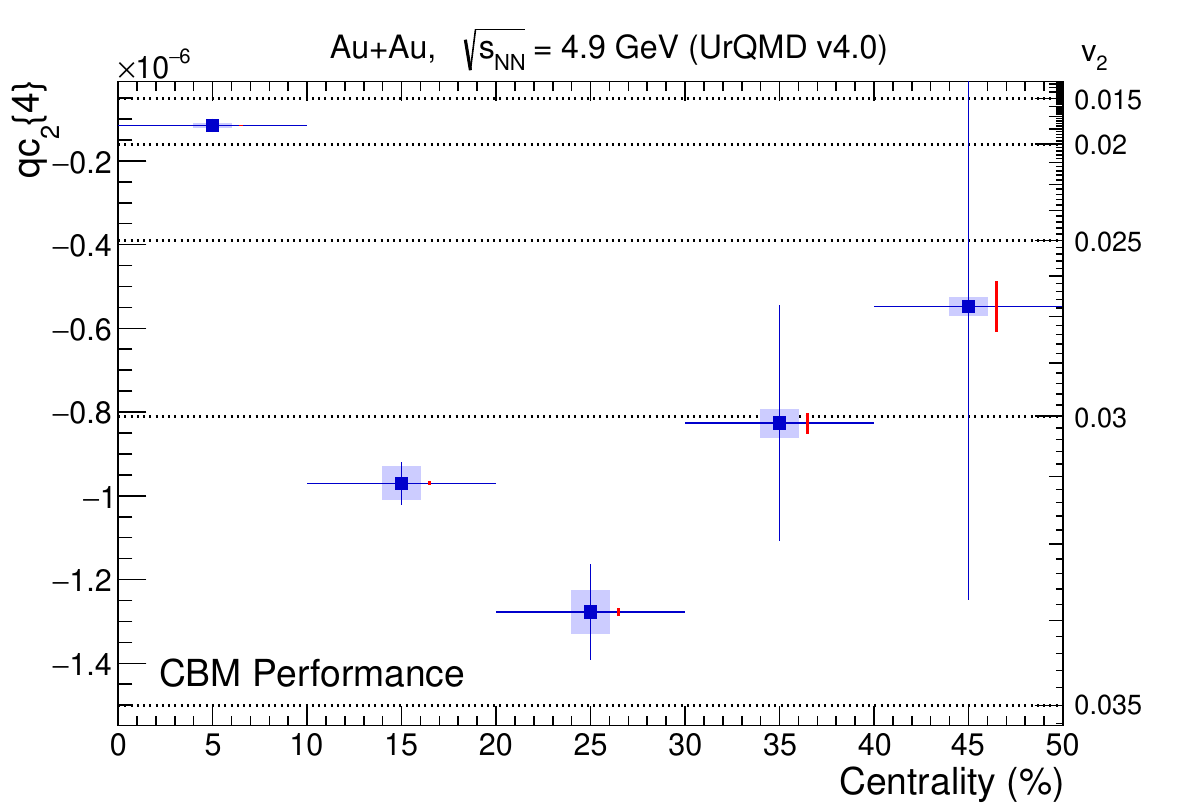}
\end{center}
\caption{Centrality dependence of elliptic flow, $v_2$, for charged particles, estimated with 4-particle cumulants, through the theoretical relation ${\rm qc}_{2}\{4\} = -v_2^4$. 
After all selection criteria, around $0.62$~M events per 10\% centrality bin width have been analysed. The statistical (systematic) uncertainties are represented by vertical bars (boxes). Shown as a red bar on the right side of the symbols are extrapolations for statistical uncertainties corresponding to a Minimum Bias dataset of $10^{9}$~events.}
\label{fig:flow}
\end{figure}

Anisotropic flow degrees of freedom, which appear in the Fourier series expansion of particle azimuthal distribution in the plane transverse to the beam direction, amplitudes $v_n$ and symmetry planes $\Psi_n$, cannot be measured directly in an experiment. Instead, they are determined with multiparticle azimuthal correlations. To suppress unwanted systematic biases arising from few-particle non-collective correlations (e.g., from resonance decays), improved estimators are obtained using multiparticle cumulants of azimuthal angles. Cumulants were introduced into analyses of collective flow in Refs.~\cite {Borghini:2000sa,Borghini:2001vi}, and have been widely used afterwards. However, they are a precision technique only in collisions characterized by a very large number of produced particles (multiplicity), because both statistical and systematic uncertainties scale as inverse powers of multiplicity (the power in the scaling is determined by the order of the correlator). In \figref{fig:flow}, the feasibility study for $v_2$ measurements in the CBM environment using 4-particle cumulants is presented, using Au+Au collisions at $\sqrt{s_{\rm NN}} = 4.9$~GeV simulated with the UrQMD~v4.0 model~\cite{Bass:1998ca, Bleicher:1999xi}. After all the particle and event selection criteria are applied, around $0.62$~M events per 10\% centrality bin width have been analysed. Cumulants and corresponding corrections for detector inefficiencies were implemented using the analytic approach from Refs.~\cite{Bilandzic:2010jr,Bilandzic:2013kga}. This study demonstrates that in the most central collisions, characterized by the largest multiplicities, 4-particle cumulants can be used reliably as a precision technique in the CBM environment, given the fact that a signal in $v_2$ even smaller than 0.02 can be estimated with negligible statistical and systematic uncertainty. For other centralities where multiplicities are smaller and uncertainties are larger, the red vertical lines show extrapolations of the statistical uncertainties to a dataset consisting of $10^9$~events.

\subsection{Femtoscopy} \label{sec:femto}
Femtoscopy is a powerful technique that exploits momentum correlations of particles emitted close in momentum phase space to probe the space–time structure of the particle-emitting source created in relativistic heavy-ion collisions \cite{LisaPratt10decades}. By analyzing two-particle correlation functions at small relative momenta, one gains access to characteristic source sizes, emission durations, and collective dynamics at the femtometer scale, providing direct insight into the evolution of strongly interacting matter at extreme densities and temperatures. Complementary to collective flow measurements, femtoscopy provides access to the space-time evolution of the system and thus offers valuable insight into the properties of baryon-rich matter created at FAIR energies.

\begin{figure}[htb]
\centering\includegraphics[width=0.6\linewidth]{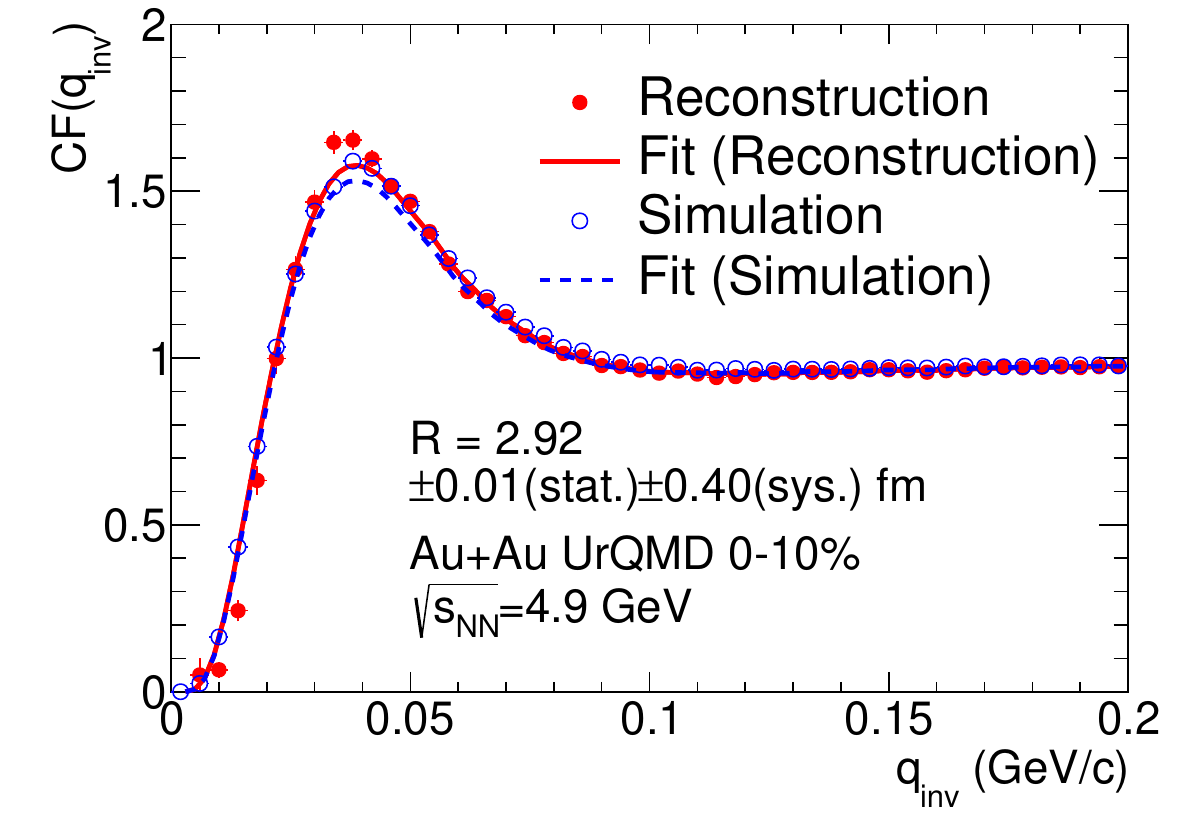}
\caption{Proton–proton correlation functions for simulated and reconstructed particles for Au+Au collisions at $\sqrtsNN = 4.9$~GeV from UrQMD v4.0 transport model, up to 10\% of the most central collisions. The fitted Lednicky-Lyuboshitz \cite{Lednicky} formula which includes quantum-statistical effects and final-state interactions (Coulomb and strong), assumes a Gaussian shape of the source function and provides a good description of the reconstructed correlation function. "Simulation" refers to the pure Monte Carlo correlation function, whereas "Reconstruction" denotes the detector configuration foreseen at the start of experimental operation.}
\label{fig:ppcorr}
\end{figure}

Correlation measurements of both non-identical and identical particle pairs will play a key role in the CBM physics program. As an example (see \figref{fig:ppcorr}), we consider proton–proton correlation functions calculated within the UrQMD v4.0 transport model for central (0–10\%) Au+Au heavy-ion collisions at $\sqrtsNN = 4.9$~GeV. Such correlations, driven by quantum statistics and final-state interactions, are sensitive to the spatial extent of the emitting source and to its collective expansion. A comparison of the correlation functions obtained directly from the pure model output with those reconstructed after detector simulation allows one to quantify the influence of finite acceptance, momentum resolution, and track reconstruction efficiency on the extracted femtoscopic observables. The systematic uncertainties account for particle identification impurities and two-track effects related to track splitting and track merging. Track splitting occurs when a single physical particle is mistakenly reconstructed as two separate tracks in the detector, which can produce artificial pairs with very small relative momentum and distort the correlation function. Track merging, on the other hand, happens when two nearby particles are reconstructed as a single track because their trajectories are too close to be resolved by the detector. Both effects are detector - related artifacts that can bias femtoscopic measurements and therefore must be carefully identified and corrected or removed through track quality and pair selection cuts.

\subsection{Event-by-Event multiplicity fluctuations} \label{sec:fluct}

Event-by-event fluctuations of conserved quantities, such as baryon number, electric charge, and strangeness, quantified by the cumulants of the corresponding multiplicity distributions, constitute a powerful and sensitive experimental probe of the phase structure of QCD~\cite{Rustamov:2022hdi, Braun-Munzinger:2022bkc, Gross:2022hyw}. In particular, they provide a promising avenue for the search for critical phenomena. Near a critical point, the correlation length $\xi$ increases, causing larger regions of the system to fluctuate coherently. Since multiplicity cumulants are related to integrals of correlation functions, they grow with the correlation length. Higher-order cumulants exhibit an especially strong dependence on $\xi$ and are therefore particularly sensitive probes of critical phenomena, including the possible QCD critical point~\cite{Stephanov:1999zu, Stephanov:2008qz}. This topic has gained particular importance in light of recent measurements by the STAR~\cite{STAR:2025zdq} and HADES~\cite{Nabroth:2025elh} experiments, as well as their interpretation within different phenomenological and theoretical frameworks; see Refs.~\cite{Braun-Munzinger:2022bkc, Friman:2025swg} for a detailed discussion. To draw firm conclusions, high-precision experimental data are required at center-of-mass energies per nucleon pair around 5 GeV and below. The large data samples expected from the CBM experiment will provide unprecedented opportunities to study higher-order multiplicity cumulants with high precision and to search for signatures of critical behavior in strongly interacting matter. In the present study  we focus on the feasibility of measuring proton number fluctuations with the CBM experiment.
Experimentally, such fluctuations are studied by measuring the number of particles produced in each individual nuclear collision (event) and constructing the cumulants of their multiplicity distributions~\cite{Rustamov:2020ekv}. For high-precision experimental measurements of fluctuation signals, several key challenges must be addressed:
(i) Incomplete particle identification arising from overlapping detector signals used for particle identification~\cite{Rustamov:2012bx, Arslandok:2018pcu, Rustamov:2024hvq}.
(ii) Non-critical contributions to the measured fluctuation signals, such as those originating from participant (volume) fluctuations or correlations induced by global conservation laws~\cite{Skokov:2012ds, Braun-Munzinger:2016yjz, Rustamov:2022sqm, Holzmann:2024wyd, Braun-Munzinger:2020jbk, Braun-Munzinger:2018yru, Braun-Munzinger:2023gsd, Friman:2025swg}.
(iii) Non-trivial efficiency correction procedures~\cite{Bzdak:2012ab, Nonaka:2017kko, Nonaka:2018mgw}.
The analysis is based on events generated with the PHQMD v5.2 model, transported through a realistic CBM detector simulation using the GEANT3 software package, and subsequently reconstructed with the standard CBM analysis chain. For this analysis, the collision centrality is determined using the charged-pion multiplicity. This approach suppresses autocorrelations between the centrality selection and the proton sample used to construct the fluctuation observables.
Particle identification is performed using reconstructed inclusive mass distributions of different particle species, obtained by accumulating the reconstructed-mass values of all tracks from all events. To address the issue of particle misidentification, which in this case refers to overlapping mass distributions corresponding to different particles, we employ both a standard cut-based analysis~\cite{STAR:2025zdq} and a recently developed method based on fuzzy logic~\cite{Rustamov:2024hvq, Nabroth:2025elh}. Using these approaches, we reconstruct the cumulants $\kappa_n$ of proton number distributions up to fourth order, defined as
\begin{align*}
\kappa_1 &= \langle N\rangle,\\
\kappa_2 &= \langle N^2\rangle-\langle N\rangle^2,\\
\kappa_3 &= \langle N^3\rangle-3\langle N^2\rangle\langle N\rangle+2\langle N\rangle^3,\\
\kappa_4 &= \langle N^4\rangle-4\langle N^3\rangle\langle N\rangle
          -3\langle N^2\rangle^2+12\langle N^2\rangle\langle N\rangle^2-6\langle N\rangle^4.
\end{align*}

\begin{figure}[htb]
\centering\includegraphics[width=0.85\linewidth]{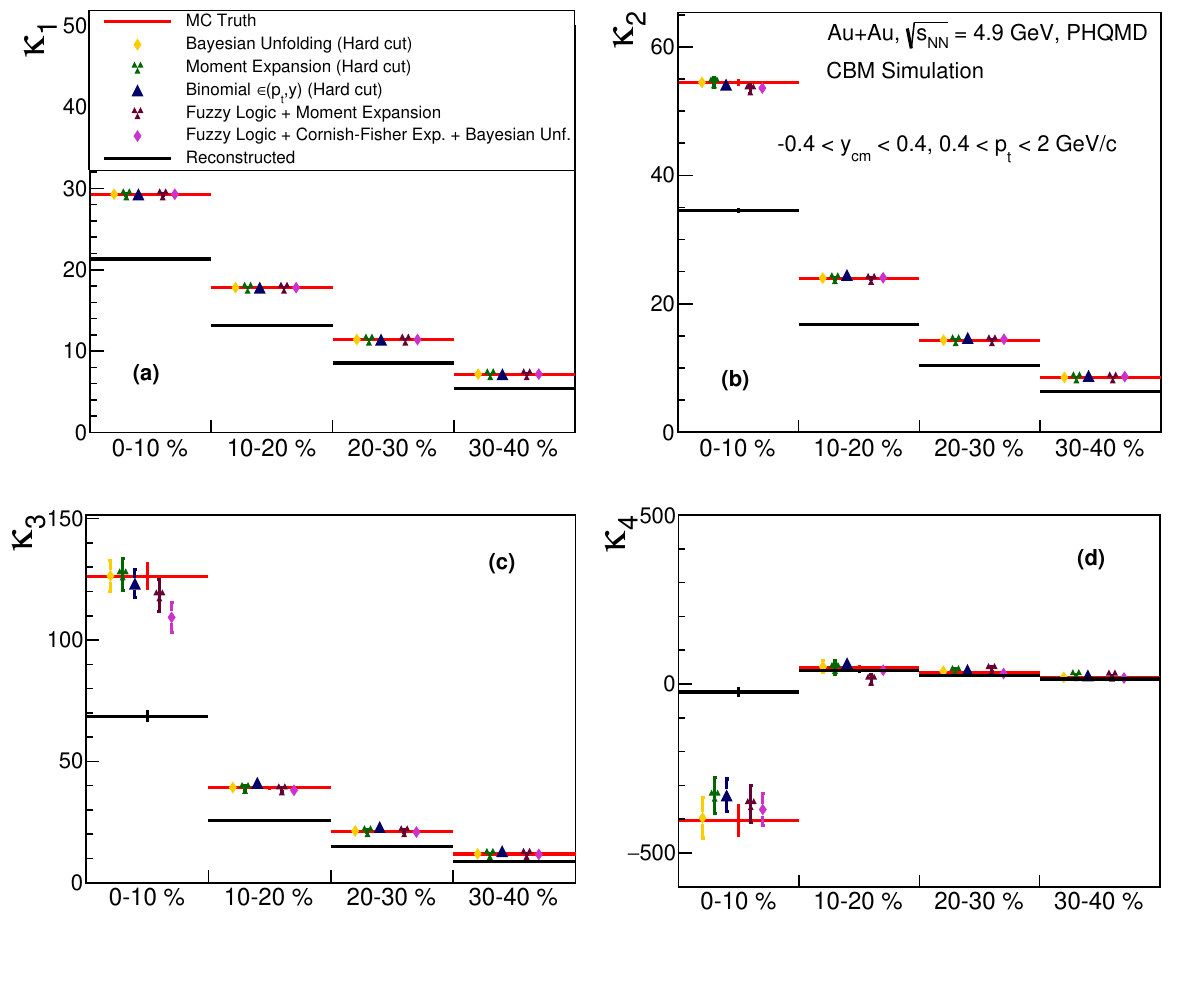}
\caption{Cumulants of the proton-number distributions for events generated with the PHQMD v5.2 model. Panels (a)–(d) display the first through fourth-order cumulants. The red lines represent the generated (true) cumulant values, while the black lines indicate the reconstructed values before efficiency correction. The error bars denote statistical uncertainties. The various colored markers correspond to results obtained using the cut-based and fuzzy-logic approaches after applying different efficiency-correction procedures. See text for details.}
\label{fig:fluc}
\end{figure}

The fuzzy logic approach constitutes a generalization of the Identity Method~\cite{Rustamov:2012bx, Arslandok:2018pcu}, which has already been applied in several experimental studies for the reconstruction of cumulants of multiplicity distributions up to third order~\cite{Anticic:2013htn, ALICE:2017jsh, ALICE:2019nbs, ALICE:2022xpf}.
For the efficiency correction, we use different approaches depending on the analysis method. For the cut-based analysis, we apply iterative Bayesian unfolding~\cite{DAgostini:2010hil}, the moment expansion technique~\cite{Nonaka:2018mgw}, and binomial unfolding~\cite{Nonaka:2017kko}. For the fuzzy logic approach, we employ the moment expansion technique and the Cornish–Fisher expansion~\cite{CornishFisher1938}, followed by Bayesian unfolding.

The obtained results are presented in \figref{fig:fluc}, where panels (a)–(d) show the proton number cumulants of first through fourth order. The generated cumulant values are denoted by red lines, while the reconstructed values prior to efficiency correction are indicated by black lines. The various colored symbols correspond to results obtained with the cut-based and fuzzy-logic approaches after applying different efficiency correction procedures. The error bars shown represent statistical uncertainties estimated using the subsampling technique~\cite{Anticic:2013htn}. Overall, a good agreement is achieved for all the cumulant orders between the reconstructed and generated results for both the cut-based and the fuzzy-logic approaches. This demonstrates that proton identification in CBM is rather clean, and applying cuts around the proton mass distribution does not lead to significant efficiency losses. The situation may, however, change when considering fluctuations of strange particles, such as kaons and/or light nuclei. In such cases, the fuzzy-logic approach is expected to provide significantly improved precision compared to the cut-based analysis~\cite{Nabroth:2025elh}. 

In summary, we have demonstrated that the CBM detector systems are well suited for high-precision measurements of proton-number fluctuations, a key prerequisite for exploring the phase structure of QCD in the high net-baryon density regime, where definitive conclusions have yet to be reached.

\subsection{Physics opportunities with elementary reactions}

Baseline measurements using proton or deuteron beams incident on hydrogen targets constitute a cornerstone of the CBM experimental program. Through high-precision and differential analyses of observables in proton-proton and neutron-proton interactions, these studies establish essential reference points, enabling a microscopic study of elementary hadron production processes. Furthermore, these elementary reactions create an ideal environment for studying the structure and spectroscopy of strange and charmed hadrons, and enable the systematic extraction of low-energy parameters through hadron-hadron final-state interactions. Particularly, the reconstruction of exclusive final states in elementary
proton-proton collisions at SIS100 energies has an exceptional physics potential.
 The extensive range of research opportunities available in this domain has been thoroughly outlined in a recent White Paper~\cite{qcdatfair2025}, which serves as the foundation for a dedicated hadron-physics research program currently under development and overarching various fields and their applications in the domain of non-perturbative QCD. 

\section{Conclusion}
The status and performance of the CBM detector, which is in construction at FAIR/GSI Darmstadt, was presented. We have illustrated the expected performance for the core CBM physics observables, namely multi-strange hyperons, hypernuclei, dileptons, collective flow, femtoscopy, and fluctuations. While the detector description in Monte Carlo simulation will be steadily improved, the current studies demonstrate that CBM will be able to bring a significant and unique contribution to the characterization of the hot and dense baryonic matter, including the discovery of its expected critical endpoint, which will be a landmark of hot QCD physics.

\section*{Acknowledgments}
The CBM Collaboration would like to thank the GSI/FAIR accelerator teams for excellent beam conditions within the FAIR Phase-0 program and all GSI/FAIR infrastructure departments for providing resources and excellent support.
This work was supported in part by the European Union’s Horizon 2020 research and innovation program EURIZON, the Bundesministerium für Forschung, Technologie und Raumfahrt (BMFTR, Germany), the GSI Helmholtzzentrum für Schwerionenforschung GmbH (Germany),the GSI R\&D Program with the Universities of Darmstadt, Frankfurt, Gießen, Heidelberg, Münster and Wuppertal (Germany), the Deutsche Forschungsgemeinschaft (DFG, Germany), the Helmholtz Forschungsakademie Hessen für FAIR (HFHF,Germany), the Helmholtz International Centre for FAIR (Germany), the ``Netzwerke 2021'', an initiative of the Ministry of Culture and Science of the State of Northrine Westphalia (Germany), the Frankfurt Institute for Advanced Studies (FIAS,Germany), the FAIR-CZ Project infrastructure program (Czech Republic), the FAIR-CZ Innovation program (Czech Republic),the Facility for Antiproton and Ion Research-participation of the Czech Republic (OPII" MEYSOP VVV and OPIII" MEYS OPVVV JAK program), the National Research, Development and Innovation Office (NKFIH, Hungary), the Hungarian OTKA fund, the National Research Foundation of Korea, the Department of Science and Technology (DST, India), the Japan Society for the Promotion of Science,the Polish Ministry of Education and Science program ``Premia na Horyzoncie 2'', the statutory funds of the Institute of Electronic Systems, Warsaw (Poland), the Warsaw University of Technology under the Excellence Initiative - Research University (ID-UB) program (Poland), the National Science Center (Poland), the Romanian Ministry of Education and Research/ National Research Authority, and the Institute of Atomic Physics (IFA, Romania).

\bibliography{CBM-overview}


\begin{thebibliography}{103}
\ifx \bisbn   \undefined \def \bisbn  #1{ISBN #1}\fi
\ifx \binits  \undefined \def \binits#1{#1}\fi
\ifx \bauthor  \undefined \def \bauthor#1{#1}\fi
\ifx \batitle  \undefined \def \batitle#1{#1}\fi
\ifx \bjtitle  \undefined \def \bjtitle#1{#1}\fi
\ifx \bvolume  \undefined \def \bvolume#1{\textbf{#1}}\fi
\ifx \byear  \undefined \def \byear#1{#1}\fi
\ifx \bissue  \undefined \def \bissue#1{#1}\fi
\ifx \bfpage  \undefined \def \bfpage#1{#1}\fi
\ifx \blpage  \undefined \def \blpage #1{#1}\fi
\ifx \burl  \undefined \def \burl#1{\textsf{#1}}\fi
\ifx \doiurl  \undefined \def \doiurl#1{\url{https://doi.org/#1}}\fi
\ifx \betal  \undefined \def \betal{\textit{et al.}}\fi
\ifx \binstitute  \undefined \def \binstitute#1{#1}\fi
\ifx \binstitutionaled  \undefined \def \binstitutionaled#1{#1}\fi
\ifx \bctitle  \undefined \def \bctitle#1{#1}\fi
\ifx \beditor  \undefined \def \beditor#1{#1}\fi
\ifx \bpublisher  \undefined \def \bpublisher#1{#1}\fi
\ifx \bbtitle  \undefined \def \bbtitle#1{#1}\fi
\ifx \bedition  \undefined \def \bedition#1{#1}\fi
\ifx \bseriesno  \undefined \def \bseriesno#1{#1}\fi
\ifx \blocation  \undefined \def \blocation#1{#1}\fi
\ifx \bsertitle  \undefined \def \bsertitle#1{#1}\fi
\ifx \bsnm \undefined \def \bsnm#1{#1}\fi
\ifx \bsuffix \undefined \def \bsuffix#1{#1}\fi
\ifx \bparticle \undefined \def \bparticle#1{#1}\fi
\ifx \barticle \undefined \def \barticle#1{#1}\fi
\bibcommenthead
\ifx \bconfdate \undefined \def \bconfdate #1{#1}\fi
\ifx \botherref \undefined \def \botherref #1{#1}\fi
\ifx \url \undefined \def \url#1{\textsf{#1}}\fi
\ifx \bchapter \undefined \def \bchapter#1{#1}\fi
\ifx \bbook \undefined \def \bbook#1{#1}\fi
\ifx \bcomment \undefined \def \bcomment#1{#1}\fi
\ifx \oauthor \undefined \def \oauthor#1{#1}\fi
\ifx \citeauthoryear \undefined \def \citeauthoryear#1{#1}\fi
\ifx \endbibitem  \undefined \def \endbibitem {}\fi
\ifx \bconflocation  \undefined \def \bconflocation#1{#1}\fi
\ifx \arxivurl  \undefined \def \arxivurl#1{\textsf{#1}}\fi
\csname PreBibitemsHook\endcsname

\bibitem[\protect\citeauthoryear{Ablyazimov et~al.}{2017}]{CBM:2016kpk}
\begin{barticle}
\bauthor{\bsnm{Ablyazimov}, \binits{T.}}, \betal:
\batitle{{Challenges in QCD matter physics --The scientific programme of the
  Compressed Baryonic Matter experiment at FAIR}}.
\bjtitle{Eur. Phys. J. A}
\bvolume{53}(\bissue{3}),
\bfpage{60}
(\byear{2017})
\doiurl{10.1140/epja/i2017-12248-y}
{\href{https://arxiv.org/abs/1607.01487}{{arXiv:1607.01487}}}
{[nucl-ex]}
\end{barticle}
\endbibitem

\bibitem[\protect\citeauthoryear{Andronic et~al.}{2018}]{Andronic:2017pug}
\begin{barticle}
\bauthor{\bsnm{Andronic}, \binits{A.}},
\bauthor{\bsnm{Braun-Munzinger}, \binits{P.}},
\bauthor{\bsnm{Redlich}, \binits{K.}},
\bauthor{\bsnm{Stachel}, \binits{J.}}:
\batitle{{Decoding the phase structure of QCD via particle production at high
  energy}}.
\bjtitle{Nature}
\bvolume{561}(\bissue{7723}),
\bfpage{321}--\blpage{330}
(\byear{2018})
\doiurl{10.1038/s41586-018-0491-6}
{\href{https://arxiv.org/abs/1710.09425}{{arXiv:1710.09425}}}
{[nucl-th]}
\end{barticle}
\endbibitem

\bibitem[\protect\citeauthoryear{Du et~al.}{2024}]{Du:2024wjm}
\begin{barticle}
\bauthor{\bsnm{Du}, \binits{L.}},
\bauthor{\bsnm{Sorensen}, \binits{A.}},
\bauthor{\bsnm{Stephanov}, \binits{M.}}:
\batitle{{The QCD phase diagram and Beam Energy Scan physics: A theory
  overview}}.
\bjtitle{Int. J. Mod. Phys. E}
\bvolume{33}(\bissue{07}),
\bfpage{2430008}
(\byear{2024})
\doiurl{10.1142/9789811294679_0007}
{\href{https://arxiv.org/abs/2402.10183}{{arXiv:2402.10183}}}
{[nucl-th]}
\end{barticle}
\endbibitem

\bibitem[\protect\citeauthoryear{Borsanyi and Parotto}{2025}]{Borsanyi:2025ttb}
\begin{botherref}
\oauthor{\bsnm{Borsanyi}, \binits{S.}},
\oauthor{\bsnm{Parotto}, \binits{P.}}:
{The QCD phase diagram}
(2025)
{\href{https://arxiv.org/abs/2512.08843}{{arXiv:2512.08843}}}
{[nucl-th]}
\end{botherref}
\endbibitem

\bibitem[\protect\citeauthoryear{Friman et~al.}{2011}]{Friman:2011zz}
\begin{bbook}
\beditor{\bsnm{Friman}, \binits{B.}},
\beditor{\bsnm{Hohne}, \binits{C.}},
\beditor{\bsnm{Knoll}, \binits{J.}},
\beditor{\bsnm{Leupold}, \binits{S.}},
\beditor{\bsnm{Randrup}, \binits{J.}},
\beditor{\bsnm{Rapp}, \binits{R.}},
\beditor{\bsnm{Senger}, \binits{P.}} (eds.):
\bbtitle{{The CBM Physics Book: Compressed Baryonic Matter in Laboratory
  Experiments}}
vol. \bseriesno{814},
pp. \bfpage{1}--\blpage{980}
(\byear{2011}).
\doiurl{10.1007/978-3-642-13293-3}
\end{bbook}
\endbibitem

\bibitem[\protect\citeauthoryear{Bzdak et~al.}{2020}]{Bzdak:2019pkr}
\begin{barticle}
\bauthor{\bsnm{Bzdak}, \binits{A.}},
\bauthor{\bsnm{Esumi}, \binits{S.}},
\bauthor{\bsnm{Koch}, \binits{V.}},
\bauthor{\bsnm{Liao}, \binits{J.}},
\bauthor{\bsnm{Stephanov}, \binits{M.}},
\bauthor{\bsnm{Xu}, \binits{N.}}:
\batitle{{Mapping the Phases of Quantum Chromodynamics with Beam Energy Scan}}.
\bjtitle{Phys. Rept.}
\bvolume{853},
\bfpage{1}--\blpage{87}
(\byear{2020})
\doiurl{10.1016/j.physrep.2020.01.005}
{\href{https://arxiv.org/abs/1906.00936}{{arXiv:1906.00936}}}
{[nucl-th]}
\end{barticle}
\endbibitem

\bibitem[\protect\citeauthoryear{Sorensen et~al.}{2024}]{Sorensen:2023zkk}
\begin{barticle}
\bauthor{\bsnm{Sorensen}, \binits{A.}}, \betal:
\batitle{{Dense nuclear matter equation of state from heavy-ion collisions}}.
\bjtitle{Prog. Part. Nucl. Phys.}
\bvolume{134},
\bfpage{104080}
(\byear{2024})
\doiurl{10.1016/j.ppnp.2023.104080}
{\href{https://arxiv.org/abs/2301.13253}{{arXiv:2301.13253}}}
{[nucl-th]}
\end{barticle}
\endbibitem

\bibitem[\protect\citeauthoryear{Rapp and Wambach}{2000}]{Rapp:1999ej}
\begin{barticle}
\bauthor{\bsnm{Rapp}, \binits{R.}},
\bauthor{\bsnm{Wambach}, \binits{J.}}:
\batitle{{Chiral symmetry restoration and dileptons in relativistic heavy ion
  collisions}}.
\bjtitle{Adv. Nucl. Phys.}
\bvolume{25},
\bfpage{1}
(\byear{2000})
\doiurl{10.1007/0-306-47101-9_1}
{\href{https://arxiv.org/abs/hep-ph/9909229}{{arXiv:hep-ph/9909229}}}
\end{barticle}
\endbibitem

\bibitem[\protect\citeauthoryear{Salabura and Stroth}{2021}]{Salabura:2020tou}
\begin{barticle}
\bauthor{\bsnm{Salabura}, \binits{P.}},
\bauthor{\bsnm{Stroth}, \binits{J.}}:
\batitle{{Dilepton radiation from strongly interacting systems}}.
\bjtitle{Prog. Part. Nucl. Phys.}
\bvolume{120},
\bfpage{103869}
(\byear{2021})
\doiurl{10.1016/j.ppnp.2021.103869}
{\href{https://arxiv.org/abs/2005.14589}{{arXiv:2005.14589}}}
{[nucl-ex]}
\end{barticle}
\endbibitem

\bibitem[\protect\citeauthoryear{Geurts and Tripolt}{2023}]{Geurts:2022xmk}
\begin{barticle}
\bauthor{\bsnm{Geurts}, \binits{F.}},
\bauthor{\bsnm{Tripolt}, \binits{R.-A.}}:
\batitle{{Electromagnetic probes: Theory and experiment}}.
\bjtitle{Prog. Part. Nucl. Phys.}
\bvolume{128},
\bfpage{104004}
(\byear{2023})
\doiurl{10.1016/j.ppnp.2022.104004}
{\href{https://arxiv.org/abs/2210.01622}{{arXiv:2210.01622}}}
{[hep-ph]}
\end{barticle}
\endbibitem

\bibitem[\protect\citeauthoryear{Fischer and Pawlowski}{2026}]{Fischer:2026uni}
\begin{botherref}
\oauthor{\bsnm{Fischer}, \binits{C.S.}},
\oauthor{\bsnm{Pawlowski}, \binits{J.M.}}:
{Phase structure and observables at high densities from first principles QCD}
(2026)
{\href{https://arxiv.org/abs/2603.11135}{{arXiv:2603.11135}}}
{[hep-ph]}
\end{botherref}
\endbibitem

\bibitem[\protect\citeauthoryear{Braun-Munzinger
  et~al.}{2026}]{Braun-Munzinger:2026krf}
\begin{botherref}
\oauthor{\bsnm{Braun-Munzinger}, \binits{P.}},
\oauthor{\bsnm{Rustamov}, \binits{A.}},
\oauthor{\bsnm{Xu}, \binits{N.}}:
{The phase structure of QCD: Fluctuations and Correlations}
(2026)
\doiurl{10.1146/annurev-nucl-100324-014902}
{\href{https://arxiv.org/abs/2601.18666}{{arXiv:2601.18666}}}
{[nucl-ex]}
\end{botherref}
\endbibitem

\bibitem[\protect\citeauthoryear{Andronic et~al.}{2011}]{Andronic:2010qu}
\begin{barticle}
\bauthor{\bsnm{Andronic}, \binits{A.}},
\bauthor{\bsnm{Braun-Munzinger}, \binits{P.}},
\bauthor{\bsnm{Stachel}, \binits{J.}},
\bauthor{\bsnm{Stocker}, \binits{H.}}:
\batitle{{Production of light nuclei, hypernuclei and their antiparticles in
  relativistic nuclear collisions}}.
\bjtitle{Phys. Lett. B}
\bvolume{697},
\bfpage{203}--\blpage{207}
(\byear{2011})
\doiurl{10.1016/j.physletb.2011.01.053}
{\href{https://arxiv.org/abs/1010.2995}{{arXiv:1010.2995}}}
{[nucl-th]}
\end{barticle}
\endbibitem

\bibitem[\protect\citeauthoryear{Balassa and Wolf}{2023}]{Balassa2023}
\begin{barticle}
\bauthor{\bsnm{Balassa}, \binits{G.}},
\bauthor{\bsnm{Wolf}, \binits{G.}}:
\batitle{Hypernuclei production with a modified coalescence model in buu
  transport calculations}.
\bjtitle{The European Physical Journal A}
\bvolume{59}(\bissue{4}),
\bfpage{89}
(\byear{2023})
\doiurl{10.1140/epja/s10050-023-01014-7}
\end{barticle}
\endbibitem

\bibitem[\protect\citeauthoryear{Project}{2020}]{FAIR:operation_modes}
\begin{botherref}
\oauthor{\bsnm{Project}, \binits{F.}}:
FAIR Operation Modes -- Reference Modes for the Modularized Start Version.
EDMS-2374493 (requires CERN login)
(2020).
\url{https://edms.cern.ch/document/2374493}
\end{botherref}
\endbibitem

\bibitem[\protect\citeauthoryear{{Akishina, Valentina}
  et~al.}{2018}]{cbm:time-based}
\begin{barticle}
\bauthor{\bsnm{{Akishina, Valentina}}},
\bauthor{\bsnm{{Kisel, Ivan}}},
\bauthor{\bsnm{{Vassiliev, Iouri}}},
\bauthor{\bsnm{{Zyzak, Maksym}}}:
\batitle{Time-based reconstruction of free-streaming data in cbm}.
\bjtitle{EPJ Web Conf.}
\bvolume{173},
\bfpage{04002}
(\byear{2018})
\doiurl{10.1051/epjconf/201817304002}
\end{barticle}
\endbibitem

\bibitem[\protect\citeauthoryear{{Akishina, Valentina} and {Kisel,
  Ivan}}{2018}]{cbm:online-ev-reco}
\begin{barticle}
\bauthor{\bsnm{{Akishina, Valentina}}},
\bauthor{\bsnm{{Kisel, Ivan}}}:
\batitle{Online event reconstruction in the cbm experiment at fair}.
\bjtitle{EPJ Web Conf.}
\bvolume{173},
\bfpage{01002}
(\byear{2018})
\doiurl{10.1051/epjconf/201817301002}
\end{barticle}
\endbibitem

\bibitem[\protect\citeauthoryear{Friese and for~the
  CBM~Collaboration}{2017}]{Friese_2017}
\begin{barticle}
\bauthor{\bsnm{Friese}, \binits{V.}},
\bauthor{\bsnm{CBM~Collaboration}}:
\batitle{The high-rate data challenge: computing for the cbm experiment}.
\bjtitle{Journal of Physics: Conference Series}
\bvolume{898}(\bissue{11}),
\bfpage{112003}
(\byear{2017})
\doiurl{10.1088/1742-6596/898/11/112003}
\end{barticle}
\endbibitem

\bibitem[\protect\citeauthoryear{Pietraszko et~al.}{2010}]{bmon1}
\begin{barticle}
\bauthor{\bsnm{Pietraszko}, \binits{J.}},
\bauthor{\bsnm{Fabbietti}, \binits{L.}},
\bauthor{\bsnm{Koenig}, \binits{W.}},
\bauthor{\bsnm{Weber}, \binits{M.}}:
\batitle{Diamonds as timing detectors for minimum-ionizing particles: The hades
  proton-beam monitor and start signal detectors for time of flight
  measurements}.
\bjtitle{Nuclear Instruments and Methods in Physics Research Section A:
  Accelerators, Spectrometers, Detectors and Associated Equipment}
\bvolume{618}(\bissue{1}),
\bfpage{121}--\blpage{123}
(\byear{2010})
\doiurl{10.1016/j.nima.2010.02.113}
\end{barticle}
\endbibitem

\bibitem[\protect\citeauthoryear{Pietraszko et~al.}{2014}]{bmon2}
\begin{botherref}
\oauthor{\bsnm{Pietraszko}, \binits{J.}},
\oauthor{\bsnm{Koenig}, \binits{W.}},
\oauthor{\bsnm{HADES-Collaboration}}:
Radiation damage in single crystal cvd diamond material investigated with a
  high current au beam.
Verhandlungen der Deutschen Physikalischen Gesellschaft
(Frankfurt 2014 issue),
1
(2014)
\end{botherref}
\endbibitem

\bibitem[\protect\citeauthoryear{Pietraszko et~al.}{2025}]{Pietraszko:2025kxv}
\begin{barticle}
\bauthor{\bsnm{Pietraszko}, \binits{J.}}, \betal:
\batitle{{LGAD technology for heavy ion detection and radiation damage
  diagnostics in diamond}}.
\bjtitle{JINST}
\bvolume{20}(\bissue{08}),
\bfpage{08007}
(\byear{2025})
\doiurl{10.1088/1748-0221/20/08/C08007}
\end{barticle}
\endbibitem

\bibitem[\protect\citeauthoryear{Deveaux et~al.}{2022}]{tdr:mvd}
\begin{bbook}
\beditor{\bsnm{Deveaux}, \binits{M.}},
\beditor{\bsnm{Klaus}, \binits{P.}},
\beditor{\bsnm{Koziel}, \binits{M.}},
\beditor{\bsnm{Michel}, \binits{J.}},
\beditor{\bsnm{Müntz}, \binits{C.}},
\beditor{\bsnm{Stroth}, \binits{J.}} (eds.):
\bbtitle{{T}echnical {D}esign {R}eport for the {CBM} {M}icro {V}ertex
  {D}etector}.
\bsertitle{CBM Technical Design Reports},
p. \bfpage{157}.
\bpublisher{GSI}, \blocation{???}
(\byear{2022}).
\burl{https://repository.gsi.de/record/246516}
\end{bbook}
\endbibitem

\bibitem[\protect\citeauthoryear{Matejcek et~al.}{2025}]{Matejcek:2025fdh}
\begin{barticle}
\bauthor{\bsnm{Matejcek}, \binits{F.}}, \betal:
\batitle{{Integration concept of the CBM Micro Vertex Detector}}.
\bjtitle{JINST}
\bvolume{20}(\bissue{06}),
\bfpage{06024}
(\byear{2025})
\doiurl{10.1088/1748-0221/20/06/C06024}
{\href{https://arxiv.org/abs/2502.04858}{{arXiv:2502.04858}}}
{[physics.ins-det]}
\end{barticle}
\endbibitem

\bibitem[\protect\citeauthoryear{Deveaux et~al.}{2025}]{Deveaux:2025sxl}
\begin{barticle}
\bauthor{\bsnm{Deveaux}, \binits{M.}}, \betal:
\batitle{{In beam performances of the MIMOSIS-2.1 CMOS Monolithic Active Pixel
  Sensor}}.
\bjtitle{JINST}
\bvolume{20}(\bissue{07}),
\bfpage{07056}
(\byear{2025})
\doiurl{10.1088/1748-0221/20/07/C07056}
{\href{https://arxiv.org/abs/2502.05303}{{arXiv:2502.05303}}}
{[physics.ins-det]}
\end{barticle}
\endbibitem

\bibitem[\protect\citeauthoryear{Darwish et~al.}{2025}]{Darwish:2025ega}
\begin{barticle}
\bauthor{\bsnm{Darwish}, \binits{H.}}, \betal:
\batitle{{In-beam performance of neutron irradiated MIMOSIS-1 CMOS Monolithic
  Active Pixel Sensors}}.
\bjtitle{JINST}
\bvolume{20}(\bissue{08}),
\bfpage{08009}
(\byear{2025})
\doiurl{10.1088/1748-0221/20/08/C08009}
\end{barticle}
\endbibitem

\bibitem[\protect\citeauthoryear{Heuser et~al.}{2013}]{Heuser:54798}
\begin{bbook}
\beditor{\bsnm{Heuser}, \binits{J.}},
\beditor{\bsnm{Müller}, \binits{W.}},
\beditor{\bsnm{Pugatch}, \binits{V.}},
\beditor{\bsnm{Senger}, \binits{P.}},
\beditor{\bsnm{Schmidt}, \binits{C.J.}},
\beditor{\bsnm{Sturm}, \binits{C.}},
\beditor{\bsnm{Frankenfeld}, \binits{U.}} (eds.):
\bbtitle{[{GSI} {R}eport 2013-4] {T}echnical {D}esign {R}eport for the {CBM}
  {S}ilicon {T}racking {S}ystem ({STS})},
p. \bfpage{167}.
\bpublisher{GSI},
\blocation{Darmstadt}
(\byear{2013}).
\burl{https://repository.gsi.de/record/54798}
\end{bbook}
\endbibitem

\bibitem[\protect\citeauthoryear{Teklishyn et~al.}{2025}]{CBM:2025mnp}
\begin{barticle}
\bauthor{\bsnm{Teklishyn}, \binits{M.}}, \betal:
\batitle{{Minimal material, maximum coverage: Silicon Tracking System for
  high-occupancy conditions}}.
\bjtitle{Nucl. Instrum. Meth. A}
\bvolume{1080},
\bfpage{170714}
(\byear{2025})
\doiurl{10.1016/j.nima.2025.170714}
{\href{https://arxiv.org/abs/2503.15721}{{arXiv:2503.15721}}}
{[physics.ins-det]}
\end{barticle}
\endbibitem

\bibitem[\protect\citeauthoryear{Agarwal et~al.}{2026}]{CBM:2025voh}
\begin{barticle}
\bauthor{\bsnm{Agarwal}, \binits{A.}}, \betal:
\batitle{{Performance of the prototype Silicon Tracking System of the CBM
  experiment tested with heavy-ion beams at SIS18}}.
\bjtitle{Nucl. Instrum. Meth. A}
\bvolume{1082},
\bfpage{171059}
(\byear{2026})
\doiurl{10.1016/j.nima.2025.171059}
{\href{https://arxiv.org/abs/2505.20517}{{arXiv:2505.20517}}}
{[physics.ins-det]}
\end{barticle}
\endbibitem

\bibitem[\protect\citeauthoryear{Collaboration}{2025}]{cbmSTS2025}
\begin{botherref}
\oauthor{\bsnm{Collaboration}, \binits{T.C.}}:
Performance of the prototype Silicon Tracking System of the CBM experiment
  tested with heavy-ion beams at SIS18
(2025).
\url{https://arxiv.org/abs/2505.20517}
\end{botherref}
\endbibitem

\bibitem[\protect\citeauthoryear{Becker et~al.}{2024}]{TRB:2023asr}
\begin{barticle}
\bauthor{\bsnm{Becker}, \binits{M.}}, \betal:
\batitle{{Status of the development of the RICH detector for CBM including a
  mRICH prototype in mCBM}}.
\bjtitle{Nucl. Instrum. Meth. A}
\bvolume{1059},
\bfpage{168832}
(\byear{2024})
\doiurl{10.1016/j.nima.2023.168832}
\end{barticle}
\endbibitem

\bibitem[\protect\citeauthoryear{Michel et~al.}{2017}]{TRB:2017xkj}
\begin{barticle}
\bauthor{\bsnm{Michel}, \binits{J.}}, \betal:
\batitle{{Electronics for the RICH detectors of the HADES and CBM
  experiments}}.
\bjtitle{JINST}
\bvolume{12}(\bissue{01}),
\bfpage{01072}
(\byear{2017})
\doiurl{10.1088/1748-0221/12/01/C01072}
\end{barticle}
\endbibitem

\bibitem[\protect\citeauthoryear{Meyer-Ahrens}{2025}]{MeyerAhrens:2025}
\begin{botherref}
\oauthor{\bsnm{Meyer-Ahrens}, \binits{A.}}:
Dielectron performance of the {CBM} experiment.
Phd thesis,
Universität Münster
(2025)
\end{botherref}
\endbibitem

\bibitem[\protect\citeauthoryear{Subramani}{2025}]{Subramani:2025}
\begin{botherref}
\oauthor{\bsnm{Subramani}, \binits{P.}}:
Performance qualification of the {DIRICH} readout system and development of a
  novel electron reconstruction scheme for the {CBM} experiment.
Phd thesis,
Bergische Universität Wuppertal
(2025)
\end{botherref}
\endbibitem

\bibitem[\protect\citeauthoryear{Dubey et~al.}{2013}]{Dubey:2013hva}
\begin{barticle}
\bauthor{\bsnm{Dubey}, \binits{A.K.}},
\bauthor{\bsnm{Prakash}, \binits{A.}},
\bauthor{\bsnm{Chattopadhyay}, \binits{S.}},
\bauthor{\bsnm{Ganti}, \binits{M.S.}},
\bauthor{\bsnm{Singaraju}, \binits{R.}},
\bauthor{\bsnm{Saini}, \binits{J.}},
\bauthor{\bsnm{Singh}, \binits{B.K.}},
\bauthor{\bsnm{Viyogi}, \binits{Y.P.}}:
\batitle{{GEM detector development for CBM experiment at FAIR}}.
\bjtitle{Nucl. Instrum. Meth. A}
\bvolume{718},
\bfpage{418}--\blpage{420}
(\byear{2013})
\doiurl{10.1016/j.nima.2012.10.043}
\end{barticle}
\endbibitem

\bibitem[\protect\citeauthoryear{Ghosh et~al.}{2025}]{Ghosh:2025hmc}
\begin{barticle}
\bauthor{\bsnm{Ghosh}, \binits{C.}},
\bauthor{\bsnm{Sharma}, \binits{A.K.}},
\bauthor{\bsnm{Gudla}, \binits{S.}},
\bauthor{\bsnm{Saini}, \binits{J.}},
\bauthor{\bsnm{Dubey}, \binits{A.K.}},
\bauthor{\bsnm{Bandyopadhyay}, \binits{A.}}:
\batitle{{First prototype of MuCh second station GEM module: design,
  fabrication and testing for CBM experiment at GSI, Germany}}.
\bjtitle{JINST}
\bvolume{20}(\bissue{09}),
\bfpage{09005}
(\byear{2025})
\doiurl{10.1088/1748-0221/20/09/P09005}
\end{barticle}
\endbibitem

\bibitem[\protect\citeauthoryear{Agarwal et~al.}{2025}]{Agarwal:2025kle}
\begin{barticle}
\bauthor{\bsnm{Agarwal}, \binits{A.}}, \betal:
\batitle{{Testing a large size triple GEM detector for the first station of the
  CBM-Muon Chambers with a high-intensity gamma source at GIF++ under
  large-area illumination}}.
\bjtitle{JINST}
\bvolume{20}(\bissue{04}),
\bfpage{04022}
(\byear{2025})
\doiurl{10.1088/1748-0221/20/04/P04022}
{\href{https://arxiv.org/abs/2504.18445}{{arXiv:2504.18445}}}
{[hep-ex]}
\end{barticle}
\endbibitem

\bibitem[\protect\citeauthoryear{Tolba et~al.}{2022}]{Tolba:2022oxn}
\begin{barticle}
\bauthor{\bsnm{Tolba}, \binits{T.}},
\bauthor{\bsnm{Grzonka}, \binits{D.}},
\bauthor{\bsnm{Sefzick}, \binits{T.}},
\bauthor{\bsnm{Ritman}, \binits{J.}}:
\batitle{{Irradiation studies of silicon photomultipliers with proton beam from
  the JULIC cyclotron}}.
\bjtitle{Nucl. Instrum. Meth. A}
\bvolume{1025},
\bfpage{166130}
(\byear{2022})
\doiurl{10.1016/j.nima.2021.166130}
\end{barticle}
\endbibitem

\bibitem[\protect\citeauthoryear{Bondorf et~al.}{1995}]{Bondorf:1995ua}
\begin{barticle}
\bauthor{\bsnm{Bondorf}, \binits{J.P.}},
\bauthor{\bsnm{Botvina}, \binits{A.S.}},
\bauthor{\bsnm{Ilinov}, \binits{A.S.}},
\bauthor{\bsnm{Mishustin}, \binits{I.N.}},
\bauthor{\bsnm{Sneppen}, \binits{K.}}:
\batitle{{Statistical multifragmentation of nuclei}}.
\bjtitle{Phys. Rept.}
\bvolume{257},
\bfpage{133}--\blpage{221}
(\byear{1995})
\doiurl{10.1016/0370-1573(94)00097-M}
\end{barticle}
\endbibitem

\bibitem[\protect\citeauthoryear{Baznat et~al.}{2020}]{Baznat:2019iom}
\begin{barticle}
\bauthor{\bsnm{Baznat}, \binits{M.}},
\bauthor{\bsnm{Botvina}, \binits{A.}},
\bauthor{\bsnm{Musulmanbekov}, \binits{G.}},
\bauthor{\bsnm{Toneev}, \binits{V.}},
\bauthor{\bsnm{Zhezher}, \binits{V.}}:
\batitle{{Monte-Carlo Generator of Heavy Ion Collisions DCM-SMM}}.
\bjtitle{Phys. Part. Nucl. Lett.}
\bvolume{17}(\bissue{3}),
\bfpage{303}--\blpage{324}
(\byear{2020})
\doiurl{10.1134/S1547477120030024}
{\href{https://arxiv.org/abs/1912.09277}{{arXiv:1912.09277}}}
{[nucl-th]}
\end{barticle}
\endbibitem

\bibitem[\protect\citeauthoryear{Cuveland et~al.}{2023}]{ONLINE_TDR}
\begin{botherref}
{T}echnical {D}esign {R}eport for the {CBM} {O}nline {S}ystems – {P}art {I},
  {DAQ} and {FLES} {E}ntry {S}tage.
Technical Report~-,
Darmstadt
(2023).
\doiurl{10.15120/GSI-2023-00739} .
FAIR Technical Design Report; ccby4.
\url{https://repository.gsi.de/record/340597}
\end{botherref}
\endbibitem

\bibitem[\protect\citeauthoryear{Moreira et~al.}{2009}]{Moreira:1235836}
\begin{botherref}
\oauthor{\bsnm{Moreira}, \binits{P.}}, et al.:
{The GBT Project}
(2009)
\doiurl{10.5170/CERN-2009-006.342}
\end{botherref}
\endbibitem

\bibitem[\protect\citeauthoryear{}{2017}]{mCBM_TDR}
\begin{botherref}
m{CBM}@{SIS}18.
Technical Report CBM,
Darmstadt
(2017).
\doiurl{10.15120/GSI-2019-00977} .
\url{https://repository.gsi.de/record/220072}
\end{botherref}
\endbibitem

\bibitem[\protect\citeauthoryear{Agarwal et~al.}{2026}]{CBM:2026kqm}
\begin{botherref}
\oauthor{\bsnm{Agarwal}, \binits{A.}}, et al.:
{Demonstrating CBM Capabilities by $\Lambda$ Baryon Reconstruction in Ni+Ni
  Collisions with the mCBM Experiment at SIS18 of GSI/FAIR}
(2026)
{\href{https://arxiv.org/abs/2606.01971}{{arXiv:2606.01971}}}
{[physics.ins-det]}
\end{botherref}
\endbibitem

\bibitem[\protect\citeauthoryear{Bass et~al.}{1998}]{Bass:1998ca}
\begin{barticle}
\bauthor{\bsnm{Bass}, \binits{S.}}, \betal:
\batitle{{Microscopic models for ultrarelativistic heavy ion collisions}}.
\bjtitle{Prog. Part. Nucl. Phys.}
\bvolume{41},
\bfpage{255}--\blpage{369}
(\byear{1998})
\doiurl{10.1016/S0146-6410(98)00058-1}
{\href{https://arxiv.org/abs/nucl-th/9803035}{{arXiv:nucl-th/9803035}}}
\end{barticle}
\endbibitem

\bibitem[\protect\citeauthoryear{Aichelin et~al.}{2020}]{Aichelin:2019tnk}
\begin{barticle}
\bauthor{\bsnm{Aichelin}, \binits{J.}},
\bauthor{\bsnm{Bratkovskaya}, \binits{E.}},
\bauthor{\bsnm{Le~F{\`e}vre}, \binits{A.}},
\bauthor{\bsnm{Kireyeu}, \binits{V.}},
\bauthor{\bsnm{Kolesnikov}, \binits{V.}},
\bauthor{\bsnm{Leifels}, \binits{Y.}},
\bauthor{\bsnm{Voronyuk}, \binits{V.}},
\bauthor{\bsnm{Coci}, \binits{G.}}:
\batitle{{Parton-hadron-quantum-molecular dynamics: A novel microscopic $n$
  -body transport approach for heavy-ion collisions, dynamical cluster
  formation, and hypernuclei production}}.
\bjtitle{Phys. Rev. C}
\bvolume{101}(\bissue{4}),
\bfpage{044905}
(\byear{2020})
\doiurl{10.1103/PhysRevC.101.044905}
{\href{https://arxiv.org/abs/1907.03860}{{arXiv:1907.03860}}}
{[nucl-th]}
\end{barticle}
\endbibitem

\bibitem[\protect\citeauthoryear{Bleicher and
  Bratkovskaya}{2022}]{Bleicher:2022kcu}
\begin{barticle}
\bauthor{\bsnm{Bleicher}, \binits{M.}},
\bauthor{\bsnm{Bratkovskaya}, \binits{E.}}:
\batitle{{Modelling relativistic heavy-ion collisions with dynamical transport
  approaches}}.
\bjtitle{Prog. Part. Nucl. Phys.}
\bvolume{122},
\bfpage{103920}
(\byear{2022})
\doiurl{10.1016/j.ppnp.2021.103920}
\end{barticle}
\endbibitem

\bibitem[\protect\citeauthoryear{Rafelski and Muller}{1982}]{Rafelski:1982pu}
\begin{barticle}
\bauthor{\bsnm{Rafelski}, \binits{J.}},
\bauthor{\bsnm{Muller}, \binits{B.}}:
\batitle{{Strangeness Production in the Quark - Gluon Plasma}}.
\bjtitle{Phys. Rev. Lett.}
\bvolume{48},
\bfpage{1066}
(\byear{1982})
\doiurl{10.1103/PhysRevLett.48.1066} .
\bcomment{[Erratum: Phys.Rev.Lett. 56, 2334 (1986)]}
\end{barticle}
\endbibitem

\bibitem[\protect\citeauthoryear{Koch et~al.}{1983}]{Koch:1982ij}
\begin{barticle}
\bauthor{\bsnm{Koch}, \binits{P.}},
\bauthor{\bsnm{Rafelski}, \binits{J.}},
\bauthor{\bsnm{Greiner}, \binits{W.}}:
\batitle{{Strange hadron in hot nuclear matter}}.
\bjtitle{Phys. Lett. B}
\bvolume{123},
\bfpage{151}--\blpage{154}
(\byear{1983})
\doiurl{10.1016/0370-2693(83)90411-2}
\end{barticle}
\endbibitem

\bibitem[\protect\citeauthoryear{Koch et~al.}{1986}]{Koch:1986ud}
\begin{barticle}
\bauthor{\bsnm{Koch}, \binits{P.}},
\bauthor{\bsnm{Muller}, \binits{B.}},
\bauthor{\bsnm{Rafelski}, \binits{J.}}:
\batitle{{Strangeness in Relativistic Heavy Ion Collisions}}.
\bjtitle{Phys. Rept.}
\bvolume{142},
\bfpage{167}--\blpage{262}
(\byear{1986})
\doiurl{10.1016/0370-1573(86)90096-7}
\end{barticle}
\endbibitem

\bibitem[\protect\citeauthoryear{Blume}{2005}]{Blume:2005ru}
\begin{barticle}
\bauthor{\bsnm{Blume}, \binits{C.}}:
\batitle{{Energy dependence of hadronic observables}}.
\bjtitle{J. Phys. G}
\bvolume{31},
\bfpage{57}--\blpage{68}
(\byear{2005})
\doiurl{10.1088/0954-3899/31/4/008}
\end{barticle}
\endbibitem

\bibitem[\protect\citeauthoryear{Blume and Markert}{2011}]{Blume:2011sb}
\begin{barticle}
\bauthor{\bsnm{Blume}, \binits{C.}},
\bauthor{\bsnm{Markert}, \binits{C.}}:
\batitle{{Strange hadron production in heavy ion collisions from SPS to RHIC}}.
\bjtitle{Prog. Part. Nucl. Phys.}
\bvolume{66},
\bfpage{834}--\blpage{879}
(\byear{2011})
\doiurl{10.1016/j.ppnp.2011.05.001}
{\href{https://arxiv.org/abs/1105.2798}{{arXiv:1105.2798}}}
{[nucl-ex]}
\end{barticle}
\endbibitem

\bibitem[\protect\citeauthoryear{Piasecki and
  Piotrowski}{2023}]{Piasecki:2023hoo}
\begin{barticle}
\bauthor{\bsnm{Piasecki}, \binits{K.}},
\bauthor{\bsnm{Piotrowski}, \binits{P.}}:
\batitle{{Systematics of yields of strange hadrons from heavy-ion collisions
  around threshold energies}}.
\bjtitle{Eur. Phys. J. A}
\bvolume{59}(\bissue{11}),
\bfpage{272}
(\byear{2023})
\doiurl{10.1140/epja/s10050-023-01182-6}
{\href{https://arxiv.org/abs/2305.13760}{{arXiv:2305.13760}}}
{[nucl-ex]}
\end{barticle}
\endbibitem

\bibitem[\protect\citeauthoryear{Gl{\"a}{\ss}el et~al.}{2022}]{Glassel:2021rod}
\begin{barticle}
\bauthor{\bsnm{Gl{\"a}{\ss}el}, \binits{S.}},
\bauthor{\bsnm{Kireyeu}, \binits{V.}},
\bauthor{\bsnm{Voronyuk}, \binits{V.}},
\bauthor{\bsnm{Aichelin}, \binits{J.}},
\bauthor{\bsnm{Blume}, \binits{C.}},
\bauthor{\bsnm{Bratkovskaya}, \binits{E.}},
\bauthor{\bsnm{Coci}, \binits{G.}},
\bauthor{\bsnm{Kolesnikov}, \binits{V.}},
\bauthor{\bsnm{Winn}, \binits{M.}}:
\batitle{{Cluster and hypercluster production in relativistic heavy-ion
  collisions within the parton-hadron-quantum-molecular-dynamics approach}}.
\bjtitle{Phys. Rev. C}
\bvolume{105}(\bissue{1}),
\bfpage{014908}
(\byear{2022})
\doiurl{10.1103/PhysRevC.105.014908}
{\href{https://arxiv.org/abs/2106.14839}{{arXiv:2106.14839}}}
{[nucl-th]}
\end{barticle}
\endbibitem

\bibitem[\protect\citeauthoryear{Messchendorp et~al.}{2025}]{qcdatfair2025}
\begin{botherref}
\oauthor{\bsnm{Messchendorp}, \binits{J.G.}},
\oauthor{\bsnm{Nerling}, \binits{F.}}, et al.:
{Hadron Physics Opportunities at FAIR}
(2025).
\url{https://arxiv.org/abs/2512.15986}
\end{botherref}
\endbibitem

\bibitem[\protect\citeauthoryear{Reichert et~al.}{2023}]{Reichert:2022mek}
\begin{barticle}
\bauthor{\bsnm{Reichert}, \binits{T.}},
\bauthor{\bsnm{Steinheimer}, \binits{J.}},
\bauthor{\bsnm{Vovchenko}, \binits{V.}},
\bauthor{\bsnm{D{\"o}nigus}, \binits{B.}},
\bauthor{\bsnm{Bleicher}, \binits{M.}}:
\batitle{{Energy dependence of light hypernuclei production in heavy-ion
  collisions from a coalescence and statistical-thermal model perspective}}.
\bjtitle{Phys. Rev. C}
\bvolume{107}(\bissue{1}),
\bfpage{014912}
(\byear{2023})
\doiurl{10.1103/PhysRevC.107.014912}
{\href{https://arxiv.org/abs/2210.11876}{{arXiv:2210.11876}}}
{[nucl-th]}
\end{barticle}
\endbibitem

\bibitem[\protect\citeauthoryear{Schnedermann~E. and
  Heinz}{1993}]{Schnedermann:1993}
\begin{barticle}
\bauthor{\bsnm{Schnedermann~E.}, \binits{S.J.}},
\bauthor{\bsnm{Heinz}, \binits{U.}}:
\batitle{{Thermal phenomenology of hadrons from 200 A GeV S+S collisions}}.
\bjtitle{Phys. ReV. C}
\bvolume{48},
\bfpage{2462}
(\byear{1993})
\doiurl{10.1103/PhysRevC.48.2462}
{\href{https://arxiv.org/abs/9307020}{{arXiv:9307020}}}
{[nucl-th]}
\end{barticle}
\endbibitem

\bibitem[\protect\citeauthoryear{Rapp and van Hees}{2016}]{Rapp:2016xzw}
\begin{barticle}
\bauthor{\bsnm{Rapp}, \binits{R.}},
\bauthor{\bsnm{Hees}, \binits{H.}}:
\batitle{{Thermal Electromagnetic Radiation in Heavy-Ion Collisions}}.
\bjtitle{Eur. Phys. J. A}
\bvolume{52}(\bissue{8}),
\bfpage{257}
(\byear{2016})
\doiurl{10.1140/epja/i2016-16257-0}
{\href{https://arxiv.org/abs/1608.05279}{{arXiv:1608.05279}}}
{[hep-ph]}
\end{barticle}
\endbibitem

\bibitem[\protect\citeauthoryear{Jung et~al.}{2017}]{Jung:2016yxl}
\begin{barticle}
\bauthor{\bsnm{Jung}, \binits{C.}},
\bauthor{\bsnm{Rennecke}, \binits{F.}},
\bauthor{\bsnm{Tripolt}, \binits{R.-A.}},
\bauthor{\bsnm{Smekal}, \binits{L.}},
\bauthor{\bsnm{Wambach}, \binits{J.}}:
\batitle{{In-Medium Spectral Functions of Vector- and Axial-Vector Mesons from
  the Functional Renormalization Group}}.
\bjtitle{Phys. Rev. D}
\bvolume{95}(\bissue{3}),
\bfpage{036020}
(\byear{2017})
\doiurl{10.1103/PhysRevD.95.036020}
{\href{https://arxiv.org/abs/1610.08754}{{arXiv:1610.08754}}}
{[hep-ph]}
\end{barticle}
\endbibitem

\bibitem[\protect\citeauthoryear{Rapp and van Hees}{2016}]{Rapp:2014hha}
\begin{barticle}
\bauthor{\bsnm{Rapp}, \binits{R.}},
\bauthor{\bsnm{Hees}, \binits{H.}}:
\batitle{{Thermal Dileptons as Fireball Thermometer and Chronometer}}.
\bjtitle{Phys. Lett. B}
\bvolume{753},
\bfpage{586}--\blpage{590}
(\byear{2016})
\doiurl{10.1016/j.physletb.2015.12.065}
{\href{https://arxiv.org/abs/1411.4612}{{arXiv:1411.4612}}}
{[hep-ph]}
\end{barticle}
\endbibitem

\bibitem[\protect\citeauthoryear{Arnaldi et~al.}{2009}]{NA60:2008dcb}
\begin{barticle}
\bauthor{\bsnm{Arnaldi}, \binits{R.}}, \betal:
\batitle{{Evidence for the production of thermal-like muon pairs with masses
  above 1-GeV/c**2 in 158-A-GeV Indium-Indium Collisions}}.
\bjtitle{Eur. Phys. J. C}
\bvolume{59},
\bfpage{607}--\blpage{623}
(\byear{2009})
\doiurl{10.1140/epjc/s10052-008-0857-2}
{\href{https://arxiv.org/abs/0810.3204}{{arXiv:0810.3204}}}
{[nucl-ex]}
\end{barticle}
\endbibitem

\bibitem[\protect\citeauthoryear{Adamczewski-Musch
  et~al.}{2019}]{HADES:2019auv}
\begin{barticle}
\bauthor{\bsnm{Adamczewski-Musch}, \binits{J.}}, \betal:
\batitle{{Probing dense baryon-rich matter with virtual photons}}.
\bjtitle{Nature Phys.}
\bvolume{15}(\bissue{10}),
\bfpage{1040}--\blpage{1045}
(\byear{2019})
\doiurl{10.1038/s41567-019-0583-8}
\end{barticle}
\endbibitem

\bibitem[\protect\citeauthoryear{Agakishiev et~al.}{2011}]{HADES:2011nqx}
\begin{barticle}
\bauthor{\bsnm{Agakishiev}, \binits{G.}}, \betal:
\batitle{{Dielectron production in Ar+KCl collisions at 1.76A GeV}}.
\bjtitle{Phys. Rev. C}
\bvolume{84},
\bfpage{014902}
(\byear{2011})
\doiurl{10.1103/PhysRevC.84.014902}
{\href{https://arxiv.org/abs/1103.0876}{{arXiv:1103.0876}}}
{[nucl-ex]}
\end{barticle}
\endbibitem

\bibitem[\protect\citeauthoryear{Fr{\"o}hlich et~al.}{2007}]{Frohlich:2007bi}
\begin{barticle}
\bauthor{\bsnm{Fr{\"o}hlich}, \binits{I.}}, \betal:
\batitle{{Pluto: A Monte Carlo Simulation Tool for Hadronic Physics}}.
\bjtitle{PoS}
\bvolume{ACAT},
\bfpage{076}
(\byear{2007})
\doiurl{10.22323/1.050.0076}
{\href{https://arxiv.org/abs/0708.2382}{{arXiv:0708.2382}}}
{[nucl-ex]}
\end{barticle}
\endbibitem

\bibitem[\protect\citeauthoryear{Galatyuk}{2009}]{Galatyuk:2009}
\begin{botherref}
\oauthor{\bsnm{Galatyuk}, \binits{T.}}:
{D}i-electron spectroscopy in {HADES} and {CBM}: from p+p and n+p collisions at
  {GSI} to {A}u+{A}u collisions at {FAIR}.
Phd thesis,
Johann Wolfgang Goethe Universität Frankfurt a. M.
(2009)
\end{botherref}
\endbibitem

\bibitem[\protect\citeauthoryear{Danielewicz and
  Odyniec}{1985}]{Danielewicz:1985hn}
\begin{barticle}
\bauthor{\bsnm{Danielewicz}, \binits{P.}},
\bauthor{\bsnm{Odyniec}, \binits{G.}}:
\batitle{{Transverse Momentum Analysis of Collective Motion in Relativistic
  Nuclear Collisions}}.
\bjtitle{Phys. Lett. B}
\bvolume{157},
\bfpage{146}--\blpage{150}
(\byear{1985})
\doiurl{10.1016/0370-2693(85)91535-7}
{\href{https://arxiv.org/abs/2109.05308}{{arXiv:2109.05308}}}
{[nucl-th]}
\end{barticle}
\endbibitem

\bibitem[\protect\citeauthoryear{Ollitrault}{1992}]{Ollitrault:1992bk}
\begin{barticle}
\bauthor{\bsnm{Ollitrault}, \binits{J.-Y.}}:
\batitle{{Anisotropy as a signature of transverse collective flow}}.
\bjtitle{Phys. Rev. D}
\bvolume{46},
\bfpage{229}--\blpage{245}
(\byear{1992})
\doiurl{10.1103/PhysRevD.46.229}
\end{barticle}
\endbibitem

\bibitem[\protect\citeauthoryear{Pinkenburg et~al.}{1999}]{E895:1999ldn}
\begin{barticle}
\bauthor{\bsnm{Pinkenburg}, \binits{C.}}, \betal:
\batitle{{Elliptic flow: Transition from out-of-plane to in-plane emission in
  Au + Au collisions}}.
\bjtitle{Phys. Rev. Lett.}
\bvolume{83},
\bfpage{1295}--\blpage{1298}
(\byear{1999})
\doiurl{10.1103/PhysRevLett.83.1295}
{\href{https://arxiv.org/abs/nucl-ex/9903010}{{arXiv:nucl-ex/9903010}}}
\end{barticle}
\endbibitem

\bibitem[\protect\citeauthoryear{Steinheimer
  et~al.}{2025}]{Steinheimer:2025trr}
\begin{botherref}
\oauthor{\bsnm{Steinheimer}, \binits{J.}},
\oauthor{\bsnm{Reichert}, \binits{T.}},
\oauthor{\bsnm{Bleicher}, \binits{M.}}:
{Multi-strange and charmed hadrons: A novel probe for the QCD equation of state
  at high baryon densities}
(2025)
{\href{https://arxiv.org/abs/2510.23353}{{arXiv:2510.23353}}}
{[nucl-th]}
\end{botherref}
\endbibitem

\bibitem[\protect\citeauthoryear{Borghini et~al.}{2001a}]{Borghini:2000sa}
\begin{barticle}
\bauthor{\bsnm{Borghini}, \binits{N.}},
\bauthor{\bsnm{Dinh}, \binits{P.M.}},
\bauthor{\bsnm{Ollitrault}, \binits{J.-Y.}}:
\batitle{{A New method for measuring azimuthal distributions in nucleus-nucleus
  collisions}}.
\bjtitle{Phys. Rev. C}
\bvolume{63},
\bfpage{054906}
(\byear{2001})
\doiurl{10.1103/PhysRevC.63.054906}
{\href{https://arxiv.org/abs/nucl-th/0007063}{{arXiv:nucl-th/0007063}}}
\end{barticle}
\endbibitem

\bibitem[\protect\citeauthoryear{Borghini et~al.}{2001b}]{Borghini:2001vi}
\begin{barticle}
\bauthor{\bsnm{Borghini}, \binits{N.}},
\bauthor{\bsnm{Dinh}, \binits{P.M.}},
\bauthor{\bsnm{Ollitrault}, \binits{J.-Y.}}:
\batitle{{Flow analysis from multiparticle azimuthal correlations}}.
\bjtitle{Phys. Rev. C}
\bvolume{64},
\bfpage{054901}
(\byear{2001})
\doiurl{10.1103/PhysRevC.64.054901}
{\href{https://arxiv.org/abs/nucl-th/0105040}{{arXiv:nucl-th/0105040}}}
\end{barticle}
\endbibitem

\bibitem[\protect\citeauthoryear{Bleicher et~al.}{1999}]{Bleicher:1999xi}
\begin{barticle}
\bauthor{\bsnm{Bleicher}, \binits{M.}}, \betal:
\batitle{{Relativistic hadron hadron collisions in the ultrarelativistic
  quantum molecular dynamics model}}.
\bjtitle{J. Phys. G}
\bvolume{25},
\bfpage{1859}--\blpage{1896}
(\byear{1999})
\doiurl{10.1088/0954-3899/25/9/308}
{\href{https://arxiv.org/abs/hep-ph/9909407}{{arXiv:hep-ph/9909407}}}
\end{barticle}
\endbibitem

\bibitem[\protect\citeauthoryear{Bilandzic et~al.}{2011}]{Bilandzic:2010jr}
\begin{barticle}
\bauthor{\bsnm{Bilandzic}, \binits{A.}},
\bauthor{\bsnm{Snellings}, \binits{R.}},
\bauthor{\bsnm{Voloshin}, \binits{S.}}:
\batitle{{Flow analysis with cumulants: Direct calculations}}.
\bjtitle{Phys. Rev. C}
\bvolume{83},
\bfpage{044913}
(\byear{2011})
\doiurl{10.1103/PhysRevC.83.044913}
{\href{https://arxiv.org/abs/1010.0233}{{arXiv:1010.0233}}}
{[nucl-ex]}
\end{barticle}
\endbibitem

\bibitem[\protect\citeauthoryear{Bilandzic et~al.}{2014}]{Bilandzic:2013kga}
\begin{barticle}
\bauthor{\bsnm{Bilandzic}, \binits{A.}},
\bauthor{\bsnm{Christensen}, \binits{C.H.}},
\bauthor{\bsnm{Gulbrandsen}, \binits{K.}},
\bauthor{\bsnm{Hansen}, \binits{A.}},
\bauthor{\bsnm{Zhou}, \binits{Y.}}:
\batitle{{Generic framework for anisotropic flow analyses with multiparticle
  azimuthal correlations}}.
\bjtitle{Phys. Rev. C}
\bvolume{89}(\bissue{6}),
\bfpage{064904}
(\byear{2014})
\doiurl{10.1103/PhysRevC.89.064904}
{\href{https://arxiv.org/abs/1312.3572}{{arXiv:1312.3572}}}
{[nucl-ex]}
\end{barticle}
\endbibitem

\bibitem[\protect\citeauthoryear{Lisa et~al.}{2005}]{LisaPratt10decades}
\begin{barticle}
\bauthor{\bsnm{Lisa}, \binits{M.}},
\bauthor{\bsnm{Pratt}, \binits{S.}},
\bauthor{\bsnm{Soltz}, \binits{R.}},
\bauthor{\bsnm{Wiedemann}, \binits{U.}}:
\batitle{{Femtoscopy in Relativistic Heavy Ion Collisions: Two Decades of
  Progress}}.
\bjtitle{Ann. Rev. Nucl. Part. Sci.}
\bvolume{55},
\bfpage{357}
(\byear{2005})
\doiurl{10.1146/annurev.nucl.55.090704.151533}
{\href{https://arxiv.org/abs/0505014}{{arXiv:0505014}}}
{[nucl-ex]}
\end{barticle}
\endbibitem

\bibitem[\protect\citeauthoryear{Lednicky and Lyuboshitz}{1982}]{Lednicky}
\begin{barticle}
\bauthor{\bsnm{Lednicky}, \binits{R.}},
\bauthor{\bsnm{Lyuboshitz}, \binits{V.L.}}:
\batitle{{Final State Interaction of Particles with Small Relative Momenta}}.
\bjtitle{Yad.Fiz.}
\bvolume{35},
\bfpage{1316}
(\byear{1982})
\end{barticle}
\endbibitem

\bibitem[\protect\citeauthoryear{Rustamov}{2023}]{Rustamov:2022hdi}
\begin{barticle}
\bauthor{\bsnm{Rustamov}, \binits{A.}}:
\batitle{{Deciphering the phases of QCD matter with fluctuations and
  correlations of conserved charges}}.
\bjtitle{EPJ Web Conf.}
\bvolume{276},
\bfpage{01007}
(\byear{2023})
\doiurl{10.1051/epjconf/202327601007}
{\href{https://arxiv.org/abs/2210.14810}{{arXiv:2210.14810}}}
{[hep-ph]}
\end{barticle}
\endbibitem

\bibitem[\protect\citeauthoryear{Braun-Munzinger
  et~al.}{2022}]{Braun-Munzinger:2022bkc}
\begin{botherref}
\oauthor{\bsnm{Braun-Munzinger}, \binits{P.}},
\oauthor{\bsnm{Rustamov}, \binits{A.}},
\oauthor{\bsnm{Stachel}, \binits{J.}}:
{QCD under extreme conditions}
(2022)
{\href{https://arxiv.org/abs/2211.08819}{{arXiv:2211.08819}}}
{[hep-ph]}
\end{botherref}
\endbibitem

\bibitem[\protect\citeauthoryear{Gross et~al.}{2023}]{Gross:2022hyw}
\begin{barticle}
\bauthor{\bsnm{Gross}, \binits{F.}}, \betal:
\batitle{{50 Years of Quantum Chromodynamics}}.
\bjtitle{Eur. Phys. J. C}
\bvolume{83},
\bfpage{1125}
(\byear{2023})
\doiurl{10.1140/epjc/s10052-023-11949-2}
{\href{https://arxiv.org/abs/2212.11107}{{arXiv:2212.11107}}}
{[hep-ph]}
\end{barticle}
\endbibitem

\bibitem[\protect\citeauthoryear{Stephanov et~al.}{1999}]{Stephanov:1999zu}
\begin{barticle}
\bauthor{\bsnm{Stephanov}, \binits{M.A.}},
\bauthor{\bsnm{Rajagopal}, \binits{K.}},
\bauthor{\bsnm{Shuryak}, \binits{E.V.}}:
\batitle{{Event-by-event fluctuations in heavy ion collisions and the QCD
  critical point}}.
\bjtitle{Phys. Rev. D}
\bvolume{60},
\bfpage{114028}
(\byear{1999})
\doiurl{10.1103/PhysRevD.60.114028}
{\href{https://arxiv.org/abs/hep-ph/9903292}{{arXiv:hep-ph/9903292}}}
\end{barticle}
\endbibitem

\bibitem[\protect\citeauthoryear{Stephanov}{2009}]{Stephanov:2008qz}
\begin{barticle}
\bauthor{\bsnm{Stephanov}, \binits{M.A.}}:
\batitle{{Non-Gaussian fluctuations near the QCD critical point}}.
\bjtitle{Phys. Rev. Lett.}
\bvolume{102},
\bfpage{032301}
(\byear{2009})
\doiurl{10.1103/PhysRevLett.102.032301}
{\href{https://arxiv.org/abs/0809.3450}{{arXiv:0809.3450}}}
{[hep-ph]}
\end{barticle}
\endbibitem

\bibitem[\protect\citeauthoryear{Aboona et~al.}{2025}]{STAR:2025zdq}
\begin{barticle}
\bauthor{\bsnm{Aboona}, \binits{B.E.}}, \betal:
\batitle{{Precision Measurement of Net-Proton-Number Fluctuations in Au+Au
  Collisions at RHIC}}.
\bjtitle{Phys. Rev. Lett.}
\bvolume{135}(\bissue{14}),
\bfpage{142301}
(\byear{2025})
\doiurl{10.1103/9l69-2d7p}
{\href{https://arxiv.org/abs/2504.00817}{{arXiv:2504.00817}}}
{[nucl-ex]}
\end{barticle}
\endbibitem

\bibitem[\protect\citeauthoryear{Nabroth}{2025}]{Nabroth:2025elh}
\begin{bchapter}
\bauthor{\bsnm{Nabroth}, \binits{M.}}:
\bctitle{{Cumulants of the multiplicity distributions of identified particles
  measured in heavy-ion collisions by HADES}}.
In: \bbtitle{{31st International Conference on Ultra-relativistic
  Nucleus-Nucleus Collisions}}
(\byear{2025})
\end{bchapter}
\endbibitem

\bibitem[\protect\citeauthoryear{Friman et~al.}{2025}]{Friman:2025swg}
\begin{botherref}
\oauthor{\bsnm{Friman}, \binits{B.}},
\oauthor{\bsnm{Redlich}, \binits{K.}},
\oauthor{\bsnm{Rustamov}, \binits{A.}}:
{Baselines for Abelian Charge Fluctuations in Nuclear Collisions:Theory and
  Comparison with Experimental Data}
(2025)
{\href{https://arxiv.org/abs/2508.18879}{{arXiv:2508.18879}}}
{[nucl-th]}
\end{botherref}
\endbibitem

\bibitem[\protect\citeauthoryear{Rustamov}{2021}]{Rustamov:2020ekv}
\begin{barticle}
\bauthor{\bsnm{Rustamov}, \binits{A.}}:
\batitle{{Overview of fluctuation and correlation measurements}}.
\bjtitle{Nucl. Phys. A}
\bvolume{1005},
\bfpage{121858}
(\byear{2021})
\doiurl{10.1016/j.nuclphysa.2020.121858}
{\href{https://arxiv.org/abs/2005.13398}{{arXiv:2005.13398}}}
{[nucl-ex]}
\end{barticle}
\endbibitem

\bibitem[\protect\citeauthoryear{Rustamov and
  Gorenstein}{2012}]{Rustamov:2012bx}
\begin{barticle}
\bauthor{\bsnm{Rustamov}, \binits{A.}},
\bauthor{\bsnm{Gorenstein}, \binits{M.I.}}:
\batitle{{Identity Method for Moments of Multiplicity Distribution}}.
\bjtitle{Phys. Rev. C}
\bvolume{86},
\bfpage{044906}
(\byear{2012})
\doiurl{10.1103/PhysRevC.86.044906}
{\href{https://arxiv.org/abs/1204.6632}{{arXiv:1204.6632}}}
{[nucl-th]}
\end{barticle}
\endbibitem

\bibitem[\protect\citeauthoryear{Arslandok and
  Rustamov}{2019}]{Arslandok:2018pcu}
\begin{barticle}
\bauthor{\bsnm{Arslandok}, \binits{M.}},
\bauthor{\bsnm{Rustamov}, \binits{A.}}:
\batitle{{TIdentity module for the reconstruction of the moments of
  multiplicity distributions}}.
\bjtitle{Nucl. Instrum. Meth. A}
\bvolume{946},
\bfpage{162622}
(\byear{2019})
\doiurl{10.1016/j.nima.2019.162622}
{\href{https://arxiv.org/abs/1807.06370}{{arXiv:1807.06370}}}
{[hep-ex]}
\end{barticle}
\endbibitem

\bibitem[\protect\citeauthoryear{Rustamov}{2024}]{Rustamov:2024hvq}
\begin{barticle}
\bauthor{\bsnm{Rustamov}, \binits{A.}}:
\batitle{{Fuzzy logic for reconstructing arbitrary moments of multiplicity
  distributions}}.
\bjtitle{Phys. Rev. C}
\bvolume{110}(\bissue{6}),
\bfpage{064910}
(\byear{2024})
\doiurl{10.1103/PhysRevC.110.064910}
{\href{https://arxiv.org/abs/2409.09814}{{arXiv:2409.09814}}}
{[nucl-th]}
\end{barticle}
\endbibitem

\bibitem[\protect\citeauthoryear{Skokov et~al.}{2013}]{Skokov:2012ds}
\begin{barticle}
\bauthor{\bsnm{Skokov}, \binits{V.}},
\bauthor{\bsnm{Friman}, \binits{B.}},
\bauthor{\bsnm{Redlich}, \binits{K.}}:
\batitle{{Volume Fluctuations and Higher Order Cumulants of the Net Baryon
  Number}}.
\bjtitle{Phys. Rev. C}
\bvolume{88},
\bfpage{034911}
(\byear{2013})
\doiurl{10.1103/PhysRevC.88.034911}
{\href{https://arxiv.org/abs/1205.4756}{{arXiv:1205.4756}}}
{[hep-ph]}
\end{barticle}
\endbibitem

\bibitem[\protect\citeauthoryear{Braun-Munzinger
  et~al.}{2017}]{Braun-Munzinger:2016yjz}
\begin{barticle}
\bauthor{\bsnm{Braun-Munzinger}, \binits{P.}},
\bauthor{\bsnm{Rustamov}, \binits{A.}},
\bauthor{\bsnm{Stachel}, \binits{J.}}:
\batitle{{Bridging the gap between event-by-event fluctuation measurements and
  theory predictions in relativistic nuclear collisions}}.
\bjtitle{Nucl. Phys. A}
\bvolume{960},
\bfpage{114}--\blpage{130}
(\byear{2017})
\doiurl{10.1016/j.nuclphysa.2017.01.011}
{\href{https://arxiv.org/abs/1612.00702}{{arXiv:1612.00702}}}
{[nucl-th]}
\end{barticle}
\endbibitem

\bibitem[\protect\citeauthoryear{Rustamov et~al.}{2023}]{Rustamov:2022sqm}
\begin{barticle}
\bauthor{\bsnm{Rustamov}, \binits{A.}},
\bauthor{\bsnm{Stroth}, \binits{J.}},
\bauthor{\bsnm{Holzmann}, \binits{R.}}:
\batitle{{A model-free procedure to correct for volume fluctuations in E-by-E
  analyses of particle multiplicities}}.
\bjtitle{Nucl. Phys. A}
\bvolume{1034},
\bfpage{122641}
(\byear{2023})
\doiurl{10.1016/j.nuclphysa.2023.122641}
{\href{https://arxiv.org/abs/2211.14849}{{arXiv:2211.14849}}}
{[nucl-th]}
\end{barticle}
\endbibitem

\bibitem[\protect\citeauthoryear{Holzmann et~al.}{2024}]{Holzmann:2024wyd}
\begin{barticle}
\bauthor{\bsnm{Holzmann}, \binits{R.}},
\bauthor{\bsnm{Koch}, \binits{V.}},
\bauthor{\bsnm{Rustamov}, \binits{A.}},
\bauthor{\bsnm{Stroth}, \binits{J.}}:
\batitle{{Controlling volume fluctuations for studies of critical phenomena in
  nuclear collisions}}.
\bjtitle{Nucl. Phys. A}
\bvolume{1050},
\bfpage{122924}
(\byear{2024})
\doiurl{10.1016/j.nuclphysa.2024.122924}
{\href{https://arxiv.org/abs/2403.03598}{{arXiv:2403.03598}}}
{[nucl-th]}
\end{barticle}
\endbibitem

\bibitem[\protect\citeauthoryear{Braun-Munzinger
  et~al.}{2021}]{Braun-Munzinger:2020jbk}
\begin{barticle}
\bauthor{\bsnm{Braun-Munzinger}, \binits{P.}},
\bauthor{\bsnm{Friman}, \binits{B.}},
\bauthor{\bsnm{Redlich}, \binits{K.}},
\bauthor{\bsnm{Rustamov}, \binits{A.}},
\bauthor{\bsnm{Stachel}, \binits{J.}}:
\batitle{{Relativistic nuclear collisions: Establishing a non-critical baseline
  for fluctuation measurements}}.
\bjtitle{Nucl. Phys. A}
\bvolume{1008},
\bfpage{122141}
(\byear{2021})
\doiurl{10.1016/j.nuclphysa.2021.122141}
{\href{https://arxiv.org/abs/2007.02463}{{arXiv:2007.02463}}}
{[nucl-th]}
\end{barticle}
\endbibitem

\bibitem[\protect\citeauthoryear{Braun-Munzinger
  et~al.}{2019}]{Braun-Munzinger:2018yru}
\begin{barticle}
\bauthor{\bsnm{Braun-Munzinger}, \binits{P.}},
\bauthor{\bsnm{Rustamov}, \binits{A.}},
\bauthor{\bsnm{Stachel}, \binits{J.}}:
\batitle{{Experimental results on fluctuations of conserved charges confronted
  with predictions from canonical thermodynamics}}.
\bjtitle{Nucl. Phys. A}
\bvolume{982},
\bfpage{307}--\blpage{310}
(\byear{2019})
\doiurl{10.1016/j.nuclphysa.2018.09.074}
{\href{https://arxiv.org/abs/1807.08927}{{arXiv:1807.08927}}}
{[nucl-th]}
\end{barticle}
\endbibitem

\bibitem[\protect\citeauthoryear{Braun-Munzinger
  et~al.}{2024}]{Braun-Munzinger:2023gsd}
\begin{barticle}
\bauthor{\bsnm{Braun-Munzinger}, \binits{P.}},
\bauthor{\bsnm{Redlich}, \binits{K.}},
\bauthor{\bsnm{Rustamov}, \binits{A.}},
\bauthor{\bsnm{Stachel}, \binits{J.}}:
\batitle{{The imprint of conservation laws on correlated particle production}}.
\bjtitle{JHEP}
\bvolume{08},
\bfpage{113}
(\byear{2024})
\doiurl{10.1007/JHEP08(2024)113}
{\href{https://arxiv.org/abs/2312.15534}{{arXiv:2312.15534}}}
{[nucl-th]}
\end{barticle}
\endbibitem

\bibitem[\protect\citeauthoryear{Bzdak and Koch}{2012}]{Bzdak:2012ab}
\begin{barticle}
\bauthor{\bsnm{Bzdak}, \binits{A.}},
\bauthor{\bsnm{Koch}, \binits{V.}}:
\batitle{{Acceptance corrections to net baryon and net charge cumulants}}.
\bjtitle{Phys. Rev. C}
\bvolume{86},
\bfpage{044904}
(\byear{2012})
\doiurl{10.1103/PhysRevC.86.044904}
{\href{https://arxiv.org/abs/1206.4286}{{arXiv:1206.4286}}}
{[nucl-th]}
\end{barticle}
\endbibitem

\bibitem[\protect\citeauthoryear{Nonaka et~al.}{2017}]{Nonaka:2017kko}
\begin{barticle}
\bauthor{\bsnm{Nonaka}, \binits{T.}},
\bauthor{\bsnm{Kitazawa}, \binits{M.}},
\bauthor{\bsnm{Esumi}, \binits{S.}}:
\batitle{{More efficient formulas for efficiency correction of cumulants and
  effect of using averaged efficiency}}.
\bjtitle{Phys. Rev. C}
\bvolume{95}(\bissue{6}),
\bfpage{064912}
(\byear{2017})
\doiurl{10.1103/PhysRevC.95.064912}
{\href{https://arxiv.org/abs/1702.07106}{{arXiv:1702.07106}}}
{[physics.data-an]}.
\bcomment{[Erratum: Phys.Rev.C 103, 029901 (2021)]}
\end{barticle}
\endbibitem

\bibitem[\protect\citeauthoryear{Nonaka et~al.}{2018}]{Nonaka:2018mgw}
\begin{barticle}
\bauthor{\bsnm{Nonaka}, \binits{T.}},
\bauthor{\bsnm{Kitazawa}, \binits{M.}},
\bauthor{\bsnm{Esumi}, \binits{S.}}:
\batitle{{A general procedure for detector{\textendash}response correction of
  higher order cumulants}}.
\bjtitle{Nucl. Instrum. Meth. A}
\bvolume{906},
\bfpage{10}--\blpage{17}
(\byear{2018})
\doiurl{10.1016/j.nima.2018.08.013}
{\href{https://arxiv.org/abs/1805.00279}{{arXiv:1805.00279}}}
{[physics.data-an]}
\end{barticle}
\endbibitem

\bibitem[\protect\citeauthoryear{Anticic et~al.}{2014}]{Anticic:2013htn}
\begin{barticle}
\bauthor{\bsnm{Anticic}, \binits{T.}}, \betal:
\batitle{{Phase-space dependence of particle-ratio fluctuations in Pb + Pb
  collisions from 20 A to 158 A GeV beam energy}}.
\bjtitle{Phys. Rev. C}
\bvolume{89}(\bissue{5}),
\bfpage{054902}
(\byear{2014})
\doiurl{10.1103/PhysRevC.89.054902}
{\href{https://arxiv.org/abs/1310.3428}{{arXiv:1310.3428}}}
{[nucl-ex]}
\end{barticle}
\endbibitem

\bibitem[\protect\citeauthoryear{Acharya et~al.}{2019}]{ALICE:2017jsh}
\begin{barticle}
\bauthor{\bsnm{Acharya}, \binits{S.}}, \betal:
\batitle{{Relative particle yield fluctuations in $\text{ Pb-Pb }$ collisions
  at $\sqrt{s_\mathrm{{NN}}} =2.76\hbox { TeV}$}}.
\bjtitle{Eur. Phys. J. C}
\bvolume{79}(\bissue{3}),
\bfpage{236}
(\byear{2019})
\doiurl{10.1140/epjc/s10052-019-6711-x}
{\href{https://arxiv.org/abs/1712.07929}{{arXiv:1712.07929}}}
{[nucl-ex]}
\end{barticle}
\endbibitem

\bibitem[\protect\citeauthoryear{Acharya et~al.}{2020}]{ALICE:2019nbs}
\begin{barticle}
\bauthor{\bsnm{Acharya}, \binits{S.}}, \betal:
\batitle{{Global baryon number conservation encoded in net-proton fluctuations
  measured in Pb{\textendash}Pb collisions at sNN=2.76 TeV}}.
\bjtitle{Phys. Lett. B}
\bvolume{807},
\bfpage{135564}
(\byear{2020})
\doiurl{10.1016/j.physletb.2020.135564}
{\href{https://arxiv.org/abs/1910.14396}{{arXiv:1910.14396}}}
{[nucl-ex]}
\end{barticle}
\endbibitem

\bibitem[\protect\citeauthoryear{Acharya et~al.}{2023}]{ALICE:2022xpf}
\begin{barticle}
\bauthor{\bsnm{Acharya}, \binits{S.}}, \betal:
\batitle{{Closing in on critical net-baryon fluctuations at LHC energies:
  Cumulants up to third order in Pb{\textendash}Pb collisions}}.
\bjtitle{Phys. Lett. B}
\bvolume{844},
\bfpage{137545}
(\byear{2023})
\doiurl{10.1016/j.physletb.2022.137545}
{\href{https://arxiv.org/abs/2206.03343}{{arXiv:2206.03343}}}
{[nucl-ex]}
\end{barticle}
\endbibitem

\bibitem[\protect\citeauthoryear{D'Agostini}{2010}]{DAgostini:2010hil}
\begin{botherref}
\oauthor{\bsnm{D'Agostini}, \binits{G.}}:
{Improved iterative Bayesian unfolding}
(2010)
{\href{https://arxiv.org/abs/1010.0632}{{arXiv:1010.0632}}}
{[physics.data-an]}
\end{botherref}
\endbibitem

\bibitem[\protect\citeauthoryear{Cornish and Fisher}{1938}]{CornishFisher1938}
\begin{barticle}
\bauthor{\bsnm{Cornish}, \binits{E.A.}},
\bauthor{\bsnm{Fisher}, \binits{R.A.}}:
\batitle{Moments and cumulants in the specification of distributions}.
\bjtitle{Revue de l'Institut International de Statistique / Review of the
  International Statistical Institute}
\bvolume{5}(\bissue{4}),
\bfpage{307}--\blpage{320}
(\byear{1938})
\doiurl{10.2307/1400905} .
\bcomment{JSTOR. Accessed 8 Dec. 2025}
\end{barticle}
\endbibitem

\end{thebibliography}

\clearpage

\nopagebreak

{\large \bf A The CBM Collaboration}

\vspace{3mm}

\noindent
A.~Agarwal$^{1}$,             
Z.~Ahammed$^{1}$,             
N.~Ahmad$^{2}$,               
L.J.~Ahrens$^{3}$,            
M.~Al-Turany$^{4}$,           
N.~Alam$^{2}$,                
F.~Alef$^{5,4}$,              
Jing An$^{4,6}$,              
J.~Andary$^{7}$,              
A.~Andronic$^{8}$
% ...

\vspace{3mm}
\noindent {$^{1}$Variable Energy Cyclotron Centre (VECC), Kolkata, India} \\
{$^{2}$Department of Physics, Aligarh Muslim University, Aligarh, India} \\
{$^{3}$Justus-Liebig-Universit\"{a}t Gie{\ss}en, Gie{\ss}en, Germany} \\
{$^{4}$GSI Helmholtzzentrum f\"{u}r Schwerionenforschung GmbH (GSI), Darmstadt, Germany} \\
{$^{5}$Institut f\"{u}r Kernphysik, Technische Universit\"{a}t Darmstadt, Darmstadt, Germany} \\
{$^{6}$College of Physical Science and Technology, Central China Normal University (CCNU), Wuhan, China} \\
{$^{7}$Institut f\"{u}r Kernphysik, Goethe-Universit\"{a}t Frankfurt, Frankfurt, Germany} \\
{$^{8}$Institut f\"{u}r Kernphysik, Universit\"{a}t M\"{u}nster, M\"{u}nster, Germany} \\
{$^{9}$Karlsruhe Institute of Technology (KIT), Karlsruhe, Germany} \\
{$^{10}$Atomic and Nuclear Physics Department, University of Bucharest, Bucharest, Romania} \\
{$^{11}$Frankfurt Institute for Advanced Studies, Goethe-Universit\"{a}t Frankfurt (FIAS), Frankfurt, Germany} \\
{$^{12}$Horia Hulubei National Institute of Physics and Nuclear Engineering (IFIN-HH), Bucharest, Romania} \\
{$^{13}$Department of Nuclear Physics, Taras Shevchenko National University of Kyiv, Kyiv, Ukraine} \\
{$^{14}$Department of Physics, University of Jammu, Jammu, India} \\
{$^{15}$Department of Physics, University of Kashmir, Srinagar, India} \\
{$^{16}$Department of Physics, Panjab University, Chandigarh, India} \\
{$^{17}$Nuclear and Radiation Physics Research Laboratory, Department of Physics, Gauhati University, Guwahati, India} \\
{$^{18}$Department of Physics and Department of Electronic Science, University of Calcutta, Kolkata, India} \\
{$^{19}$Department of Physics, Bose Institute, Kolkata, India} \\
{$^{20}$Marian Smoluchowski Institute of Physics, Jagiellonian University, Krak\'{o}w, Poland} \\
{$^{21}$Faculty of Physics, University of Warsaw, Warsaw, Poland} \\
{$^{22}$Czech Technical University in Prague (CTU), Prague, Czech Republic} \\
{$^{23}$High Energy Physics Department, Kiev Institute for Nuclear Research (KINR), Kyiv, Ukraine} \\
{$^{24}$Henryk Niewodnicza\'{n}ski Institute of Nuclear Physics Polish Academy of Sciences, Krak\'{o}w, Poland} \\
{$^{25}$Facility for Antiproton and Ion Research in Europe GmbH (FAIR), Darmstadt, Germany} \\
{$^{26}$E\"{o}tv\"{o}s Lor\'{a}nd University (ELTE), Budapest, Hungary} \\
{$^{27}$Fakult\"{a}t f\"{u}r Mathematik und Naturwissenschaften, Bergische Universit\"{a}t Wuppertal, Wuppertal, Germany} \\
{$^{28}$Physikalisches Institut, Universit\"{a}t Heidelberg, Heidelberg, Germany} \\
{$^{29}$College of Science, China Three Gorges University (CTGU), Yichang, China} \\
{$^{30}$Institut f\"{u}r Technische Informatik, Universit\"{a}t Heidelberg, Heidelberg, Germany} \\
{$^{31}$Institut f\"{u}r Experimentalphysik I, Ruhr-Universit\"{a}t Bochum, Bochum, Germany} \\
{$^{32}$Zuse Institute Berlin (ZIB), Berlin, Germany} \\
{$^{33}$Department of Engineering Physics, Tsinghua University, Beijing, China} \\
{$^{34}$Institute of Modern Physics, Chinese Academy of Sciences (IMP), Lanzhou, China} \\
{$^{35}$National Institute of Science Education and Research (NISER), Bhubaneswar, India} \\
{$^{36}$AGH University of Krak\'{o}w (AGH), Krak\'{o}w, Poland} \\
{$^{37}$Faculty of Physics, Warsaw University of Technology, Warsaw, Poland} \\
{$^{38}$Institute of Electronic Systems, Warsaw University of Technology, Warsaw, Poland} \\
{$^{39}$Nuclear Physics Institute of the Czech Academy of Sciences, \v{R}e\v{z}, Czech Republic} \\
{$^{40}$Department of Physics, Banaras Hindu University (BHU), Varanasi, India} \\
{$^{41}$Department of Modern Physics, University of Science \& Technology of China (USTC), Hefei, China} \\
{$^{42}$Institute of Physics, Bhubaneswar, India} \\
{$^{43}$Chongqing University, Chongqing, China} \\
{$^{44}$Indian Institute of Technology Madras (IITM), Chennai, India} \\
{$^{45}$Indian Institute of Technology Indore (IITI), Indore, India} \\
{$^{46}$Physikalisches Institut, Eberhard Karls Universit\"{a}t T\"{u}bingen, T\"{u}bingen, Germany} \\
{$^{47}$Indian Institute of Technology Kharagpur (IITKGP), Kharagpur, India} \\
{$^{48}$Institute for Particle and Nuclear Physics, HUN-REN Wigner RCP, Budapest, Hungary} \\
{$^{49}$Pusan National University (PNU), Busan, Korea} \\
{$^{50}$Physik Department, Technische Universität München, Munich, Germany} \\
{$^{51}$also: Helmholtz Research Academy Hesse for FAIR, Frankfurt, Germany} \\
{$^{52}$also: Institut f\"{u}r Kernphysik, Forschungszentrum J\"{u}lich, J\"{u}lich, Germany}

\end{document}